\documentclass[11pt, a4paper]{article}
\pdfoutput=1
\usepackage[T1]{fontenc}
\usepackage[utf8]{inputenc}
\usepackage[english]{babel}
\usepackage{jcappub}
\usepackage{natbib}
\usepackage{amsmath}
\usepackage{amssymb}
\usepackage{graphicx}
\usepackage[dvipsnames]{xcolor}
\usepackage{multirow}
\usepackage{tensor}
\usepackage{epstopdf}
\usepackage{siunitx}
\usepackage{grffile}
\usepackage[small]{caption}
\usepackage[normalem]{ulem}

\usepackage{cleveref}

\usepackage{url}
\usepackage{xspace}
\usepackage{subcaption}
\usepackage{bm}
\usepackage{braket}
\usepackage{amssymb}
\usepackage{amsmath}

\graphicspath{{./Immagini/}}

\title{The double Compton process in astrophysical plasmas}

\author[a,1]{Andrea Ravenni\note{Corresponding author.}}
\author[a]{and Jens Chluba}

\affiliation[a]{Jodrell Bank Centre for Astrophysics, School of Physics and Astronomy, \\\ The University of Manchester, Oxford Road, Manchester, M13 9PL, U.K.}

\emailAdd{Andrea.Ravenni@Manchester.ac.uk}
\emailAdd{Jens.Chluba@Manchester.ac.uk}

\newcommand{\pd}[2]{\frac{\partial{#1}}{\partial{#2}}}
\newcommand{\diff}{\mathop{}\!\mathrm{d}}
\newcommand\Diff[1]{\mathop{}\!\mathrm{d^#1}}

\renewcommand{\vec}[1]{\bm{#1}}
\newcommand{\vers}[1]{\vec{\hat{#1}}}

\newcommand{\Coulomb}{Coulomb\xspace}

\newcommand{\Lightman}{Lightman\xspace}
\newcommand{\Gould}{Gould\xspace}

\newcommand{\Thorne}{Thorne\xspace}

\newcommand{\Tg}{T_\gamma}

\newcommand{\sigmaT}{\sigma_\text{T}}

\newcommand{\Te}{T_\text{e}}
\newcommand{\me}{m_\text{e}}

\newcommand{\nume}{N_\text{e}}
\newcommand{\nument}{N^{\rm nt}_\text{e}}
\newcommand{\fe}{f_\text{e}}

\newcommand{\wmin}{w_i^\text{min}}

\newcommand{\wt}{w_i^\text{t}}
\newcommand{\wc}{w_i^\text{c}}
\newcommand{\wminCS}{w_\text{min}^\text{C}}
\newcommand{\mue}{\mu_{\text{e}0}}
\newcommand{\muuno}{\mu_{01}}
\newcommand{\mutwo}{\mu_{02}}
\newcommand{\phie}{\phi_\text{e}}
\newcommand{\phione}{\phi_1}
\newcommand{\phitwo}{\phi_2}

\newcommand{\omegazerocrit}{\omega_0^\text{crit}}

\newcommand{\omegamin}{\omega_\text{min}}

\makeatletter
\newcommand\ARwave{\bgroup \markoverwith{\lower3.5\p@\hbox{\sixly \textcolor{orange}{\char58}}}\ULon}
\font\sixly=lasy6 
\makeatother

\definecolor{light-gray}{gray}{0.70}

\newcommand{\ie}{{i.e.,}\xspace}

\newcommand{\expf}[1]{{{\rm e}^{#1}}}

\newcommand{\vech}[1]{\hat{\vec{#1}}}

\newcommand{\id}{{\,\rm d}}

\newcommand{\beq}{\begin{equation}}   %

\newcommand{\eeq}{\end{equation}}   %

\newcommand{\beqa}{\begin{eqnarray}}   %

\newcommand{\eeqa}{\end{eqnarray}}   %

\newcommand{\beal}{\begin{align}}

\newcommand{\enal}{\end{align}}

\newcommand{\bspl}{\begin{split}}

\newcommand{\espl}{\end{split}}

\newcommand{\bsub}{\begin{subequations}}

\newcommand{\esub}{\end{subequations}}

\newcommand{\bmulti}{\begin{multline}}   %

\newcommand{\beqm}{\begin{mathletters}}   %

\newcommand{\eeqm}{\end{mathletters}}   %

\newcommand{\Ne}{N_{\rm e}}

\newcommand{\The}{\theta_{\rm e}}

\newcommand{\Thg}{\theta_{\gamma}}

\newcommand{\sigT}{\sigma_{\rm T}}

\newcommand{\pot}[2]{#1 \times 10^{#2}}

\abstract{
We study the double Compton (DC) process for a wide range of particle energies, extending previous treatments well beyond the soft photon limit, employing both numerical and analytical methods. This allows us to investigate the physics of the DC process up to the highly relativistic regime relevant to electromagnetic particle cascades in the early Universe and photon-dominated astrophysical plasmas. Generalized exact analytic expressions for the DC emissivity in the soft photon limit are obtained.
These are compared to existing approximations, for the first time studying the ultra-relativistic regime. We also numerically integrate the full DC collision term calculating the DC emissivity at general particle energies. A careful treatment of DC infrared divergences inside astrophysical plasmas, including subtle effects related to the presence of stimulated DC emission, is discussed.
The obtained results can be efficiently represented using the code {\tt DCpack}, which also allows one to compute average emissivities for general incoming electron and photon distributions. 
This puts the modelling of the DC process inside astrophysical plasmas on a solid footing and should find applications in particular for computations of the cosmological thermalization problem in the early Universe.
}


\begin{document}
\maketitle
\flushbottom

\newpage

\section{Introduction}

The double Compton (DC) process is an $\mathcal{O}(\alpha^3)$ electron-photon interaction that can be thought of as a Compton scattering event associated with the production or destruction of an extra real photon \citep{Mandl52}. As such, DC scattering is physically similar to the Bremsstrahlung (BR) process, being the first order radiative correction to the \Coulomb interaction. Together, the DC and BR processes are responsible for the emission and absorption of photons in many astrophysical plasmas, playing a crucial role in controlling their number density. 

An important distinction between DC and BR is how their emissivity scales with the number density of and energy of the scattering particles. DC is enhanced for faster electrons, whereas BR is suppressed. Similarly, DC depends of the presence of seed photons together with free electrons to operate, while BR is present whenever there are free electrons and ions.
Consequently, DC becomes important at high temperatures and in photon-dominated plasmas \citep{Lightman1981, Thorne1981, Svensson:1984MNRAS.209..175S}.
Thanks to the large excess of photons over baryons in our Universe, at high enough redshift ($z > 3\times 10^5$ in the standard cosmological model), DC scattering is more likely to happen than BR emission \cite{Danese:1982A&A...107...39D}. The DC process is therefore highly relevant for the cosmological thermalization process in the early universe \cite{Danese:1982A&A...107...39D, Burigana:1991A&A...246...49B, Hu:1992dc, Chluba:2011hw}, and similarly, close to compact objects or broadly photon-dominated plasmas, e.g., found in gamma ray bursts and accretion flows, affecting the thermodynamics of these systems \cite{McKinney:2016voq}.

This paper aims at providing a quasi-exact description of the DC emission process, putting it on a solid footing for applications in cosmology and astrophysics. We develop the code {\tt DCpack}, which allows one to accurately represent the DC emissivity over a very wide range of photon and electron energies. Together with recent detailed works on Compton scattering \cite{Sarkar:2019har} and BR \cite{Chluba:2019ser}, this covers the most important processes relevant to the evolution of primordial spectral distortions \citep[e.g.,][]{Sunyaev:1970er, 1970CoASP...2...66S, Burigana:1991A&A...246...49B, Chluba:2011hw}, even under extreme conditions and temperatures.
For the cosmological thermalization problem, we are mainly interested in covering mildly relativistic energies (up to $\simeq 1$~MeV) that allow us to exactly describe the DC emissivity in all cosmological epochs up to well before the $\mu$-era, i.e., redshifts $z \lesssim 10^9$.
However, our general results describe interactions up to much higher energies, relevant to decaying particle showers, early-universe magnetic fields, primordial black holes and other non-standard energy injection processes. As we will see, it is actually in the latter cases that the exact solutions calculated here deviate the most from the approximations previously used in the literature.

The matrix element for the DC process was first derived by Mandl and Skyrme \cite{Mandl52}. As already noted there, a general analytic discussion of the properties of DC scattering cross section is difficult due to the large number of scattering angles involved, and a numerical study ultimately is more instructive. The improvements in the computing power make carrying out such a program possible today, as indeed anticipated by Mandl and Skyme \cite{Mandl52}. However, to build intuition it is useful to consider a wide range of limiting cases analytically. 

The cross section for close to forward scattering of the two emitted photons was first discussed by \cite{Mandl52} in relation to the infra-red catastrophe.
The {\it soft photon limit} cross section (see section~\ref{sec:DC_emissivity_soft} for more details) for resting electrons ($p=0$) was also first considered by \cite{Mandl52} and then used to derive the kinetic equation for the DC process, yielding the \Lightman-\Thorne approximation \citep{Lightman1981, Thorne1981}.
This was later extended to cases with $p>0$ but for $\omega_2\ll \omega_0, \omega_1 \wedge\omega_0, \omega_1 \ll 1$ \citep{ChlubaThesis, Chluba:2006kg} and then applied to the cosmological thermalization problem by~\citep{Chluba:2011hw}. 
Here we will provide new general soft photon limit expressions assuming only $\omega_2\ll 1$. 
Extensions beyond the soft photon limit were first analytically discussed by Gould \cite{Gould1984} for resting electrons and assuming $\omega_0\ll 1$, while allowing $\omega_2\simeq \omega_0$ (see section~\ref{sec:Gould_limit}). This was later numerically generalized to large $\omega_0$ and moving electrons in \cite{ChlubaThesis}, and will be further extended to ultra-relativistic energies here.

The paper is structured as follows.
In section \ref{sec:DC_KineticEquation} we define the double Compton collision term, and discuss some preliminary calculation and introduce some quantities that will be useful throughout the paper.
In particular we briefly review how to translate the formalism that employs the DC scattering matrix element to the DC cross section, by carrying out all integrals that can be performed analytically in the most general case.
We then derive a simplified expression for the moments of the DC scattering kernel, and finally set the stage to deal with the isotropic DC emissivity.
We approach the general case in a step by step manner:
in section \ref{sec:DC_emissivity_soft} we treat the DC scattering in the soft photon limit. We start by re-deriving the \Lightman-\Thorne emissivity \cite{Lightman1981, Thorne1981, ChlubaThesis} in our formalism, which we then generalize to a newly-found description of the soft photon emissivity for generic initial particle energies.
In section~\ref{sec:beyond_soft}, we extend the discussion beyond the soft photon limit, starting from the \Gould formula \cite{Gould1984, ChlubaThesis} to then describe the general behaviour of the exact DC collision term.
This is achieved by numerically integrating the exact DC cross section, employing the CUBA library \cite{Hahn:2004fe}.\footnote{\url{http://www.feynarts.de/cuba/}} 
In section~\ref{sec:IR_div}, we moreover study divergences related to next to leading order Compton scattering terms and DC scattering, showing both the successful explanation developed in the literature \cite{Brown:1952eu} and possibly overlooked shortcomings.
In section \ref{sec:emissivity} we present a procedure for separating emission and energy redistribution aspects due to DC scattering, discussing then at length the implication of the emission term in the scenarios most common in cosmology.
Finally we summarize our findings and draw our conclusion in section \ref{sec:conclusions}.

\section{Double Compton kinetic equation and emissivity}
\label{sec:DC_KineticEquation}
The double Compton scattering process appears as a first order radiative correction to the usual single Compton process, with an extra photon being produced in the collision. It is schematically described by \citep[see also][]{ChlubaThesis}
\bsub
\label{eq:reactions_DC}
\begin{align}
	e^-(P) + \gamma (K_0) 
	&\longleftrightarrow 
	e^-(P') + \gamma (K_1) + \gamma (K_2) \, ,
	\label{eq:reactions_DC_EP}
	\\
	e^-(P) + \gamma (K_2) 
	&\longleftrightarrow 
	e^-(P') + \gamma (K_0) + \gamma (K_1) \, ,
	\label{eq:reactions_DC_PP}
\end{align}
\esub
where $P' \equiv (\gamma', \vec{p}')$ and $K_i \equiv (\omega_i, \vec{k}_i)$ denote the four-momenta of electrons and photons respectively.%
\footnote{Energies and momenta will be expressed in units of $\me c^2$ and $\me c$ respectively. For electrons this implies, $\gamma=E/\me c^2=\sqrt{1+p^2}$ for its energy, where $\gamma$ is the Lorentz-factor and $p=p^{\rm phys}/\me c$ the electron momentum.
For photons, we use $\omega_i=h\nu_i/\me c^2$ to denote its energy, where $\nu_i$ is the photon frequency.}
Focusing on photon $\gamma(K_2)$, the first reaction in eq.~\eqref{eq:reactions_DC} describes the DC emission and absorption processes (EP), while in the second reaction $\gamma(K_2)$ plays the role of the scattering projectile particle (PP). 

The DC collision term, relevant to the time evolution of the photon occupation number, $n(\vec{k}_2)$, is then given by \cite[e.g.,][]{ChlubaThesis}
\begin{equation}
\begin{split}
	\left.
		\pd{n(\vec k_2)}{t}
	\right|_\text{DC}
	\equiv
	\frac{(2\pi)^4}{2\omega_2}
	\int 
	&
	\diff \Pi \diff \Pi' \diff \Gamma_0 \diff \Gamma_1
	\,\delta^{(4)}(P + K_0 - P' - K_1 - K_2) \,
\\
	&\qquad
	\times\left[
	|\mathcal{M}_{\gamma_0 \gamma_1 \gamma_2}|^2 \, 
	F_{\gamma_0 \gamma_1 \gamma_2}
	-
	\frac{1}{2}|\mathcal{M}_{\gamma_2 \gamma_1 \gamma_0}|^2 \, 
	F_{\gamma_2 \gamma_1 \gamma_0}
	\right]	,
\end{split}
\label{eq:dn2dt_start}
\end{equation}
where the phase space differential volumes are defined as
\begin{equation}
	\diff \Gamma_i
	\equiv
	\frac{\Diff{3} \vec{k}_i}{(2\pi)^3 \, 2\omega_i} \, ,
\qquad
	\diff \Pi 
	\equiv
	\frac{\Diff{3} \vec{p}}{(2\pi)^3 \, 2\gamma} \, ,
	\qquad
	\diff \Pi' 
	\equiv
	\frac{\Diff{3} \vec{p}'}{(2\pi)^3 \, 2\gamma'} \, .
\end{equation}
The double Compton S-matrix element, $\mathcal{M}_{\gamma_0 \gamma_1 \gamma_2}$, is given in Appendix~\ref{appendix:M2} and $F_{\gamma_i \gamma_1 \gamma_j}$ is the statistical factor of the process:
\begin{equation}
	F_{\gamma_i \gamma_1 \gamma_j}
	\equiv
	\fe n_i (1 + n_1) (1 + n_j) 
	-
	\fe' n_1 n_j (1 + n_i)
	\, ,
\end{equation}
where $\fe=\fe(\vec{p})$, $\fe=\fe(\vec{p}')$ and $n_i=n(\vec{k}_i)$ denote the electron ($\fe$) and photon ($n$) distribution functions.
Note that
we omitted the negligible Pauli blocking,
and that
the order of the photon labels in $F$ and $\mathcal{M}$ is important to distinguish the two reactions in eq.~\eqref{eq:reactions_DC}.

We point out that in full equilibrium necessarily $F_{\gamma_i \gamma_1 \gamma_j}=0$, and this relation can only be satisfied if the temperatures of photons and electrons match and the photons have a Planck distribution.
If Compton scattering were the only interaction among electrons and photons, the distributions of both would relax to the same \emph{Compton collision} temperature;
however, the photons attain a non-vanishing constant chemical potential, which prevents the DC statistical factor from vanishing.
Only through number changing interaction such as DC and BR the chemical potential can be erased, reaching full equilibrium \citep[e.g.,][]{Burigana:1991A&A...246...49B, Hu:1992dc}. 

One can further simplify the expressions by explicitating the differential DC cross section.
With the DC matrix element $X$ defined in Appendix~\ref{appendix:M2}, it is given by \citep[cf.][eq. 11-38]{Jauch1976} 
\begin{equation}
\label{eq:dsig_DC}
	\frac{\!\id^8 \sigma_{\rm DC}^{\gamma_0\gamma_1\gamma_2}}{\id^8\Lambda}
	=
	\frac{\alpha\,r_0^2}{(4\pi)^2} \,
	\frac{\omega_1\,\omega_2}{g_{\text\o 0} \gamma \,\gamma'\, \omega_0} \,
		\frac{X}{\left|
		\pd{(\omega_1 + \gamma')}{\omega_1}
	\right|} 
	=
	\frac{\alpha\,r_0^2}{(4\pi)^2} \,
	\frac{\omega_1\,\omega_2}{g_{\text\o 0} \gamma \, \omega_0} \,
	\frac{X}{\lambda_1+\omega_0 \alpha_{01}-\omega_2\alpha_{12}}
\end{equation}
for the reaction $e+\gamma_0 \leftrightarrow e'+ \gamma_1+\gamma_2$. We introduced $\id^8 \Lambda =\!\id^3 \vec{p} \id^3 \vec{k}_0 \id^2 \hat{\vec{k}}_1$ as shorthand for the total differential.
Also, $\alpha_{ij}= \hat{K}_i \cdot \hat{K}_j=1-\vech{k}_i\cdot \vech{k}_j=1-\mu_{ij}$ and $\lambda_i=P \cdot \hat{K}_i=\gamma-p\mu_{{\rm e}i}$ with $\mu_{\text e i} = \vers p \cdot \vers k_i$, and we introduced the {\it M{\o}ller relative speed}, $g_{\text\o i}=\hat{P}\cdot\hat{K}_i=\lambda_i/\gamma =1-\beta\,\mu_{\rm e\it i}$, of the incident electron and photon, with the dimensionless electron speed $\beta=|\vec{v}|/c=p/\gamma$.%
\footnote{Note that in the following an additional hat above 3- and 4- vectors indicates that they are normalized to the time-like coordinate of the corresponding 4-vector.}
The differential, $\partial_{\omega_1}(\omega_1 + \gamma') = \gamma'^{-1}
(P + K_0 - K_2)\cdot \hat{K}_1$ appears after eliminating the Dirac $\delta$-function.
Also, $\alpha=e^2/4\pi\approx 1/137$ denotes the fine structure constant and $r_0=\alpha/\me c^2\approx \SI{2.82e-13}{cm}$ is the classical electron radius.

With this definition, the kinetic equation \eqref{eq:dn2dt_start} 
can be integrated over $\diff \Pi'$ and $\id\omega_1$,
resulting in \cite[see also][]{ChlubaThesis}:
\begin{equation}
	\left.
		\pd{n(\vec k_2)}{t}
	\right|_\text{DC}
	\equiv
	\frac{1}{\omega^2_2}
	\!\int\! 
	\frac{\Diff{3} \vec{p}}{(2\pi)^3}
	\Diff{3} \vec{k}_0 
	\Diff{2} \vers k_1
	\!\left[
	g_{\text\o 0}
	\frac{\!\id^8 \sigma_{\rm DC}^{\gamma_0\gamma_1\gamma_2}}{\id^8 \Lambda}
	F_{\gamma_0 \gamma_1 \gamma_2}
	-
	\frac{\omega_2^2}{\omega_0^2}\frac{g_{\text\o 2}}{2}
	\frac{\!\id^8 \sigma_{\rm DC}^{\gamma_2\gamma_1\gamma_0}}{\id^8 \Lambda} 
	F_{\gamma_2 \gamma_1 \gamma_0}
	\right] .
\label{eq:dn2dt}
\end{equation}
Hereafter, $\vec{p}'$ and $\omega_1$ are determined by
\bsub
\label{eq:Nu1EnergyReplacement}
\begin{align}
\vec p' &= \vec p + \vec k_0 - \vec k_1 - \vec k_2 \, ,
\\[2mm]
\omega_1
&=\frac{P\cdot K_0 - P\cdot K_2 - K_0\cdot K_2}
{(P + K_0 - K_2)\cdot \hat{K}_1}
=\frac{\lambda_0\omega_0-\lambda_{2}\omega_2-\omega_0\omega_2\alpha_{02}}
{\lambda_1+\omega_0\alpha_{01}-\omega_2\alpha_{12}} \, ,
\end{align}
\esub
that simply reflect energy and momentum conservation.

It will prove useful in the following to assign some labels to the various terms that appear in eq. \eqref{eq:dn2dt_start} according to the scheme in eq. \eqref{eq:reactions_DC}:
reaction evolving from left to right will be labelled as {\it forward} processes ($\rightarrow$), and the opposite will be {\it backward} processes ($\leftarrow$).
Furthermore, in events relative to eq. \eqref{eq:reactions_DC_EP}, the tracked photon $\gamma_2$ is the emitted particle (EP), whereas in eq. \eqref{eq:reactions_DC_PP} it is the projectile particle (PP).
In practice we have
\begin{equation}
	\left.
		\pd{n(\vec k_2)}{t}
	\right|_\text{DC}
	\equiv
	\left.
		\pd{n(\vec k_2)}{t}
	\right|_\text{DC} ^{\text{EP }\rightarrow}
	-
	\left.
		\pd{n(\vec k_2)}{t}
	\right|_\text{DC} ^{\text{EP } \leftarrow}
	-
	\left.
		\pd{n(\vec k_2)}{t}
	\right|_\text{DC} ^{\text{PP } \rightarrow}
	+
	\left.
		\pd{n(\vec k_2)}{t}
	\right|_\text{DC} ^{\text{PP } \leftarrow} ,
\label{eq:dn2dt_split_terms}
\end{equation}
where we introduced the individual terms
\begin{alignat}{2}
	\pd{n(\vec k_2)}{t}
	&
	\bigg|_\text{DC} ^{\text{EP } \rightarrow}
	&&	
	\equiv
	\frac{1}{\omega^2_2}
	\int 
	\frac{\Diff{3} \vec{p}}{(2\pi)^3}
	\Diff{3} \vec{k}_0 
	\Diff{2} \vers k_1
	g_{\text\o 0}
	\frac{\!\id^8 \sigma_{\rm DC}^{\gamma_0\gamma_1\gamma_2}}{\id^8 \Lambda}
	\fe n_0 (1 + n_1) (1 + n_2)  \, ,
\label{eq:dn2dt_EP_frw}
\\
	\pd{n(\vec k_2)}{t}
	&
	\bigg|_\text{DC} ^{\text{EP } \leftarrow}
	&&
	\equiv
	\frac{1}{\omega^2_2}
	\int 
	\frac{\Diff{3} \vec{p}}{(2\pi)^3}
	\Diff{3} \vec{k}_0 
	\Diff{2} \vers k_1
	g_{\text\o 0}
	\frac{\!\id^8 \sigma_{\rm DC}^{\gamma_0\gamma_1\gamma_2}}{\id^8 \Lambda}
	\fe' n_1 n_2 (1 + n_0) \, ,
\label{eq:dn2dt_EP_bkw}
\\
	\pd{n(\vec k_2)}{t}
	&
	\bigg|_\text{DC} ^{\text{PP } \rightarrow}
	&&
	\equiv
	\frac{1}{2}
	\int 
	\frac{\Diff{3} \vec{p}}{(2\pi)^3}
	\Diff{3} \vec{k}_0 
	\Diff{2} \vers k_1
	\frac{g_{\text\o 2}}{\omega_0^2}
	\frac{\!\id^8 \sigma_{\rm DC}^{\gamma_2\gamma_1\gamma_0}}{\id^8 \Lambda}
	\fe n_2 (1 + n_0) (1 + n_1) \, ,
\label{eq:dn2dt_PP_frw}
\\
	\pd{n(\vec k_2)}{t}
	&
	\bigg|_\text{DC} ^{\text{PP } \leftarrow}
	&&
	\equiv
	\frac{1}{2}
	\int 
	\frac{\Diff{3} \vec{p}}{(2\pi)^3}
	\Diff{3} \vec{k}_0 
	\Diff{2} \vers k_1
	\frac{g_{\text\o 2}}{\omega_0^2}
	\frac{\!\id^8 \sigma_{\rm DC}^{\gamma_2\gamma_1\gamma_0}}{\id^8 \Lambda}
	\fe' n_0 n_1 (1 + n_2) \, .
\label{eq:dn2dt_PP_bkw}
\end{alignat}
Studying the DC collision terms in all generality is quite difficult.
However, we can simplify the situation by assuming that the particle distributions are isotropic. Then all reactions only affect the spectrum of the average photon distribution, $n(\omega)=\int n(\vec{k})\frac{\Diff{2} \vers k}{4\pi}$ and the average electron momentum distribution, $\fe(p)=\int \fe(\vec{p})\frac{\Diff{2} \vers p}{4\pi}$.
Caveats related to the infrared divergence of the DC process also deserve additional attention, as we explain in section~\ref{sec:IR_div}, however, we shall gloss over them until then.

\subsection{Moments of the DC collision term}
If we are interested in the net photon production rate the expressions simplify noticeably,
because we do not need to treat the EP and PP contributions in eq. \eqref{eq:dn2dt_EP_frw} to \eqref{eq:dn2dt_PP_bkw} separately.
The net photon production rate is defined by the zeroth moment of the DC collision term
\begin{equation}
	\pd{N_\text{DC}}{t}
	\equiv
	\int 
	\frac{\Diff{3} \vec k_2}{(2 \pi)^3}	
	\left.
		\pd{n(\vec k_2)}{t}
	\right|_\text{DC} \, 
	\, .
\label{eq:dNdt}
\end{equation}	
Inserting eq. \eqref{eq:dn2dt} and renaming $\vec{k}_0\rightarrow \tilde{\vec{k}}_2$, $\vec{k}_2\rightarrow \tilde{\vec{k}}_0$ in the second term in the square brackets, after dropping the tildes, yields
\begin{equation}
	\pd{N_\text{DC}}{t}
	=
	\frac{1}{2}
	\int 
	\frac{\Diff{3} \vec k_2}{(2 \pi)^3}
	\left[
		\left.
			\pd{n(\vec k_2)}{t}
		\right|_\text{DC} ^\text{EP $\rightarrow$}
		-
		\left.
			\pd{n(\vec k_2)}{t}
		\right|_\text{DC} ^\text{EP $\leftarrow$}
	\right]
	.
\label{eq:dN_drho_DC}
\end{equation}
In contrast to single Compton scattering, the DC process leads to $\partial_t N_\text{DC}\neq 0$, unless full thermal equilibrium is reached.
In that case, we have $F_{\gamma_0 \gamma_1 \gamma_2}=0$, and therefore no evolution due to the DC scattering occurs.
The overall 1/2 factor derives from the fact that in each collision two photons are created, but also one photon is destroyed: the number of {\it additional} photons is thus half of the emitted ones. Below, we can avoid needing this factor by only considering the emitting photon contribution.

For completeness, we point out that the same procedure can be applied to the DC moments, $\mathcal{M}_k$, of any order $k$:
\begin{equation}
	\mathcal{M}_k
	\equiv
	\int 
	\frac{\Diff{3} \vec k_2}{(2 \pi)^3} \,
	\omega_2^k
	\left.
		\pd{n(\vec k_2)}{t}
	\right|_\text{DC}
	\, ,
\end{equation}	
which after some rearrangements can be rewritten as
\begin{equation}
	\mathcal{M}_k
	=
	\frac{1}{2}
	\int 
	\frac{\Diff{3} \vec k_2}{(2 \pi)^3}
	\frac{\Diff{3} \vec{p}}{(2\pi)^3}
	\Diff{3} \vec{k}_0 
	\Diff{2} \vers k_1
	\left(2\omega^k_2-\omega_0^k\right)
\frac{g_{\text\o 0}}{\omega^2_2}
	\frac{\!\id^8 \sigma_{\rm DC}^{\gamma_0\gamma_1\gamma_2}}{\id^8 \Lambda}
    F_{\gamma_0 \gamma_1 \gamma_2}\,	.
\label{eq:moments}
\end{equation}
In particular, to the net energy transfer, $\mathcal{M}_1$, is given by
\begin{equation}
	\pd{\rho_\text{DC}}{t}
	=
	\frac{1}{2}
	\int 
	\frac{\Diff{3} \vec k_2}{(2 \pi)^3}
	\frac{\Diff{3} \vec{p}}{(2\pi)^3}
	\Diff{3} \vec{k}_0 
	\Diff{2} \vers k_1
	\left(2\omega_2-\omega_0\right)
\frac{g_{\text\o 0}}{\omega^2_2}
	\frac{\!\id^8 \sigma_{\rm DC}^{\gamma_0\gamma_1\gamma_2}}{\id^8 \Lambda}
    F_{\gamma_0 \gamma_1 \gamma_2}\,	.
\end{equation}
In the present paper, we will not discuss the contributions of the double Compton scattering to the energy transfer in more detail, leaving it to future work.
We anticipate, however, that expressing the moment as in eq. \eqref{eq:moments} simplifies their calculation, as it will prove useful in writing refined evolution equations using a Fokker-Planck approach.

\subsection{Isotropic DC emission without stimulated terms}
To discuss the phenomenology of the DC emissivity, it is very helpful to consider the emitted particle forward scattering, eq. \eqref{eq:dn2dt_EP_frw}, separately:
the other processes, as long as stimulated emission is negligible, can be obtained by remapping variables. In addition, they can only have significance if a bath of ambient photons spanning various energies is already present.

For isotropic media, $\fe(\vec{p})=\fe(p)$ and $n(\vec{k})=n(\omega)$, such that we can further simplify the expression. Neglecting stimulated DC emission, we find
\begin{equation}
	\left.
		\pd{n(\omega_2)}{t}
	\right|_\text{DC}^\text{EP $\rightarrow$}
	\equiv
	\frac{1}{\omega^2_2}
	\int
	\frac{p^2 \id p}{2\pi^2}\,
	\fe(p)
	\int
	\omega_0^2 \id \omega_0 \,
	n(\omega_0)
	\int
	\frac{\Diff{2} \vers p}{4\pi}
	\,\Diff{2} \vers k_0
	\,\Diff{2} \vers k_1
	\,g_{\text\o 0}\,
	\frac{\!\id^8 \sigma_{\rm DC}^{\gamma_0\gamma_1\gamma_2}}{\id^8 \Lambda}\, .
\label{eq:dn2dt_EP_frw_no_stim}
\end{equation}
If we furthermore assume that both the incident electrons and photons are monoenergetic, we can next insert
\begin{equation}
	\fe(p) = 
	2 \pi^2 \nume
	\frac{ \delta^{(1)} (p- \bar{p})}{p^2} \, ,
	\quad
	n(\omega_0) = 
	2 \pi^2 N_0
	\frac{\delta^{(1)}(\omega_0 - \bar{\omega})}{\omega_0^2} \, ,
\label{eq:isotropic_monoenergetic_distribution}
\end{equation}
normalized such that $\Ne=\int \frac{\id^3 \vec p}{(2\pi)^3} f_{\rm e}(\vec{p})$ and $N_0=\int \frac{\id^3 \vec k}{(2\pi)^3} n(\vec{k})$, into eq. \eqref{eq:dn2dt_EP_frw_no_stim}, which yields
\begin{equation}
\begin{split}
	\left.
		\pd{n(\omega_2)}{t}
	\right|_\text{DC}^\text{m}
	\equiv
	2 \pi^2\,\frac{\Ne N_0}{\omega^2_2}
	\int
	\frac{\Diff{2} \vers p}{4\pi}
	\,\Diff{2} \vers k_0
	\,\Diff{2} \vers k_1
	\,g_{\text\o 0}\,
	\frac{\!\id^6 \sigma_{\rm DC}^{\gamma_0\gamma_1\gamma_2}}{\id^6 \Lambda}\, .
\end{split}
\label{eq:dn2dt_em_m}
\end{equation}
Since for isotropic incident particle distributions the DC emission is also isotropic we are free to choose any of the particles as a reference.
In some cases discussed below, it is most convenient to align the $z$-axis with the incident electron and then carry out all the angle averages.
In this case, $\Diff{2} \vers p\rightarrow \Diff{2} \vers k_2$ without changing the final result.
This can also be thought of as additionally averaging the DC emission rate over $\Diff{2} \vers k_2$ and then dividing by $4\pi$.
With this perspective, one can always chose the best reference to reduce the dimensionality of the problem.
We also note that, although there are formally 6 remaining integrals in eq.~\eqref{eq:dn2dt_em_m}, one of the azimuthal integrals always becomes trivial, leading to a factor of $2\pi$ due to symmetries of the scattering process.
One is thus left with 5 integral over angles, that generally have to be solved numerically.

\section{DC emissivity in the soft photon limit 
\texorpdfstring{($\omega_2 \ll \omega_0, \omega_1$)}{o2<o0,o1}
}
\label{sec:DC_emissivity_soft}
Of particular importance in physical applications of the DC process is the {\it soft photon limit}, in which $\gamma(K_2)$ is very {\it soft} compared to $\gamma(K_0)$ and $\gamma(K_1)$.%
\footnote{Alternatively one could also choose $K_1$ to be the soft photon without loss of generality.}
In this regime, $\gamma(K_1)$ can be thought of as a scattered Compton photon, while $\gamma(K_2)$ is produced well outside the energy regime accessible by single Compton scattering.
For $p=0$, the DC cross section for the reaction $e+\gamma_0\rightarrow e'+ \gamma_1+\gamma_2$ factors into the usual Compton cross section $\sigma_\text{C}$ and a soft photon modulation factor \cite{Jost1947, Mandl52, Jauch1976}.
Similarly, the expression in the lab frame (where $p>0$) can be found as 
\begin{equation}
\label{eq:soft_sigma}
\frac{\!\id^8 \sigma^{\rm soft}_{\rm DC}}{\id^8\Lambda}
= \frac{\alpha}{4\pi^2 \omega_2}
\frac{\!\id^5 \sigma_{\rm C}}{\id^5\Lambda_{\rm C}}
\left[
\frac{2(1+\lambda_0\omega_0-\lambda_1\omega_1^{\rm C})}{\lambda_2 \mathcal{T}_2}-\frac{1}{\mathcal{T}_2^2}-\frac{1}{\lambda_2^2}
\right],
\end{equation}
where
$
	\mathcal{T}_2=\lambda_2+\omega_0\alpha_{02}-\omega_1^{\rm C}\alpha_{12}
$
and $\omega^{\rm C}_1=\frac{\lambda_0 \omega_0}{\lambda_1+\omega_0 \alpha_1}$ is the scattered photon energy in the single Compton limit (i.e., setting $\omega_2=0$).
We furthermore defined the Compton scattering differential cross section as \cite{Jauch1976}
\begin{equation}
\frac{\!\id^5 \sigma_{\rm C}}{\id^5\Lambda_{\rm C}}
= \frac{3\sigT}{16\pi}\left[\frac{\omega^{\rm C}_1}{\lambda_0\omega_0}\right]^2
\left[
\frac{\lambda_1\omega^{\rm C}_1}{\lambda_0\omega_0}+\frac{\lambda_0\omega_0}{\lambda_1\omega^{\rm C}_1}
-\left(\frac{2}{\lambda_1\omega^{\rm C}_1}-\frac{2}{\lambda_0\omega_0}\right)
+\left(\frac{1}{\lambda_1\omega^{\rm C}_1}-\frac{1}{\lambda_0\omega_0}\right)^2
\right],
\label{eq:sigma_Compton}
\end{equation}
with $\id^5\Lambda_{\rm C}= \!\id^3 \vec{p}\id^2 \hat{\vec{k}}_1$.
We remark that so far the only approximation used to derive eq. \eqref{eq:soft_sigma} is $\omega_2\ll \omega_0, \omega_1$.

\subsection{Lightman-Thorne approximation 
\texorpdfstring{($p=0 \wedge \omega_0 \ll 1$)}{p=0 and o0 < 1}
}
From eq.~\eqref{eq:soft_sigma} it is easy to obtain the \Lightman-Thorne approximation for the DC emissivity \cite{Lightman1981, Thorne1981}.
First, we should assume resting electrons ($p=0 \leftrightarrow \lambda_i=1$) and then only keep terms to the lowest order in $\omega_0\ll 1$. The soft photon cross section then reduces to \cite{Lightman1981, Thorne1981, ChlubaThesis, Chluba:2006kg}
\begin{equation}
\label{eq:L_sigma}
	\left.
	\frac{\!\id^8 \sigma^{\rm soft}_{\rm DC}}{\!\id^8\Lambda}
	\right|_\text{L}
	=
	\frac{\alpha \, r_0^2}{4 \pi^2}
	\,\frac{\omega_0^2}{\omega_2}\,
	(1+ \muuno^2)
	\left[1-\muuno-\frac{(\mu_{12} - \mutwo)^2}{2} \right].
\end{equation}
Using the relation $\mu_{01}=\mu_{02}\mu_{12}+\cos(\phi_{02}-\phi_{12})(1-\mu_{02}^2)^{1/2}(1-\mu_{12}^2)^{1/2}$, after integrating
over all angles, we obtain the usual \Lightman-\Thorne result for the DC emission spectrum of resting electrons and soft, monochromatic and isotropic incident photons (without stimulated scattering) \cite{ChlubaThesis}
\begin{equation}
	\left.\pd{n(\omega_2)}{t}\right|^{\rm em}_\text{L}
	= 
	\int \frac{\id^3 \vec p}{(2\pi)^3}\id^3 \vec{k}_0 \id^2 \hat{\vec{k}}_1\, 
	g_{\text\o 0}\,\frac{\!\id^8 \sigma^{\rm soft}_{\rm DC}}{\!\id^8\Lambda} \,f_{\rm e}(\vec{p}) \,n(\vec{k}_0)
	=
	2\pi^2 \,
	\frac{4 \alpha}{3 \pi} 
	\,
	\sigmaT \,
	\nume
	N_0\, 
	\frac{\omega_0^2}{\omega_2^3}
	\, .
\end{equation}
This result can also be obtained directly from eq.~\eqref{eq:dn2dt_em_m} after inserting eq.~\eqref{eq:L_sigma}.
For clarity, we point out that the prefactor $2 \pi^2$ comes from that we are considering isotropic monoenergetic photons (eq. \eqref{eq:isotropic_monoenergetic_distribution} and discussion around it).%
\footnote{Moreover, we remind the reader that the classic \Lightman-\Thorne scattering rate was expressed as
$
	\left.\pd{n(x_2)}{t}\right|^{\rm em}_\text{L}
	=
	\frac{4 \alpha}{3 \pi} 
	\,
	\sigmaT \,
	\nume
	\frac{\theta^2}{x_2^3}
	[1-n(x_2)(e^{x_2}-1)]
	\int \diff x \, x^4 n(x)[1+n(x)]
$, using the quantities we will introduce later on in section \ref{sec:emissivity}.
Such expression was derived in the original papers following the same step we discussed before,
assuming thermally distributed photons and electrons, and assuming that $\omega_2\ll \omega_0$ implies $\omega_1 \approx \omega_0$ as can be inferred from the last integral.
}

\subsection{Lightman-Thorne approximation for moving electrons
\texorpdfstring{($p\geq 0 \wedge \omega_0 \ll 1$)}{p>0 and o0 < 1}
}
Still in the limit of $\omega_2 \ll \omega_0, \omega_1$ we can generalize the previous result to take into account initial electrons with general isotropic momentum distribution. Starting from eq.~\eqref{eq:soft_sigma} and keeping only lowest order terms in $\omega_0\ll1$ (i.e., $\mathcal{O}( \omega_0^2)$) but allowing $p>0$, we obtain \cite{ChlubaThesis, Chluba:2006kg}
\begin{equation}
	\left.
	\frac{\!\id^8 \sigma^{\rm soft}_{\rm DC}}{\!\id^8\Lambda}
	\right|_\text{L}^\text{mov}
	=
	\frac{\alpha  r_0^2}{4\pi^2}
	\frac{\omega_0^2}{\omega_2}
	\Big[
		2\lambda_{0} \lambda_{1}
		\left(
			\lambda_{0} \lambda_{1}
			- \alpha_{01} 
		\right)
		+ \alpha_{01}^2
	\Big]
	\frac{
		  \lambda_{0} \lambda_{1} \lambda_{2}^2 \alpha_{01} 
		-\frac{1}{2}
		\left(
			 \lambda_{1} \alpha_{02}
			 - \lambda_{0} \alpha_{12}
		\right)^2
	}{
		\lambda^2_{0} \lambda_{1}^6 \lambda_{2}^4
	} \, .
\end{equation}
This naturally reduces to eq.~\eqref{eq:L_sigma} for $\lambda_i=1$.
Integrating over all angles, with eq.~\eqref{eq:dn2dt_em_m} we then find \cite{ChlubaThesis, Chluba:2006kg}%
\footnote{Here it is most convenient to use the electrons as a reference.}
\begin{equation}
	\left.\pd{n(\omega_2)}{t}\right|^{\rm em}_\text{L,mov}
	=
	(2p^2+1)\, \left.\pd{n(\vec{k}_2)}{t}\right|_\text{L}
	\, .
\end{equation}
As we will show below, even for $\omega_0\ll 1$ this expression breaks down once $4 p \omega_0 \gtrsim 1$, a transition that similarly happens in single Compton scattering \citep[e.g.,][]{Jones1968, Sarkar:2019har}.
However, we will now derive more general expressions, valid in those cases. 

\subsection{General soft photon formulae (arbitrary 
\texorpdfstring{$p$}{p}%
, arbitrary 
\texorpdfstring{$\omega_0$}{o0}%
)}
\label{appendix:general_omega0_p}
An exact generalization to arbitrary $\omega_0$ and $p$ was not obtained so far. In \cite{ChlubaThesis, Chluba:2006kg}, higher order correction terms were derived using explicit Taylor series expansions of eq.~\eqref{eq:soft_sigma}. The approximations converge quite slowly but the leading order terms allowed finding improved approximations using an {\it inverse Ansatz} \citep{ChlubaThesis, Chluba:2006kg}.
Here we generalize the treatment to arbitrary incident photon and electron energies by analytically carrying out additional integrals over the soft photon directions. The final general expression then only contains three additional angle averages which greatly reduces the numerical treatment.
For resting initial electrons, one can furthermore obtain a full analytic expression for the DC emission spectrum;
for ultrarelativistic electrons only one integral remains.

Aligning the $z$-axis with the incident electron, without loss of generality we can compute the average over the soft photon directions $\id^2 \hat{\vec{k}}_2$ instead of $\id^2 \hat{\vec{p}}$ for the incident electron.%
\footnote{As already stated, this is equivalent because we are considering isotropic distributions.}
This means that in eq.~\eqref{eq:soft_sigma} only the second factor is directly affected. After some admittedly tedious algebra one finds
\begin{align}
\label{eq:soft_sigma_av2}
\frac{\left<\id^8 \sigma^{\rm soft}_{\rm DC}\right>_2}{\!\id^8\Lambda}
&=\int \frac{\!\id^8 \sigma^{\rm soft}_{\rm DC}}{\!\id^8\Lambda} 
\frac{\id^2 \hat{\vec{k}}_2}{4\pi}
= \frac{\alpha}{2\pi^2 \omega_2}
\frac{\!\id^5 \sigma_{\rm C}}{\!\id^5\Lambda_{\rm C}}
\left[\frac{\Delta_0}{p_0}
\ln\left(\Delta_0 +p_0\right)-1
\right]
\end{align}
where $\Delta_0=1+\lambda_0\omega_0-\lambda_1\omega^{\rm C}_1$, $p_0=\sqrt{\Delta_0^2-1}$ and we again used the differential Compton cross section given in eq. \eqref{eq:sigma_Compton}.
For $p=0$, this directly simplifies to what was given in \cite{Mandl52} aside from an extra factor of $1/4\pi$ from our definition.
With this result, we can now compute the general DC emission spectrum with the only assumption that $\omega_2 \ll 1$.
For given $p$ and $\omega_0$, with eq.~\eqref{eq:dn2dt_em_m} and eq.~\eqref{eq:soft_sigma}, we find the explicit expression for the soft photon spectrum
\bsub
\label{eq:dn2dt_em_m_soft_general_G}
\begin{gather}
\left.
		\pd{n(\omega_2)}{t}
	\right|_\text{DC, soft}^\text{em}
= 
	G_{\rm soft}(\omega_0,p) \, \left.\pd{n(\omega_2)}{t}\right|_\text{L}\, ,
	\label{eq:dn2dt_em_m_soft_general}
\\[1ex]
G_{\rm soft}(\omega_0,p)
=\frac{9}{64\pi}\,\frac{1}{\omega^2_0}
\!\int\!
	\id \mu_{\rm e0}
	\id \mu_{\rm e1}
	\id \phi_{\rm e1}
\frac{\lambda_0}{\gamma \lambda^2_1} \frac{\mathcal{F}}{(1+\xi)^2}
\left[
\frac{1+(1+\xi)^2}{1+\xi}
-\frac{2\xi}{\lambda_0\omega_0}
+\frac{\xi^2}{\lambda^2_0 \omega_0^2}
\right]\, ,
\end{gather}
\esub
where 
$
	\mathcal{F}(\mu_{\rm e0},\mu_{\rm e1},\phi_{\rm e1})
	=
	\frac{\Delta_0}{p_0}
	\ln\left(\Delta_0+p_0 \right)-1$, $\xi=\omega_0\alpha_{01}/\lambda_1
$, and $\lambda_i=\gamma - p\mu_{{\rm e}i}$.
Here, we introduced the soft photon correction factor, $G_{\rm soft}(\omega_0,p)$, relative to the Lightman-Thorne approximation for resting electrons.
To calculate $\alpha_{01}=1-\mu_{01}$ one has to use $\mu_{01}=\mu_{{\rm e}0}\mu_{{\rm e}1}+\cos(\phi_{{\rm e}1})(1-\mu_{{\rm e}0}^2)^{1/2}(1-\mu_{{\rm e}1}^2)^{1/2}$ were we assumed $\phi_{{\rm e}0}=0$ without loss of generality. 
The remaining three integrals in general can be carried out numerically, as we discuss below.
The newly found eq.~\eqref{eq:dn2dt_em_m_soft_general_G} is one of the central results of this paper,
as it allows us to easily and reliably calculate the true scaling of the soft photon emissivity, which could have a big impact on thermalization of high energy particles recently considered by \cite{SandeepEtAl}.

\subsubsection{General expressions for resting electrons (%
\texorpdfstring{$p = 0$}{p=0}%
, arbitrary 
\texorpdfstring{$\omega_0$}{o0}%
)}
For resting electrons, the integrals in eq.~\eqref{eq:dn2dt_em_m_soft_general_G} can be carried out analytically. The derivation is tedious and requires a term-by-term procedure, but the final result can be expressed as
\begin{align}
\label{eq:dndt2_rest_general}
G_{\rm soft}(\omega_0,p=0)&=-\frac{27-45\chi+405\chi^2+45\chi^3}{128\,\omega^4_0 \chi^2}
+\Bigg[\frac{27-45\chi+306\chi^2+774\chi^3+243\chi^4-9\chi^5}{\chi^2(3+\chi)}
\nonumber\\
&\!\!\!\!\!\!\!\!\!\!\!\!\!\!\!\!\!\!\!\!\!\!\!\!\!\!\!\!\!\!\!\!\!\!\!\!\!\!
+9(1+6\chi-\chi^2)\ln \chi
\Bigg] \frac{\ln \chi}{128\,\omega_0^5}
+\frac{9\zeta(7-30\chi-24\chi^2-2\chi^3+\chi^4)}{64\,\omega_0^6(3+\chi)^2}
\Bigg[
\ln\chi \, \ln\big(2\zeta+[1+4\zeta^2]\delta_+\big)
\nonumber\\
&\!\!\!\!\!\!\!\!\!\!\!\!\!\!\!
+{\rm Re}\bigg\{{\rm Li}_2(\delta_-)-{\rm Li}_2(\delta_+)+{\rm Li}_2(\chi\delta_-)-{\rm Li}_2(\chi\delta_+)\bigg\}
\Bigg]
\end{align}
with $\chi=1+2\omega_0$, $\zeta=\sqrt{\omega_0(2+\omega_0)}$ and $\delta_\pm=1+\omega_0\pm\zeta$, and where ${\rm Li}_2(x)$ is the dilogarithm. 
It is useful to consider the limiting cases of this expression.

Taking the limit for $\omega_0\ll 1$ directly from eq.~\eqref{eq:dndt2_rest_general} turns out to be difficult. However, one can expand eq.~\eqref{eq:soft_sigma_av2} for $p=0$ into orders of $\omega_0\ll1$ and then directly carry out the average over $\id \mu_{01}$. Also multiplying by $4\pi$ for the trivial integrals, one then obtains
\begin{align}
\label{eq:dndt2_rest_general_small_omega0}
G_{\rm soft}(\omega_0,p=0)&\approx
1-\frac{21}{5}\omega_0+\frac{357}{25}\omega_0^2-\frac{7618}{175}\omega_0^3+\frac{21498}{175}\omega_0^4
-\frac{80242}{245}\omega_0^5+\frac{41000}{49}\omega_0^6
\end{align}
in agreement with \cite{ChlubaThesis, Chluba:2006kg} (although we added a couple of terms here).  For completeness we also give the simple inverse Ansatz approximation \cite{ChlubaThesis, Chluba:2006kg}
\begin{align}
\label{eq:dndt2_rest_general_small_omega0_inv}
G^{\rm inv}_{\rm soft}(\omega_0,p=0)
&\approx
\frac{1}{1+\frac{21}{5}\omega_0+\frac{84}{25}\omega_0^2-\frac{2041}{875}\omega_0^3+\frac{9663}{4375}\omega_0^4},
\end{align}
which works well up to $\omega_0\simeq 1$.  For $\omega_0 \ll 1$, these expressions can be quite useful, as the integrals in eq.~\eqref{eq:dn2dt_em_m_soft_general_G} can become numerically unstable. Additional approximations for $\omega_0, p\ll 1$ are given in \citep{ChlubaThesis, Chluba:2006kg}. These can also be directly obtained from eq.~\eqref{eq:dn2dt_em_m_soft_general_G} by expanding in orders of $p\ll 1$ and $\omega_0\ll 1$ and afterwards performing the integrals, however, we refer the reader to \citep{ChlubaThesis, Chluba:2006kg} for the corresponding analytic expressions.

For $\omega_0\gg1$, from eq. \eqref{eq:dndt2_rest_general} we find
\begin{align}
\label{eq:dndt2_rest_general_large_omega0}
G_{\rm soft}(\omega_0,p=0)&\approx\frac{3(10.6+12\ln\omega_0)\ln\omega_0-99.4}{64\omega_0^3}\, ,
\end{align}
which asymptotes to the full result a high energies. Obtaining additional correction terms $\mathcal{O}(\omega_0^{-4})$ is straightforward but does not improve the range of applicability by much.

\subsubsection{Ultra-relativistic electrons (%
\texorpdfstring{$p \gg 1$}{p>1}%
, arbitrary 
\texorpdfstring{$\omega_0$}{o0}%
)}
The integrals in eq.~\eqref{eq:dn2dt_em_m_soft_general_G} get harder to evaluate numerically for $p\gg \omega_0$
as they present sharper and sharper poles at the boundaries of the allowed angular configurations.
It can thus be beneficial to have an approximate expression 
valid for ultrarelativistic electrons.
A brute force approach trying to simply expanding the integrand around $p\rightarrow \infty$ remained unsuccessful. 
It is better to follow the procedure employed in  \citep{Jones1968} for normal Compton scattering and start in the electron rest frame where it is clear the most of the incoming photons arrive from a very narrow range of angles, $\theta \simeq 1/\gamma$, in the direction of the motion of the electron. Then only $\mu_{01}$ matters and azimuthal averages becomes trivial. After transforming back into the lab frame, the polar integral over the scattered photon, $\gamma_1$, is furthermore converted into a frequency integral, $\omega_1\in \{\omega_0, 4\omega_0 \gamma^2/(1+4\omega_0 \gamma)\}$.
After some algebra, one can find
\begin{align}
\label{eq:dn2dt_em_m_soft_UR}
G^{\rm ur}_{\rm soft}(\omega_0,p)
&\approx \frac{3}{32}\,\frac{1}{\gamma^6\omega^4_0}\,
	\int
	\frac{\id \omega_1}{\delta^2}
\Bigg\{
\omega_0
\left(\frac{\omega_1}{\omega_0}-4\gamma^2 \delta\right)
\left[\frac{\omega_1}{\omega_0}+3 \gamma^2(1+\delta^2)\right]
-2\omega_1^3\ln\left(\frac{\omega_1}{4\gamma^2 \omega_0}\right)
\nonumber\\
&\qquad+
\omega_0 \left[
\zeta
\left(
6\gamma^4\delta (1+\delta^2)-2\gamma^2(1-5\delta+\delta^2)\frac{\omega_1}{\omega_0}
-\frac{\omega^2_1}{2\omega^2_0}
\right)
-
3\gamma^2
\frac{\omega_1}{\omega_0}\delta \ln\Xi
\right]\ln\Xi
\nonumber\\
&\qquad\qquad+
\left[
7\gamma^3(3\delta -1)+6\gamma\omega_1^2+\frac{5}{2}\omega_1^3
+3\gamma^2\omega_1\delta \ln \delta
\right]\ln\delta
\Bigg\},
\end{align}
with $\delta = 1-\omega_1/\gamma$, $\zeta=\sqrt{1+\frac{1}{\omega_0\omega_1}}$ and $\Xi=1+2\omega_0\omega_1(1+\zeta)$.
This expression converges very quickly in the ultra-relativistic regime and should be preferred over eq.~\eqref{eq:dn2dt_em_m_soft_general_G} for numerical applications.
We confirmed the result with the full numerical integration enforcing high precision numbers when evaluating eq.~\eqref{eq:dn2dt_em_m_soft_general_G} (see figure \ref{fig:G_soft}).

\subsection{Illustrations of the soft photon emissivity}
\label{sec:emissivity_soft}
\begin{figure}
\begin{center}
\includegraphics[width=0.495\columnwidth]{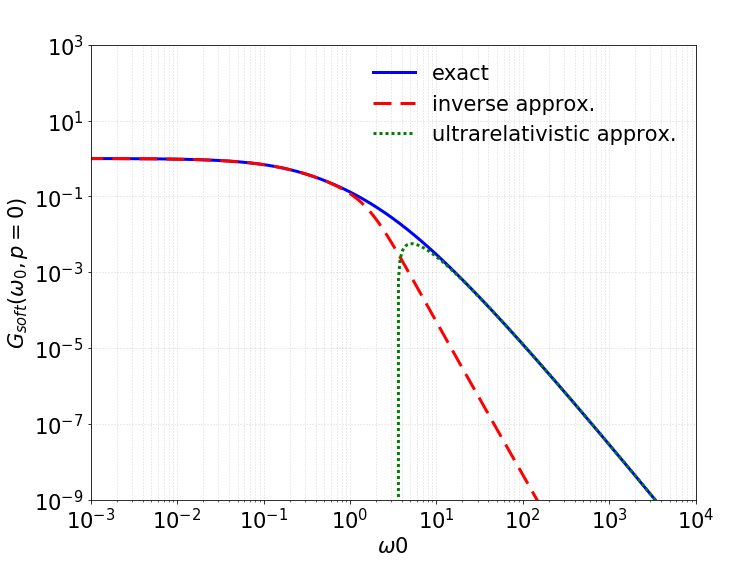}
\includegraphics[width=0.495\columnwidth]{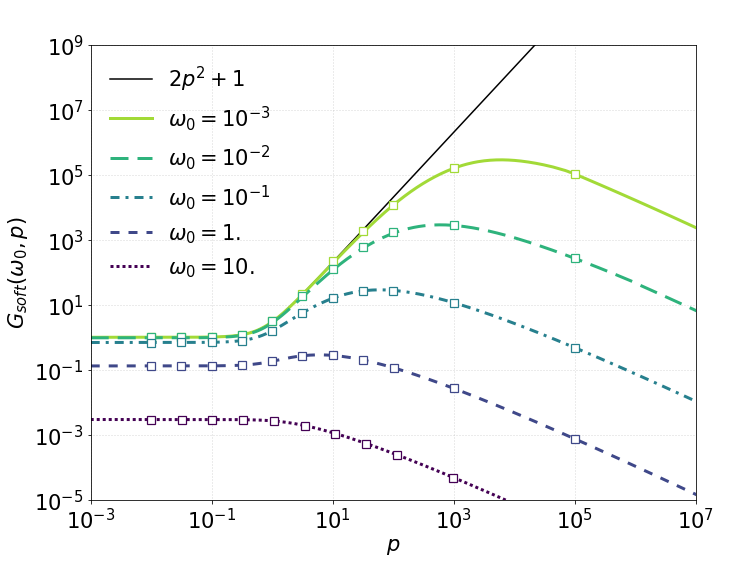}
\caption{
Soft photon double Compton emissivity for various particle energies. -- left panel: $p=0$ and varying $\omega_0$ according to eq.~\eqref{eq:dndt2_rest_general}.
We numerically confirmed the result. For comparison we also show the inverse approximation, eq.~\eqref{eq:dndt2_rest_general_small_omega0_inv}, and the ultra-relativistic approximation, eq.~\eqref{eq:dndt2_rest_general_large_omega0}, which both work very well in their respective regimes. -- right panel: general result based on eq.~\eqref{eq:dn2dt_em_m_soft_general_G} together with the simple approximation $G_{\rm soft}\simeq 2p^2+1$ valid for $4p\omega_0\ll 1$. For confirmation, the squares were numerically computed from the general DC expression, eq.~\eqref{eq:dn2dt_em_m}. At large electron momenta, the result obtained using the ultra-relativistic formula, eq.~\eqref{eq:dn2dt_em_m_soft_UR}, match exactly the general result.
}
\label{fig:G_soft}
\end{center}
\end{figure}

In figure~\ref{fig:G_soft} we illustrate the soft-photon correction factor for various combinations of the incident electron and photon energies. Assuming $p=0$, the DC emissivity drops with increasing $\omega_0$. Part of this suppression is naturally expected as also the Compton scattering probability is known to decrease in the Klein-Nishina regime. However, DC receives additional modifications due to the angular modulation factor, $\mathcal{F}$, in eq.~\eqref{eq:dn2dt_em_m_soft_general_G}.
For comparison, also the inverse formula, eq.~\eqref{eq:dndt2_rest_general_small_omega0_inv}, and the ultra-relativistic approximation, eq.~\eqref{eq:dndt2_rest_general_large_omega0}, are shown. The inverse formula underestimates the soft-photon DC emissivity at large $\omega_0$ (which is outside its limits of validity) where the ultra-relativistic approximation becomes very accurate.

In the right panel of figure~\ref{fig:G_soft} we show the soft-photon correction factor when varying the electron momentum.
At $p\ll 1$, we can observe the suppression of the overall DC emissivity with $\omega_0$, as also mentioned above. 
For $\omega_0\ll p$, one finds $G_{\rm soft}\simeq 2p^2+1$, which grows with $p$. However, at $p\simeq 1/[4\omega_0]$, additional suppression starts to dominate, as the incident photon is boosted to $\omega'_0\simeq \gamma \omega_0$ in the electron's rest frame. We confirmed that eq.~\eqref{eq:dn2dt_em_m_soft_general_G} shows excellent agreement with the full numerical result as long as the soft-photon regime is considered. Usually, this is valid at $\omega_2\lesssim [10^{-3}-10^{-2}]\, \omega_0(\gamma-p)/(\gamma+p+2\omega_0)$, i.e., well below the minimal energy accessible by the corresponding Compton event \citep{Sarkar:2019har}. For numerical applications it is beneficial to pre-tabulate the soft-photon correction factor as needed. This also accelerates the computation of thermally-averaged emission rates, should they be required.


\section{Beyond the soft photon limit (arbitrary 
\texorpdfstring{$\omega_2$}{o2}%
)}
\label{sec:beyond_soft}

In this section, we study the DC process going beyond the soft photon limit, i.e., making no simplifying assumption about the energy of the tracked photon with respect to the others. One of the first studies in this directions was carried out by \Gould \citep{Gould1984}, who assumed $p=0$ and $\omega_0\ll 1$ while allowing the whole range
$\omega_2 \in (0, \omega_0)$.
We will recap their main findings in section~\eqref{sec:Gould_limit}. This case was later generalized to moving electrons and larger photon energies \citep{ChlubaThesis}, and will be illustrated and further extended in section \ref{sec:GeneralCaseWithMovingElectrons}. 
Difficulties regarding the treatment of the infrared divergence for $\omega_1 \rightarrow 0$ will be explained in section~\ref{sec:IR_div}.

\subsection{\Gould formula (%
\texorpdfstring{$p=0 \wedge \omega_0 \ll 1$}{p=0 and o0 < 1}%
)}
\label{sec:Gould_limit}
To obtain the DC cross section that is required to compute the Gould formula \citep{Gould1984}, we assume $p=0$ and define $w_2 \equiv \omega_2/\omega_0$.
After expanding the cross section to lowest order in $\omega_0\ll1$, it becomes a polynomial in $\alpha_{01}, \alpha_{02}$ and $\alpha_{12}$.
If we use $\gamma_2$ as reference, then only $\mu_{01}$ has any azimuthal dependence and we can directly integrate over $\id^2\hat{\vec{k}}_1$.
After a few algebraic manipulations we find the required cross section
\begin{equation}
\begin{split}
\label{eq:Gould_sigma}
	\frac{\id^6 \sigma^{\rm Gould}_{\rm DC}}{\!\id^3 \vec{p} \id^3 \vec{k}_0}
	&
	=\frac{3\alpha\sigT}{160\pi}\frac{\omega_0}{y}\Bigg[
	\frac{2y^2}{w_2^2}
	\Big(
	3-7 w_2+13w_2^2-7 w_2^3+3w_2^4
	\Big)
\\
	&
	+
	\Big(
	11-21w_2+30w_2^3-21w_2^5+11w_2^6
	\Big)
\left[2-\alpha_{02}\right]\alpha_{02}
-10 w_2^3(1-\alpha_{02})\alpha^2_{02}
\Bigg] ,
\end{split}
\end{equation}
with $y=w_2(1-w_2)$. This expression will also be used to obtain a Lightman-Gould approximation for anisotropic incident photons in a forthcoming paper \cite{RavenniChlubaDCAniso}.

To obtain the DC emissivity in the Gould limit, we carrying out the remaining integrals, finding \citep{Gould1984, ChlubaThesis}%
\footnote{We used {$\int \alpha_{02}\id \mu_{02}=2$, $\int \alpha^2_{02}\id \mu_{02}=8/3$ and $\int \alpha^3_{02}\id \mu_{02}=4$}.}
\bsub
\label{eq:dn2dt_em_Gould}
\begin{align}
	\left.
		\pd{n(\omega_2)}{t}
	\right|_\text{G}
&= 
w_2 \, H_\text{G}(w_2) \, \left.\pd{n(\omega_2)}{t}\right|_\text{L}\, ,
\\
H_\text{G}(w_2)
	&= 
	\frac{1 - 3 y + \frac{3}{2} y^2 - y^3}{y} \, .
\end{align}
\esub
This shows that the Gould factor, $H_\text{G}(w_2)$, is fully symmetric under the exchange of the two emitted photons. It exhibits two poles, one as $w_2\rightarrow 0$ and the other at $w_2\rightarrow 1$. Both poles are related to the infrared divergence of the DC process, as we will discuss in section~\ref{sec:IR_div}.

Furthermore, in the Gould limit, the scattered electron is produced at rest; such that $\omega_0=\omega_1+\omega_2$. This is because recoil was neglected, but even for $\omega_0\ll 1$, this is never exact. Nevertheless, for sufficiently small $\omega_0$ a combination of the soft-photon correction term, $G_{\rm soft}(\omega_0,p)$, and the Gould-factor, $w_2 \, H_\text{G}(w_2)$, provides a good approximation for the DC emissivity \cite{ChlubaThesis}, as we also illustrate below.

\subsection{Departures from the \Gould limit (%
\texorpdfstring{$p=0$}{p=0}%
, arbitrary 
\texorpdfstring{$\omega_0$}{o0}%
)}
\label{sec:departures_Gould_limit}
Even for resting electrons, the Gould formula is only valid for $\omega_0\ll 1$. This is already evident from the soft photon correction term, $G_{\rm soft}(\omega_0,p=0)$ in eq. \eqref{eq:dn2dt_em_m_soft_general_G}, which exhibits a significant suppression of the DC emissivity with increasing $\omega_0$. In addition, recoil corrections are expected to become significant at $\omega_2\gtrsim \omega_0/(1+2\omega)$, as the scattered electrons carries away a portion of the energy \cite{ChlubaThesis}.

Treating the problem analytically proves difficult, as the pole structure of the DC cross section is rather complicated. However, eq.~\eqref{eq:dn2dt_em_m} can still be integrated numerically, with some caveats that we will now discuss.
It is useful to describe the general case as a correction to the generalized soft-photon limit. This then directly links to the \Lightman-Thorne approximation \cite{Lightman1981, Thorne1981}, such that one may write
\begin{equation}
	\left.\pd{n_2}{t}\right|^\text{rest}_\text{DC} = G_{\rm soft}(\omega_0,p=0)\,w_2 \, H(w_2, \omega_0) \left.\pd{n_2}{t}\right|_\text{L} \, ,
\label{eq:dndt_restE_G_H_L}
\end{equation}
where as before $w_2 \equiv \omega_2/\omega_0$ is the fractional energy carried by $\gamma_2$.%
\footnote{This definition differs a little from \citep{ChlubaThesis} since here we scale out the soft-photon limit, ensuring $w_2 \, H_\text{DC}(w_2, \omega_0)\rightarrow 1$ at $w_2\ll 1$.}
Notice that unlike in the \Gould limit (initial photon soft and the electron remains at rest after the scattering), in general the fractional energy of $\gamma_1$, $w_1 \equiv \omega_1/\omega_0$, 
is no longer such that $w_1=1-w_2$ but lower, since some energy is lost to the electron, 
hence, the system is no longer symmetric under the exchange $w_2 \leftrightarrow 1-w_2$.

\subsubsection{Infrared divergence of the DC process}
\label{sec:IR_div}
It is well-known that the DC cross-section diverges when the energy of either emitted photon tends to zero \citep{Brown:1952eu, Mork1965, Mork1971, Tsai:1972sg}.
This behavior finds its root in the intrinsic relation between scattering processes with different numbers of photon in the outer legs \cite{Jauch1976}, and also poses a problem when evaluating the DC collision term, as all energies of $\gamma_1$ (in our notation) are integrated over.
To consistently cure the issue, the DC process in principle has to be treated together with the next to leading order (NLO) corrections to the Compton process \citep{Brown:1952eu}. 

In this context, one is usually interested in the total Compton scattering cross section at order $\alpha^3$.
NLO corrections display a logarithmic divergence which can be shown to cancel with the corresponding DC one.
Employing a regularization --- introducing a photon mass in the standard approach \cite[e.g.][]{Brown:1952eu}--- in the calculation of both cross sections and summing the two, the dependence on the regularization parameters drops out, leaving a finite radiative correction at order $\alpha^3$.
The drawback of this approach is that the radiative correction cross section now depends on the energy resolution of the experiment, $\omega_{\rm res}$.
The argument to justify summing the two processes is that below  some $\omega_{\rm res}\ll 1$ the experiment is unable to distinguish contributions from virtual photon emission, relevant to the computation of the NLO correction, and from real photon emission by DC.
Adding all contributions to the total CS scattering cross section it was found that the correction can exceed the naive $\simeq \alpha/\pi$ level at sufficiently high energies \citep{Mork1971}.

To apply these findings to our problem, we first need to ask what plays the role of the ``energy resolution'' inside a plasma. This should have to do with the energy-scale at which the scattering process decoheres due to the presence of external perturbers, $\omega_{\rm coh}$, \cite{Heitler:1936jqw}.
Another way of saying this is to realize that the particles and their wave-functions are no longer isolated.
One guess for $\omega_{\rm coh}$ would be the energy corresponding to the plasma frequency \citep{Rybicki1979, McKinney:2016voq}, in our units expressed as $\omega_{\rm pl}\approx 9.0\,{\rm kHz}\sqrt{\Ne /{\rm cm}^{-3}}\,h/\me c^2\approx \pot{7.3}{-17}\sqrt{\Ne /{\rm cm}^{-3}}$, or, for moving electrons, the energy at which the Razin effect becomes relevant, $\omega_{\rm Razin}\approx \gamma \omega_{\rm pl}$. Both energies are usually exceedingly small, such that significant logarithmic contributions can be expected.
Another, natural energy is the self-absorption energy, $\omega_{\rm abs}$, at which the DC emission process is immediately followed by an absorption event. In this case, the photon distribution should be extremely close to a blackbody at $\omega\lesssim \omega_{\rm abs}$.
For many astrophysical plasmas, this energy is expected to be much larger than the plasma frequency (e.g., $\omega \simeq [10^{-3}-10^{-2}]\,k\Te/\me c^2$ in the cosmological thermalization problem \citep{Burigana:1991A&A...246...49B, Hu:1992dc, Chluba:2013kua}).
However, unless the absorption directly occurs on the pair of DC photons and the scattered electron, this definition can only be applied in an ensemble-averaged sense, thus leaving the fundamental problem unaltered.
To overcome these difficulties, later on in section \ref{sec:emissivity}, we will introduce a procedure that allows us --- through appropriate regrouping of terms and a categorization of the emitted photons according to their energies --- to never rely on a particular choice of energy resolution, which might be difficult to determine.

Thinking of the DC process inside plasmas, another aspect seems to becomes evident: radiative corrections can no longer be considered in vacuum.
Stimulated effects inevitably enhance the DC emissivity and in a similar manner radiative corrections should be enhanced to ensure exact cancellation of the infrared divergencies, as also discussed broadly by  \cite{Weldon:1991eg}.
To consistently add the corresponding corrections we would thus need to consider all differential contributions to the collision term at order $\alpha^3$, a problem that becomes cumbersome.

To make process, we will assume that a characteristic cutoff energy, $\omegamin$, can be defined, thus allowing us to regularize the DC collision term.
When following the evolution of $\gamma_2$, this implicitly means that we have to impose  $w_i > \wmin$ for {\it both} $i=1,2$.
This leads to conditions on the scattering angles for $\gamma_1$ that become relevant when $\omega_2$ is close to $\omega_0$.
With this approach, we can be sure that the DC contributions to the evolution equation remain finite and, in the regime of physical relevance, the result should indeed not depend much on the chosen cutoff (cf. figure \ref{fig:H_DC_rest}).\footnote{We will return to this point again later.}
In addition, radiative corrections to the Compton process can then be treated separately but shall be neglected in our main discussion as here we are interested in the production of real photons by DC.
In practice for the purposes of our qualitative description of the DC kernel we will arbitrarily choose a small value of $\wmin$ as a cut-off, also showing what is the effect of changes in this value on the kernel.
Where not stated otherwise we will take $\wmin = 10^{-4}$.

\subsubsection{Constraints on the scattering angles in the Gould limit}
\label{sec:Gould_angle_constraints}
To compute the DC emissivity, we need to ensure that $\omega_1>\omegamin$. For $p=0$, we thus have 
\begin{align}
\omega_1
=\frac{\omega_0-\omega_2-\omega_0\omega_2\alpha_{02}}
{1+\omega_0\alpha_{01}-\omega_2\alpha_{12}}
>\omega_0 \, \wmin\, .
\label{eq:Nu1_rest}
\end{align}
For sufficiently small $\omega_2$, one has $\omega_1\simeq \omega_0/[1+\omega_0\alpha_{01}]>\omegamin$, such that no constraint on the scattering angles arises as long as $\omegamin<\omega_0/[1+2\omega_0]$. Hence in the soft-photon limit one can always carry out the integrations over all directions. This statement also generalizes to the case of initially moving electrons, justifying our treatment in the previous sections.

To obtain positive $\omega_1$, the numerator and denominator of eq.~\eqref{eq:Nu1_rest} have to be both positive or both negative,
but the latter case is excluded by energy conservation (here $\omega_1+\omega_2\leq \omega_0$). For $\omega_2<\omega_0/[1+2\omega_0] \equiv \omegamin^{\rm C}$ (which incidentally is the Compton scattering emitted photon minimal energy), the numerator of eq.~\eqref{eq:Nu1_rest} is always positive without any restrictions on $\mu_{02}$, while for $\omega_2>\omegamin^{\rm C}$, constraints naturally arise. Assuming for simplicity $\omegamin=0$, directly yields $\mu^{\rm min}_{02}=\max[-1, 1+\frac{1}{\omega_0}-\frac{1}{\omega_2}]$. 
The more general conditions can be directly derived from eq.~\eqref{eq:Nu1_rest}, but are not extremely illuminating (cf. appendix \ref{app:IntDom}). We discuss some of the details in section \ref{sec:GeneralCaseWithMovingElectrons}, where also the generalization to $p>0$ is carried out.

\subsubsection{Illustrations for the general DC emissivity (%
\texorpdfstring{$p=0$}{p=0}%
)}
\label{sec:DC_emissivity_p0}
\begin{figure}
\begin{center}
\includegraphics[width=0.495\columnwidth]{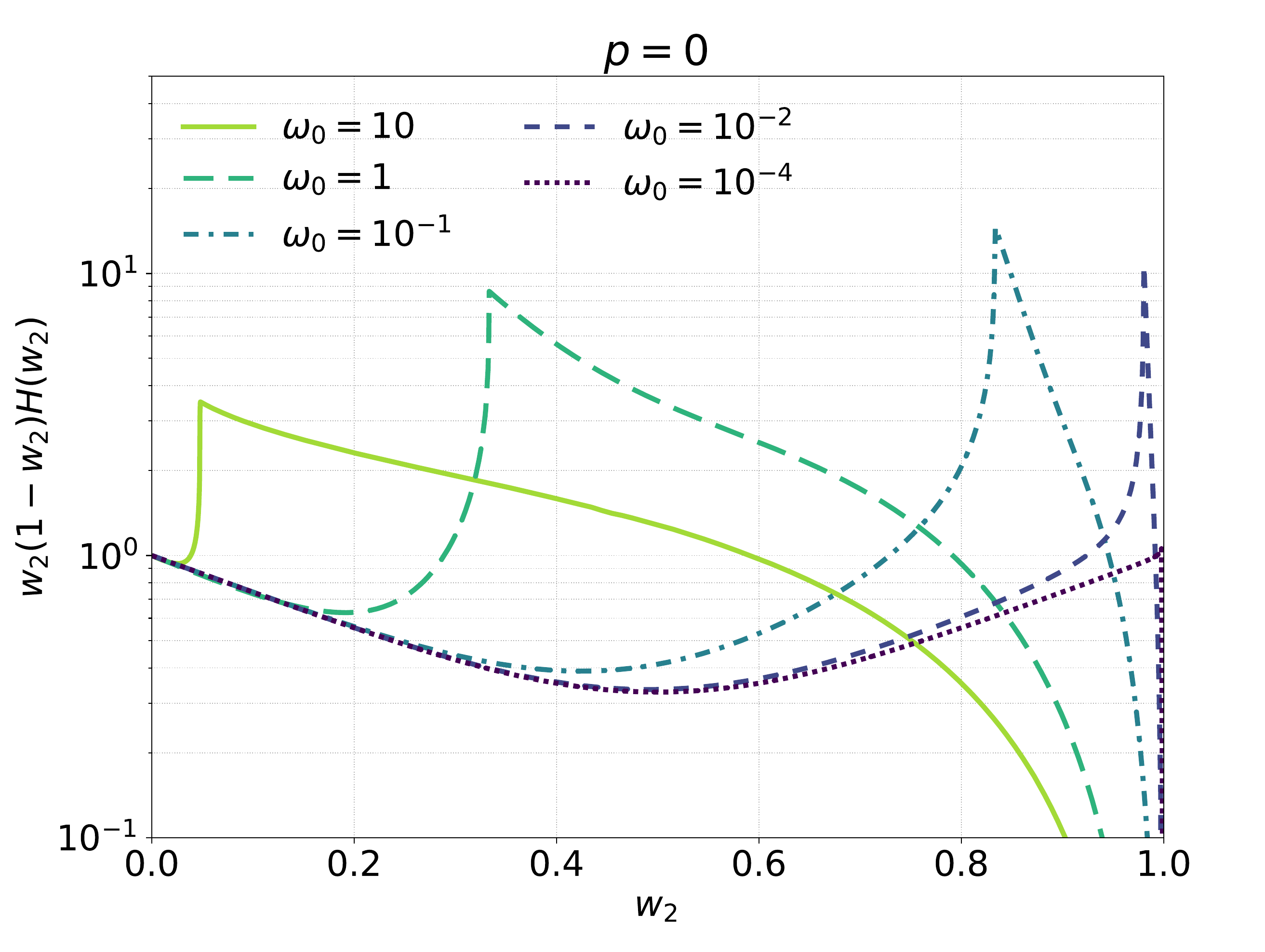}
\includegraphics[width=0.495\columnwidth]{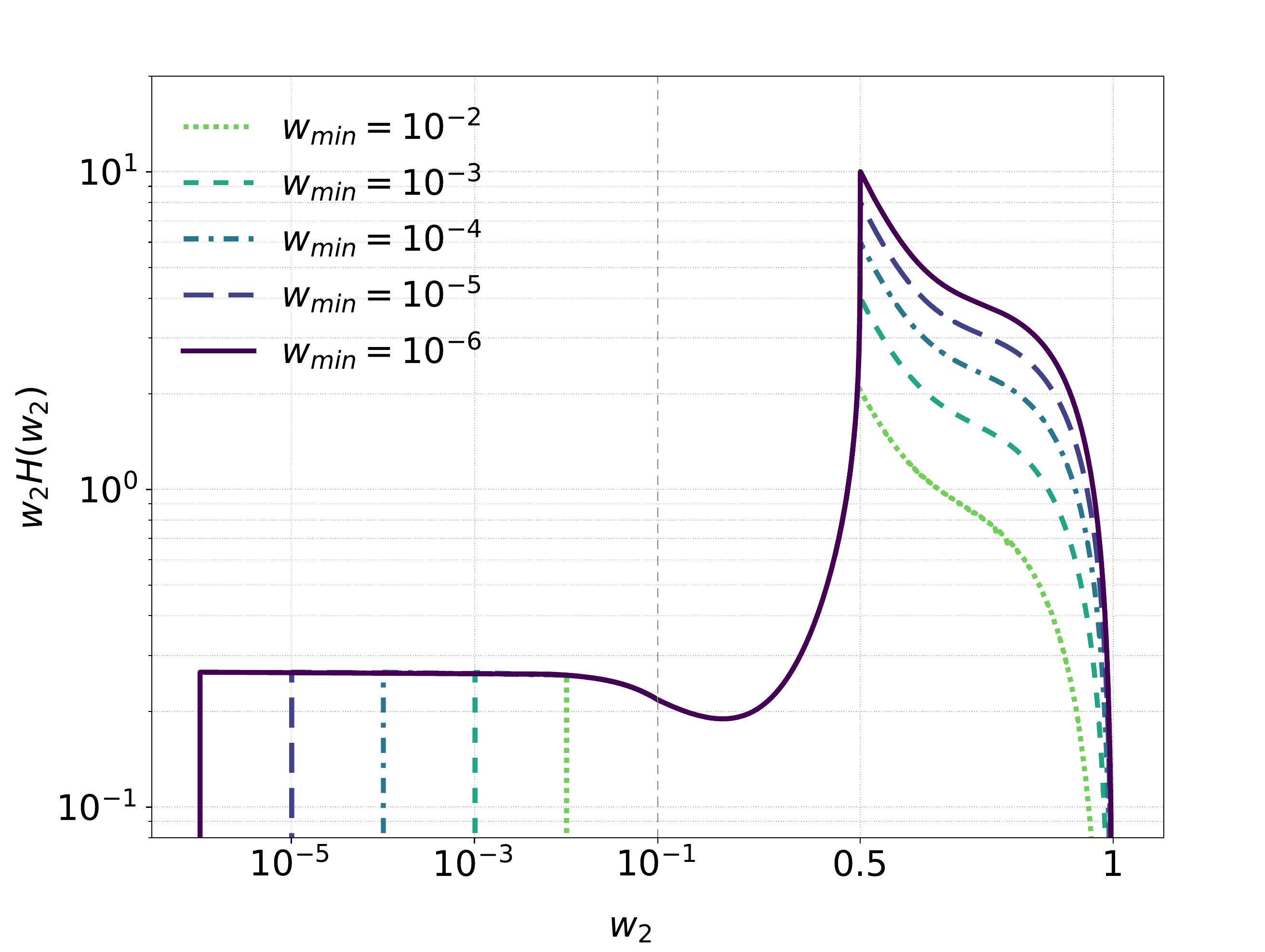}
\caption{Illustrations of the double Compton emissivity, parametrized by $w_2 (1-w_2) H_{\rm DC}$, for resting electrons. 
Notice how for low initial photon energy we recover the \Gould result.
-- Left panel: DC emissivity for various initial photon energies. -- Right panel: dependence on the low-energy cutoff, $\omegamin$, for $\omega_0 = 0.5$. Notice that the x axis switches from logarithmic to linear at $w_2= 0.1$.}
\label{fig:H_DC_rest}
\end{center}
\end{figure}

To extend our treatment beyond the soft photon limit we integrate eq. \eqref{eq:dn2dt_em_m} numerically.%
\footnote{As a sanity check of the stability of our implementation we compared the results obtained using both the VEGAS and SUAVE routines of the CUBA package, finding that the first have better performance when we consider low energy particles, while the latter is more convenient in the other cases.}
To simplify matters and improve the numerical stability we normalize the results by the \Lightman approximation multiplied by the soft photon factor, $G_{\rm soft}(\omega_0, p=0)$, so that we only have to consider the proper \Gould function correction, that correctly approaches unity at low $w_2$ according to eq.~\eqref{eq:dndt_restE_G_H_L}.
The \Gould function is shown in the left panels of figure~\ref{fig:H_DC_rest} for different $\omega_0$.
There we have isolated both the divergence at $w_2 \rightarrow 0$, and at $w_2 \rightarrow 1$, the latter being associated with the energy of the second photon $\omega_1$ approaching 0.
For small ($\omega_0 \lesssim 10^{-4}$) energies of the initial photon we approach correctly the \Gould approximation.

This parametrization makes it obvious that the symmetry $H(w) = H(1-w)$ of the classic \Gould result is lost when the energy of the initial particle is not vanishingly small, because of the appearance of a back-scattering peak.
The peak is located, as we will discuss in greater detail in section \ref{sec:ConfigurationsConstrainEnergy}, at $w_2=1/(1+2\omega_0)$, and can be therefore identified with the Compton back-scattering peak.
Incidentally, since we are considering resting electrons, the position of the back-scattering peak is degenerate with the minimum fractional energy of the Compton scattering emitted photon $\wminCS$.

In double Compton events there is the possibility of emitting \emph{one} photon with finite but lower than $\wminCS$ fractional energy:
the higher number of degrees of freedom in the system allows some configurations whose Compton scattering counterpart are kinematically forbidden.
This plays an important role when one considers the interplay of unresolved double Compton events and radiative corrections to the Compton scattering, as we discuss in section~\ref{sec:RadiativeCorrections}.

As anticipated, the choice of different infrared cutoffs has a logarithmic impact to the high energy part of the DC emission spectrum above $\wminCS$.
In the right panel of figure \ref{fig:H_DC_rest} we demonstrate the change in the \Gould function over a few different orders of magnitude of $\wmin$.
Figure \ref{fig:H_DC_rest} already gives some insight on a property that will be discussed in greater detail in section \ref{sec:emissivity}:
the cut-off frequency has a trivial impact on the low energy end of the spectrum (strictly speaking the spectrum is defined only above the imposed cut-off).
In practice, the probability of emitting a \emph{low energy} particle that can be resolved is not affected by the cut-off;
using the proper criterion to describe low energy and high energy particle, we will thus be able to describe the net DC emission rate in a cut-off independent fashion.
Moreover, we find that the emissivity plateaus at small $w_2$. Therefore, after calculating the emissivity for $\wmin$ low enough to reach the constant part, one can just set $w_2 H(w_2) = 1$ and use the expression of $G_\text{soft}$ from eq. \eqref{eq:dn2dt_em_m_soft_general_G} to fully describe the emissivity for any arbitrarily low $w_2$ and $\wmin$.

\subsection{General case with moving electrons}
\label{sec:GeneralCaseWithMovingElectrons}
As in the previous section, we can evaluate \cref{eq:dn2dt_em_m} numerically, this time assuming electrons have arbitrary energy.
In the same spirit of eq. \eqref{eq:dndt_restE_G_H_L}, also in this case we can parametrize the scattering rate as the product of the \Lightman-\Thorne result times a soft photon limit DC Gaunt factor and a energy dependent extended Gould function
\begin{equation}
	\left.\pd{n_2}{t}\right|_\text{DC} = G_{\rm soft}(\omega_0,p)\,w_2 \, H(w_2, \omega_0, p) \left.\pd{n_2}{t}\right|_\text{L} \, .
\label{eq:dndt_G_H_L}
\end{equation}
When we consider moving electrons the phase space becomes more complex, as the integrations over both the initial electron polar and the azimuthal angles are non trivial, effectively increasing the dimensionality of the problem to 5 independent variables.
Despite this, all the properties of the system that we discussed in the previous section, are retained in this limit.

Equation~\eqref{eq:dn2dt_em_m}, now recast as eq. \eqref{eq:dndt_G_H_L}, describes how a monochromatic initial photon distribution is redistributed in energy after a single DC scattering event. After multiplying by a factor of $1/2$ (to avoid double-counting of photons), it can thus be interpreted as a single-particle scattering kernel.
For this reason it is possible to find a relation with the Compton scattering kernel \cite[and references therein]{Sarkar:2019har}.
Specifically from their eq. (2) it is easy to see that the combination $H(w_2)G_\text{soft}(\omega_0, p)$ is the double Compton equivalent of their scattering kernel $P(\omega_0 \rightarrow \omega_2, p_0)$,
with the understanding that, while $P$ is independent of the photon distribution,
here $H\, G_\text{soft}$ would in principle depend on the stimulated scattering term, which we however neglected in this section.%
\footnote{In the following we will add a subscript $C$ to make clear it refers to the (normal) Compton kernel.}
To more easily shed light on the properties of the DC kernel, in the following discussion we will, when possible, draw comparisons with the Compton kernel.

Like in the case of resting electrons discussed in section \ref{sec:Gould_angle_constraints},
the phase space is again restricted by the energy conservation considerations: $\omega_1$, which is the photon we integrate over, is bounded from below by $\omegamin$ and from above by $\omega_1 < \gamma + \omega_0 - (1 + \omega_2)$.%
\footnote{We anticipate that for some energy configuration, the maximum energy that can be gained by each photon is even lower, just like for single Compton \citep{Sarkar:2019har}.}
As promised, we now discuss in greater detail the constraints on energies and angles that determine the integration boundaries for the DC collision term.

\subsubsection{Energy constrains configurations}

The energy constraint implies that some constellation are kinematically forbidden, and \textit{de facto} introduce some integration boundaries,
which can be determined studying eq. \eqref{eq:Nu1EnergyReplacement}, \ie imposing $\omega_1 > \wmin \omega_0$,
\begin{equation}
	\frac{
		P \cdot K_0 - P \cdot K_2
		- K_0\cdot K_2
	}{
		(P + K_0 - K_2) \cdot {\hat{K}_1}
	}
	>
	\wmin \omega_0\, .
\label{eq:w1condition}
\end{equation}
Some important remarks ought to be made, as this equation provide good insight into the problem.
In general, the angular boundaries depend on the chosen value of $\wmin$.
However, even for $\wmin = 0$ (which is equivalent to demanding positive energies) there are, in general, kinematically forbidden configurations.
Eq. \eqref{eq:w1condition} is satisfied if the numerator and the denominator are either both positive or both negative.
The latter branch is however to be excluded if we also consider that the energy must be conserved in the scattering.
The maximum possible fractional energy that can in principle be carried by one emitted photon is
\begin{equation}
	\wt \equiv 
	\frac{(\gamma - 1)}{\omega_0}
	+ 1 - \wmin \, ,
\label{eq:wtot}
\end{equation}
which is achieved when the emitted electron is at rest ($\gamma' = 1$) and the second emitted photon has fractional energy $\wmin$.
We will discuss in the following under which conditions this maximum-energy-transfer can take place.
We found that for any configuration in which both the numerator and denominator in eq. \eqref{eq:w1condition} are negative, $w_1 > \wt$, and as such, they are unphysical.

\subsubsection{Configurations constrain energy}
\label{sec:ConfigurationsConstrainEnergy}
To gain further insight into the properties of the double Compton kernel, it is helpful to compare it with a classical elastic scattering.
In fact, many of the properties we will highlight are shared again by the Compton kernel.

If we were consider a 2 body scattering, the minimum and maximum energy transfer occurs respectively for rear-end and head-on collisions.
What are instead the configuration that minimize or maximize the energy transferred to one of the photons?
The minimum energy is straightforward.
Regardless of the energy of the initial particles, the constellation that minimizes the RHS of eq. \eqref{eq:Nu1EnergyReplacement} is $\mue = 1$, $\muuno = \mutwo = -1$, which is a rear-end collision with both emitted photon back-scattering.
In this configuration, the fractional energy gained by the two photons together is 
\begin{equation}
	w_{1+2}^\text{min} \equiv (w_1 + w_2)\big|_\text{min}
	=
	\frac{\gamma-p}{\gamma+p + 2 \omega_0} \, 
\end{equation}
which correspond to the minimum energy of the scattering photon in a Compton scattering.
The minimum energy for the single photon is instead obviously $\wmin$.

To find other possible critical points in the scattering kernel we analyze which configurations mark a modification of the physical phase space.
From \cref{eq:w1condition} with $\wmin = 0$ we see that, for any $\muuno$ the combination $\mutwo = 1, \ \mue = -1$ is excluded for $w_2 >1$ and $\mutwo = -1, \ \mue = -1$ is excluded for $w_2 > \wc$, where $\wc$ is the frequency of the Compton back-scattering peak:
\begin{equation}
	\wc 
	\equiv
	\frac{\gamma+p}{\gamma-p + 2\omega_0}
\end{equation}
Features in the kernel, if any, ought to be found on these critical points.
We anticipate that the direct inspection of the numerical results highlight that the Compton back-scattering peak is not only present in double Compton processes too, but also a singularity.

\begin{figure}
\includegraphics[width=0.497\columnwidth]{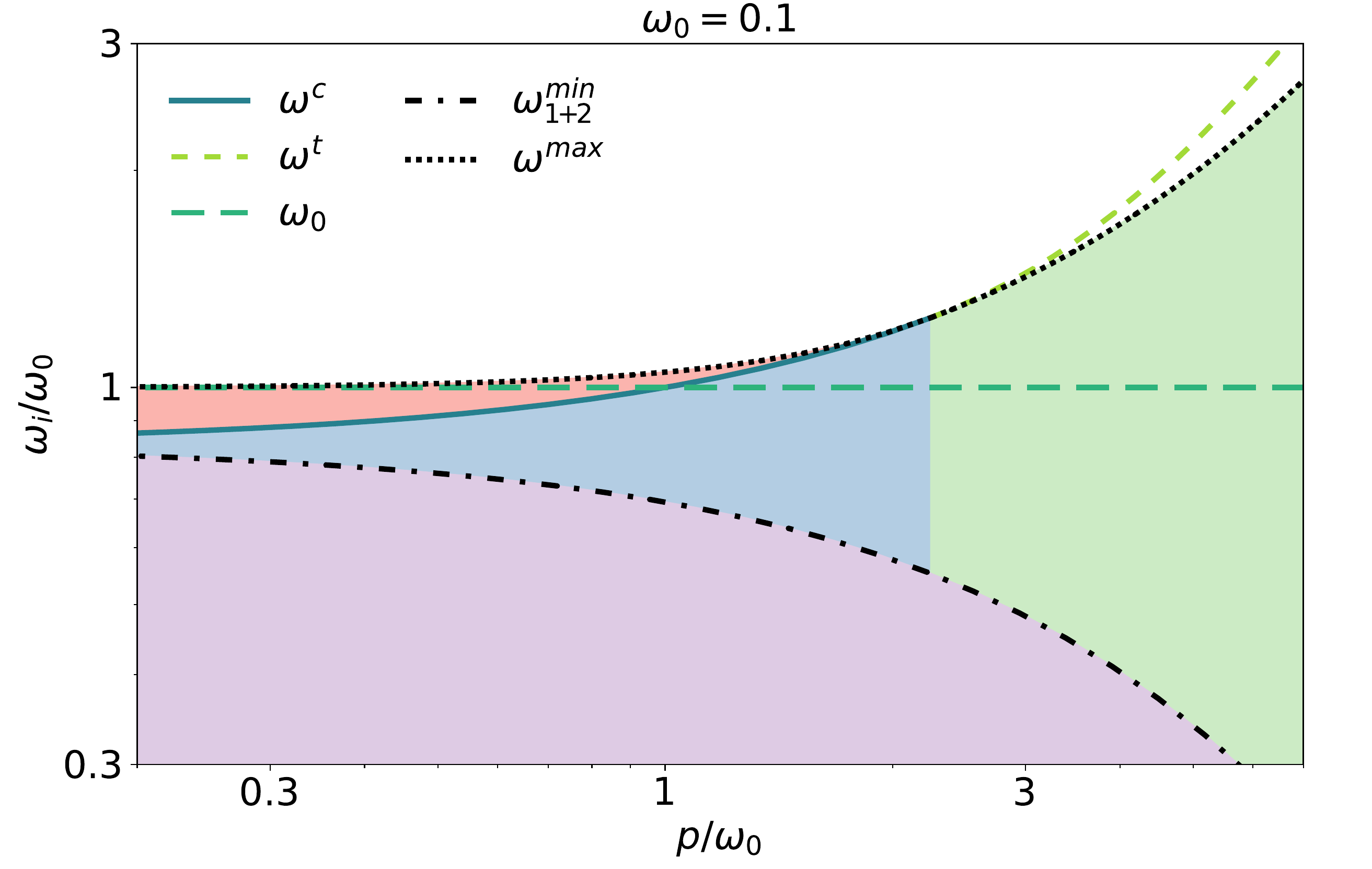}
\includegraphics[width=0.497\columnwidth]{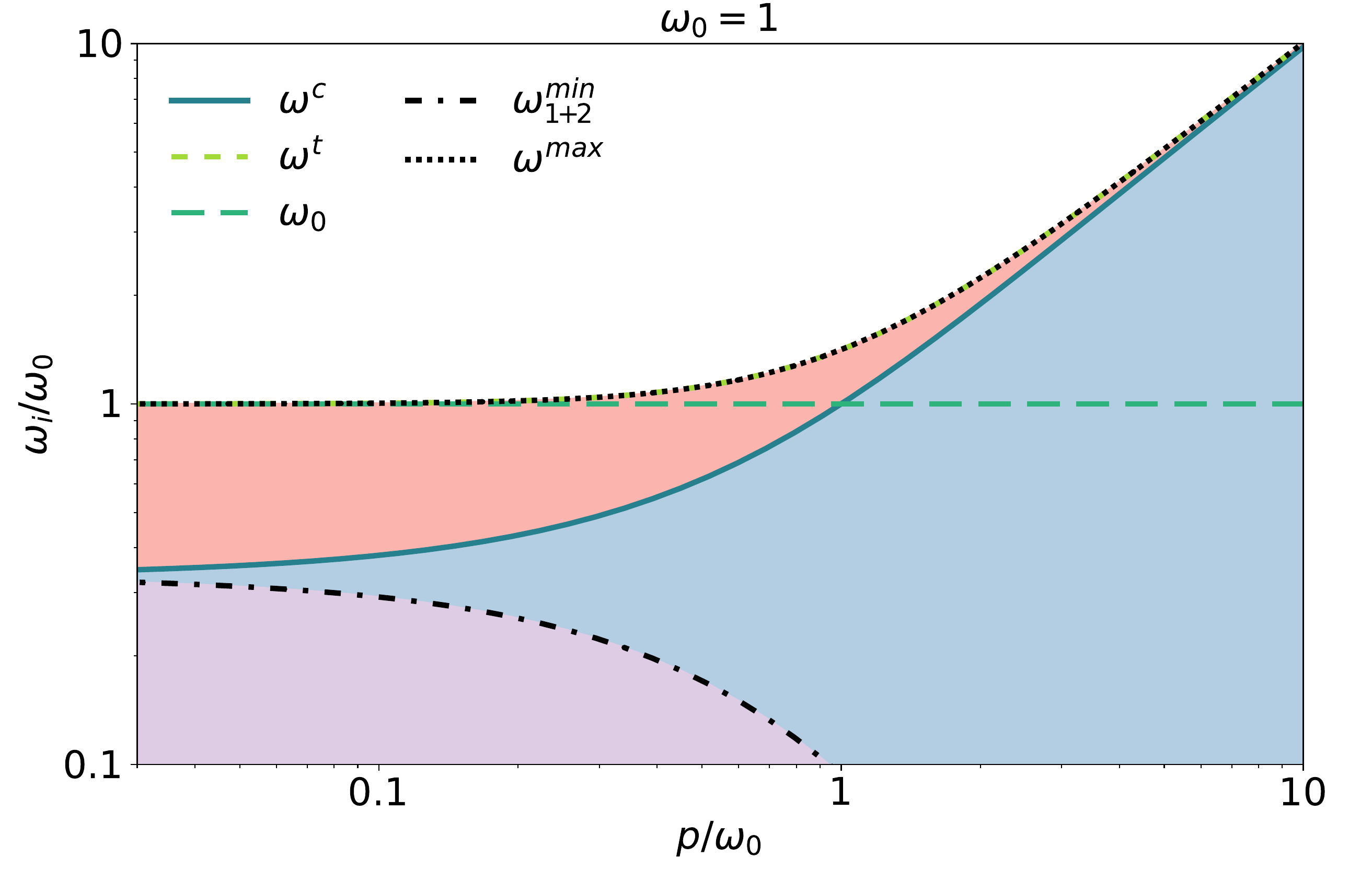}
\caption{
Different regimes of initial and final photon energies that characterize the double Compton scattering kernel.
In the left panel for $\omega_0 = 0.1$, in the right panel for $\omega_0 = 1$.
It is useful to compare this with figure 1 of \cite{Sarkar:2019har} as they only differ for the presence of a low frequency tail of emission below the accessible Compton regime at $\leq\omega^{\rm min}_{1+2}$.
}
\label{fig:DC_Regions_Eps}
\end{figure}

\begin{figure}
\begin{center}
\includegraphics[width=0.497\columnwidth]{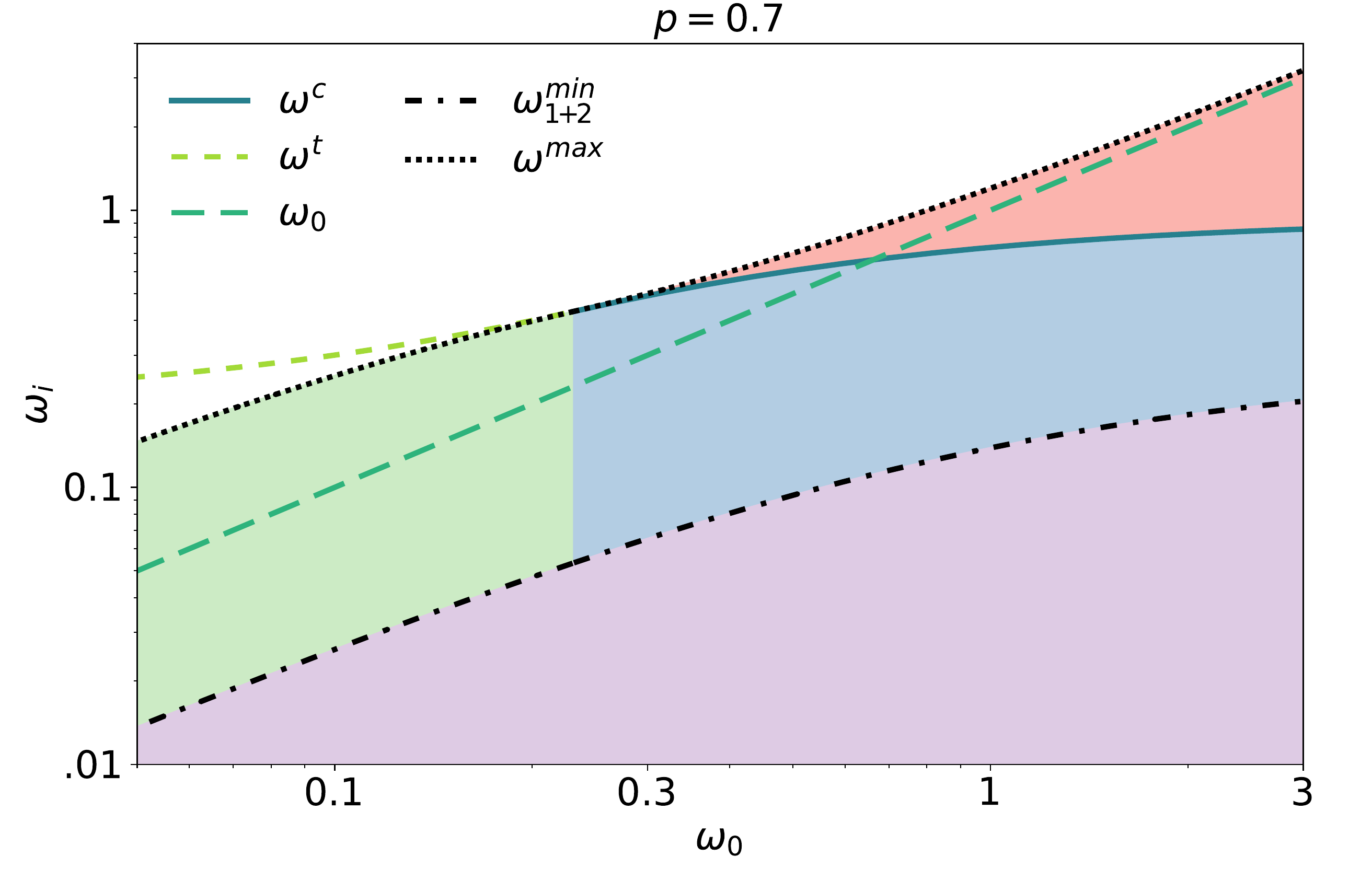}
\end{center}
\caption{
Same as figure \ref{fig:DC_Regions_Eps}, this time as a function of the initial photon energy.
The energy of the initial electron is $\epsilon = 1.2$.
}
\label{fig:DC_Regions_Omega}
\end{figure}

While $\wc$ is always smaller than $\wt$, it is not always the energy end point.
For certain initial energy configuration in fact the maximum energy transfer is not obtained in head-on scatterings.
In these cases, the energy end point coincides indeed with $\wt$.
Since the transition is smooth, we can find which combination of energies of the initial particles generate which regime by imposing $\wc = \wt$ and solving for $\omega_0$.
We find that
\begin{equation}
	\omegazerocrit \equiv \frac{1}{2}[1 + p-\gamma]
\end{equation}
is the critical value of the initial photon energy above which the maximum energy transfer (which is $\wt$) is allowed.
The different regimes are summarized in figure \ref{fig:DC_Regions_Eps} as a function of the initial electron energy, and in figure \ref{fig:DC_Regions_Omega} as a function of the initial photon energy.
In particular, we point out how figure \ref{fig:DC_Regions_Eps} traces exactly figure 1 of \cite{Sarkar:2019har}, with only one caveat:
in the purple region the tracked photon has energy smaller than the minimal one allowed by Compton scattering;
as such this region of emission is not present in the Compton scattering case.
All the other zones are shared by the two processes.
In the green region either scattered photon cannot reach energies higher than the Compton scattering peak $\wc$, while the rest of the energy is carried by the electron.
Vice versa in the red/blue regime the electron can be produced at rest in the lab frame, and the energy of either photon is bounded by energy conservation rather than the kinematic of the scattering.
The red (blue) region highlight the configurations above (below) the Compton back-scattering peak.

These consideration allow us to derive a set of boundaries for all the integration variables,  that we explicitly calculate in Appendix \ref{app:IntDom}.
Given how narrow the integration domain becomes when $\omega_0$ and $p$ are large,
this greatly simplifies the numerical evaluation, that would be straight away unmanageable otherwise.

\subsubsection{Illustration of the DC emissivity for general case}
\label{sec:DC_emissivity}

\begin{figure}
\includegraphics[width=0.497\columnwidth]{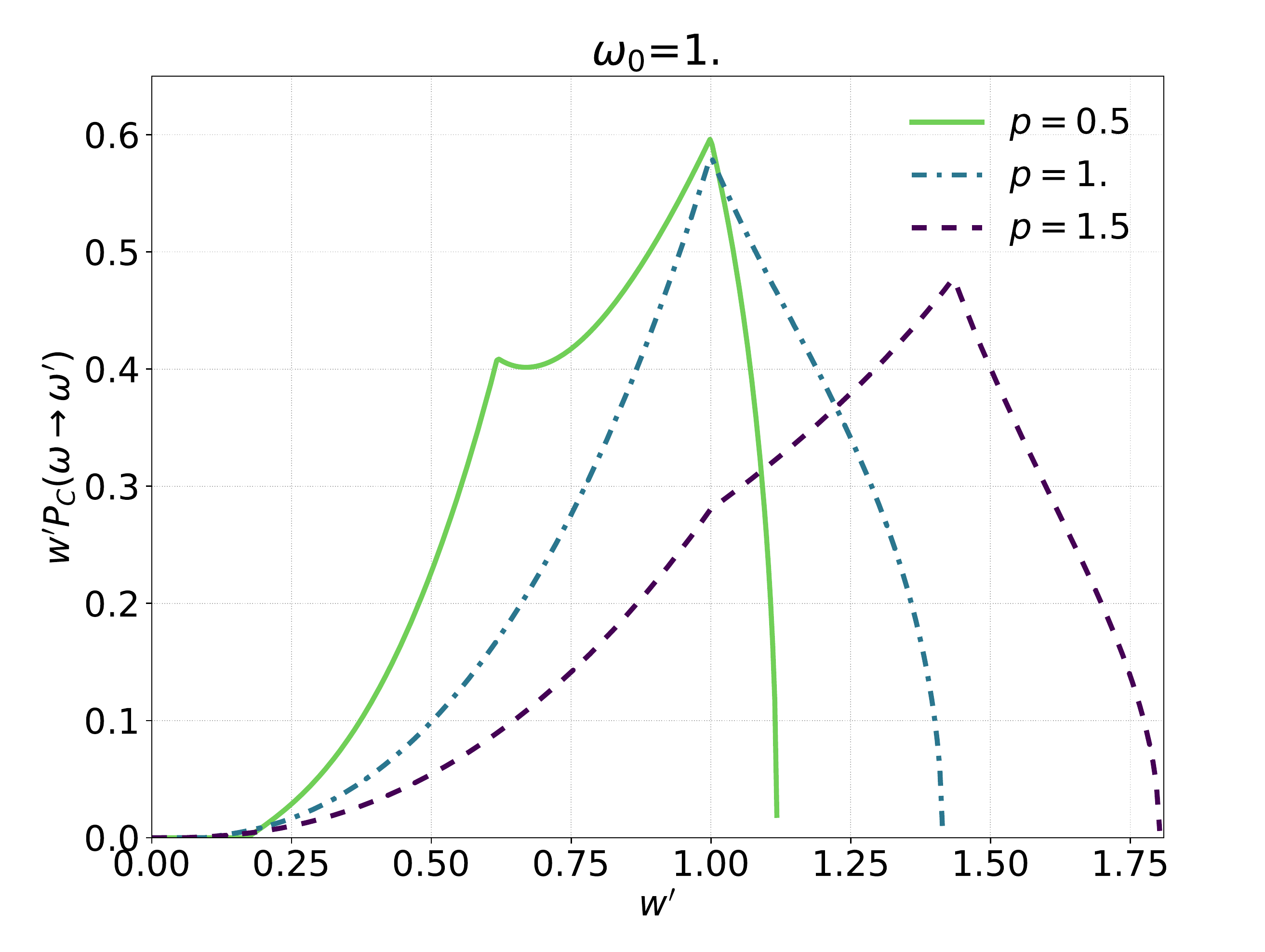}
\includegraphics[width=0.497\columnwidth]{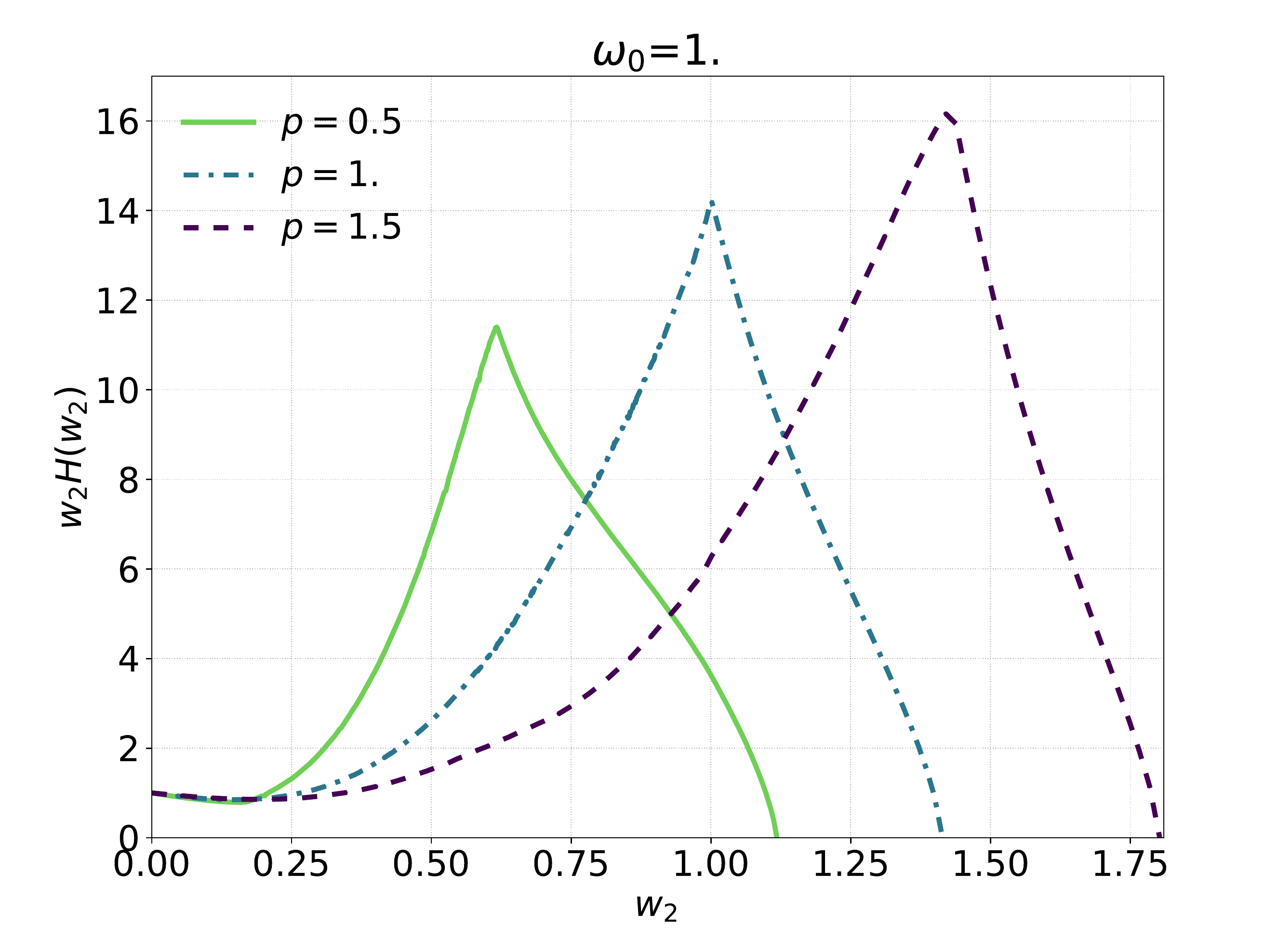}
\caption{Comparison of the Compton scattering kernel with its double Compton counterpart.
Only the Compton backscattering peak is visible in the DC case, whereas $\omega_2=\omega_0$ is not, in general, a critical point.
}
\label{fig:CS-DCS_Kernel_comparison}
\end{figure}

As we did in section \ref{sec:DC_emissivity_p0}, we integrate the double Compton kernel numerically to obtain the \Gould function in the case of energetic electrons.
To set the stage we start our discussion comparing the DC kernel with the Compton one \cite{Sarkar:2019har} in figure \ref{fig:CS-DCS_Kernel_comparison}.
As anticipated, the Compton back scattering peak is still present in the double Compton kernel, and represent a critical point much like in the Compton case.
On the contrary, $\omega_2=\omega_0$ do not show a cusp as it does in the case of Compton scattering.
Finally we highlight how the DC kernel present a low energy tail below $\omegamin^{CS}$, in the region where normal Compton scattering has zero emissivity.

\begin{figure}
\includegraphics[width=0.497\columnwidth]{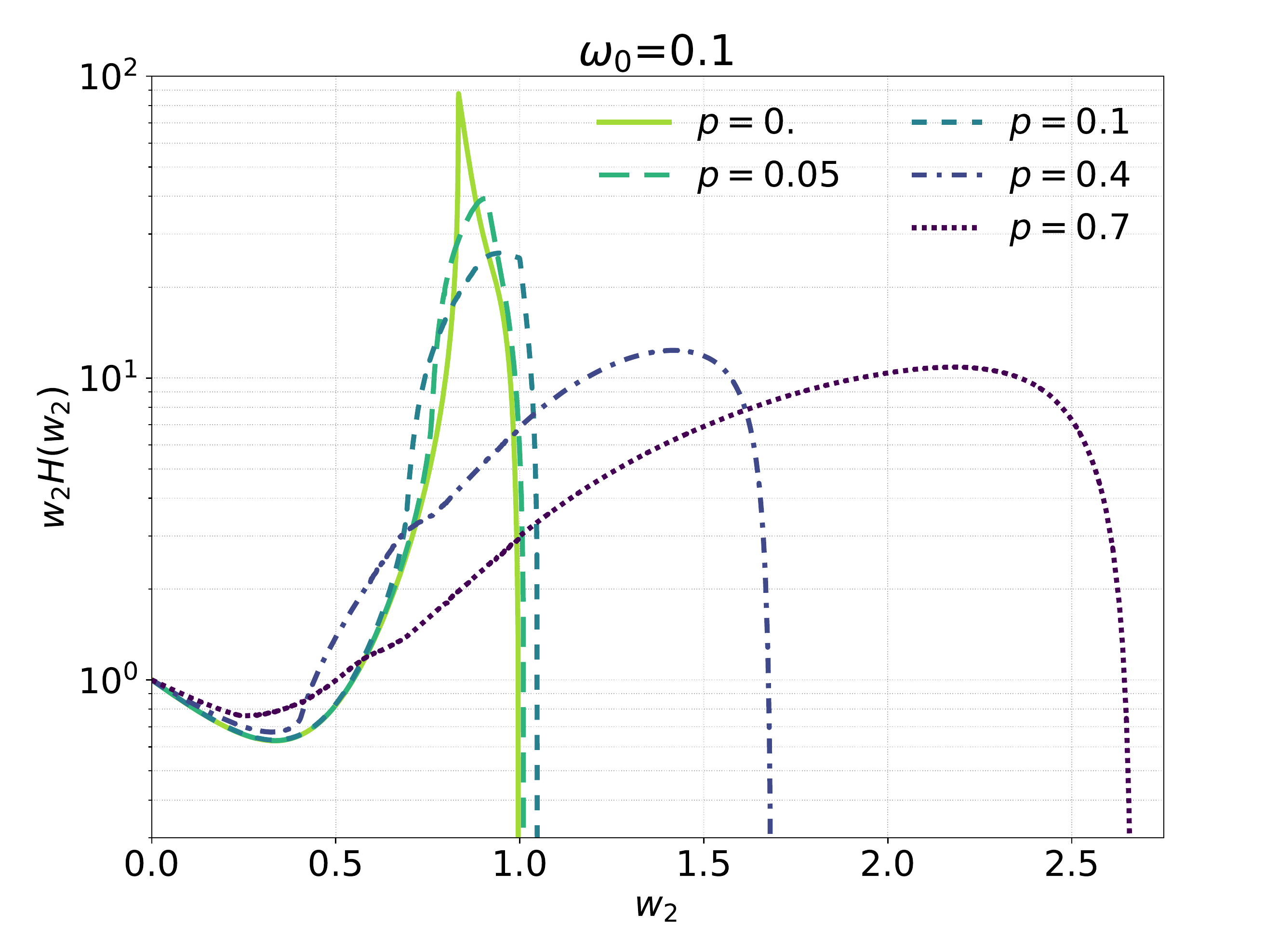}
\includegraphics[width=0.497\columnwidth]{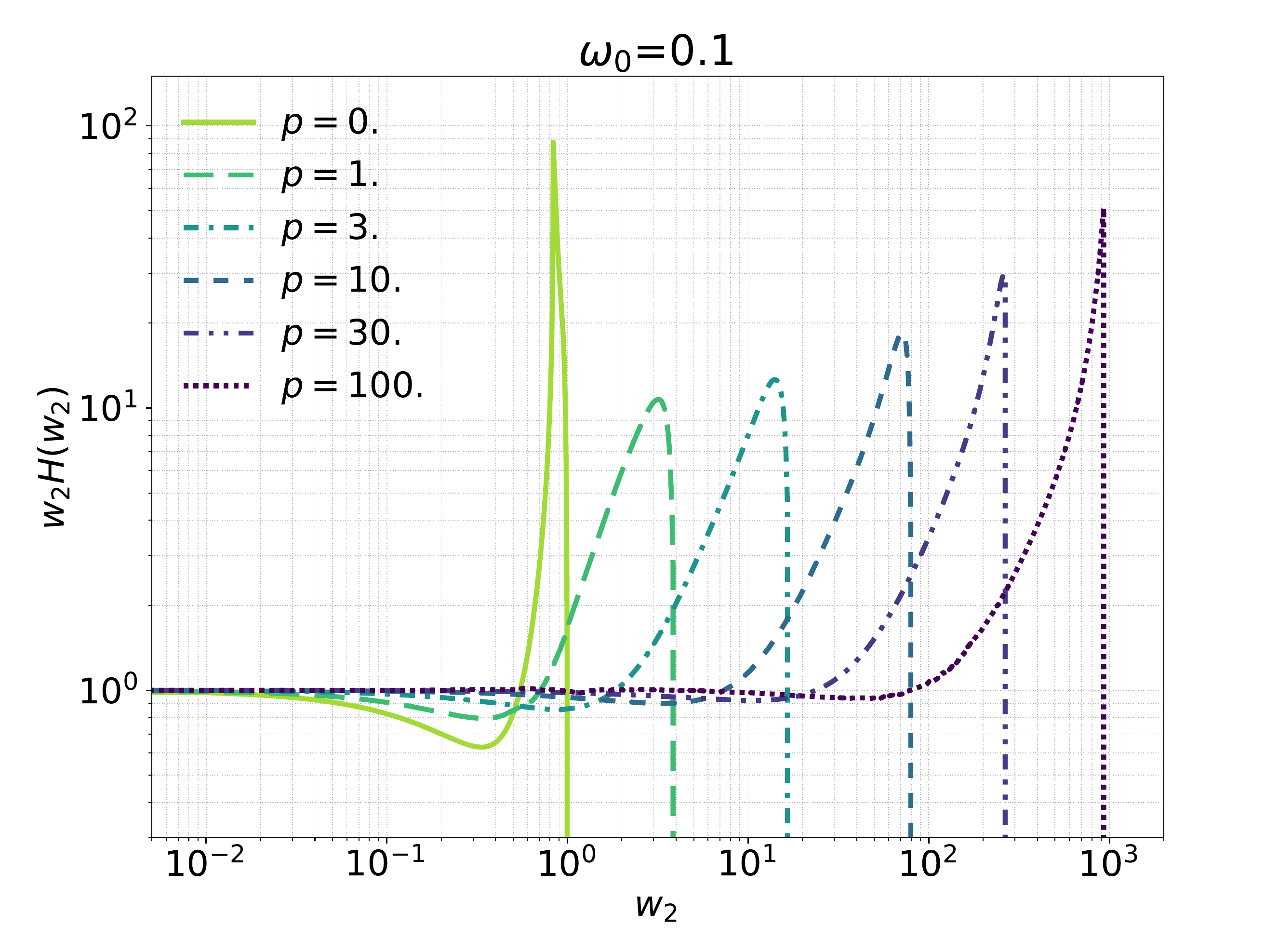}
\\
\includegraphics[width=0.497\columnwidth]{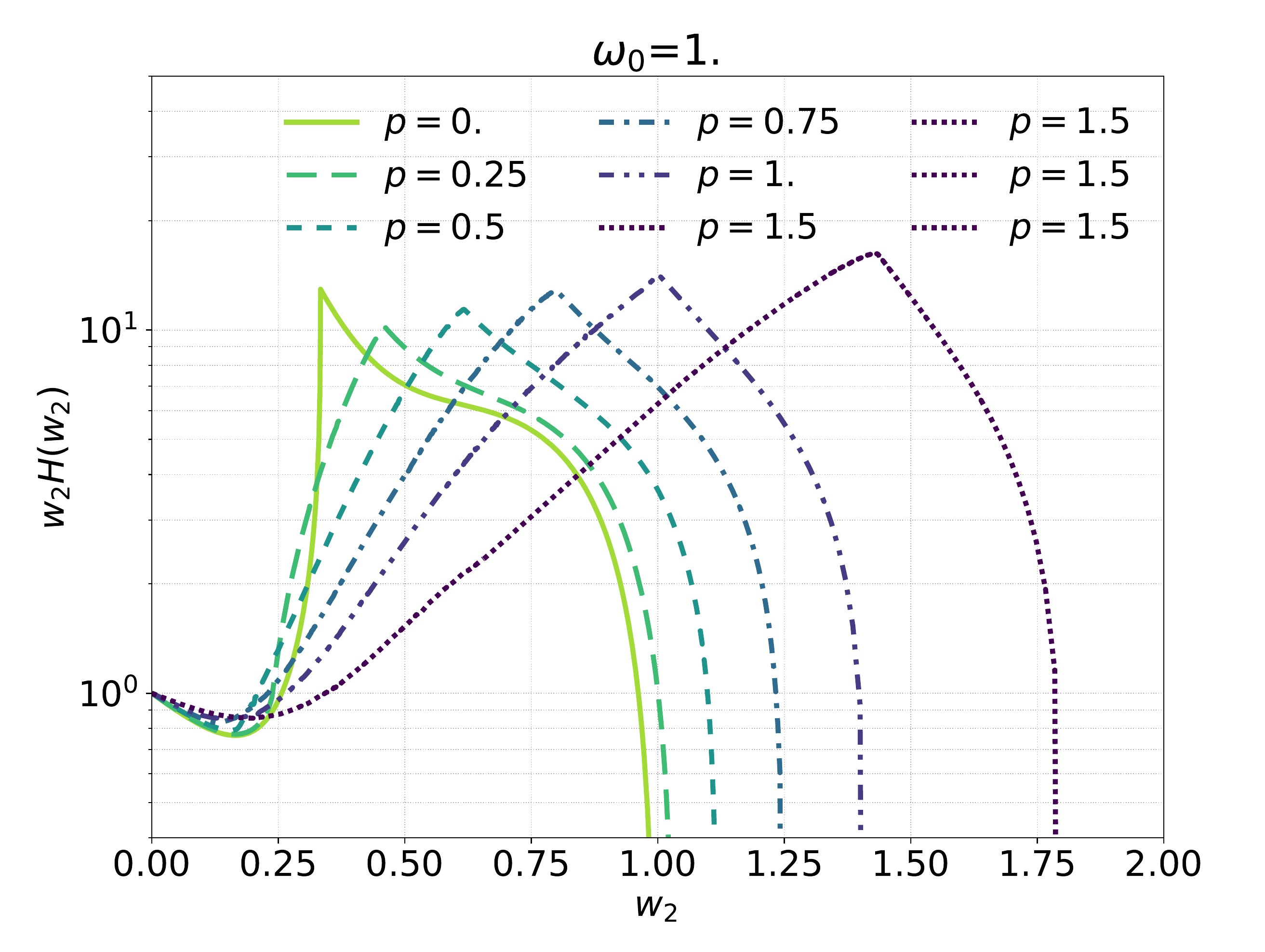}
\includegraphics[width=0.497\columnwidth]{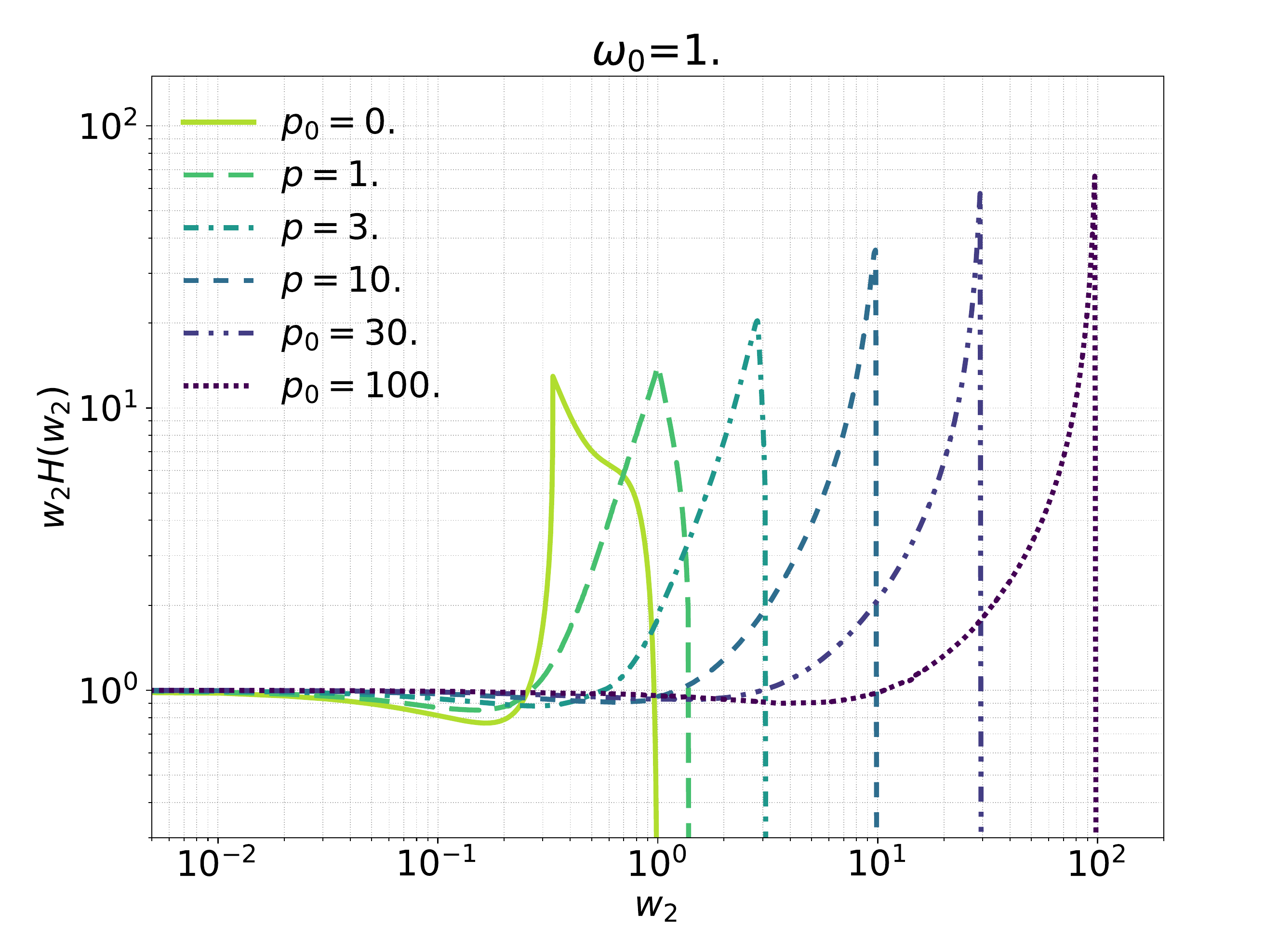}
\caption{Gould function $H(w_2)$ for different initial systems. Each panel shows the impact of electrons with different energies on the same distribution of photons.}
\label{fig:HG_varying_p}
\end{figure}

\begin{figure}
\includegraphics[width=0.497\columnwidth]{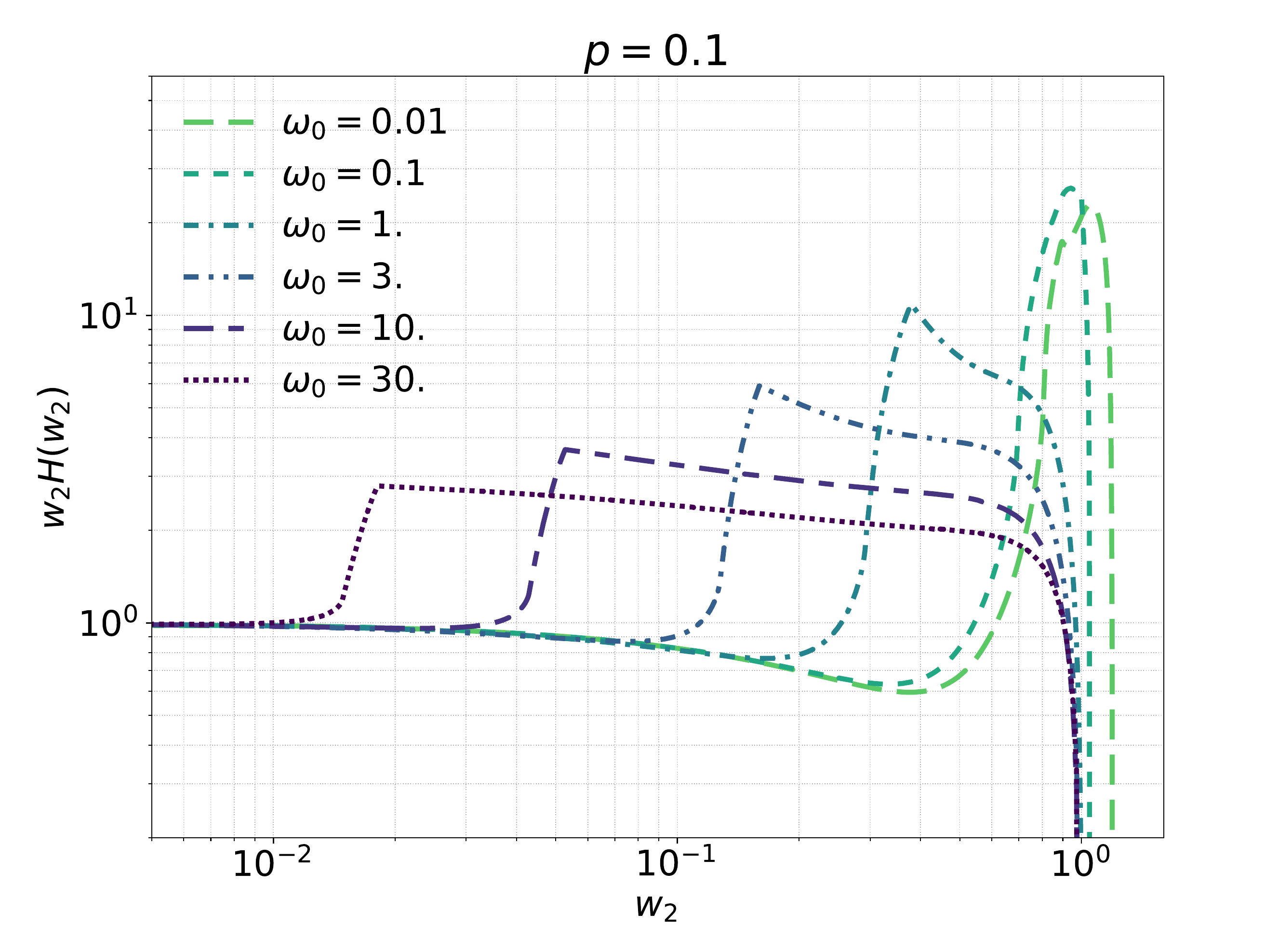}
\includegraphics[width=0.497\columnwidth]{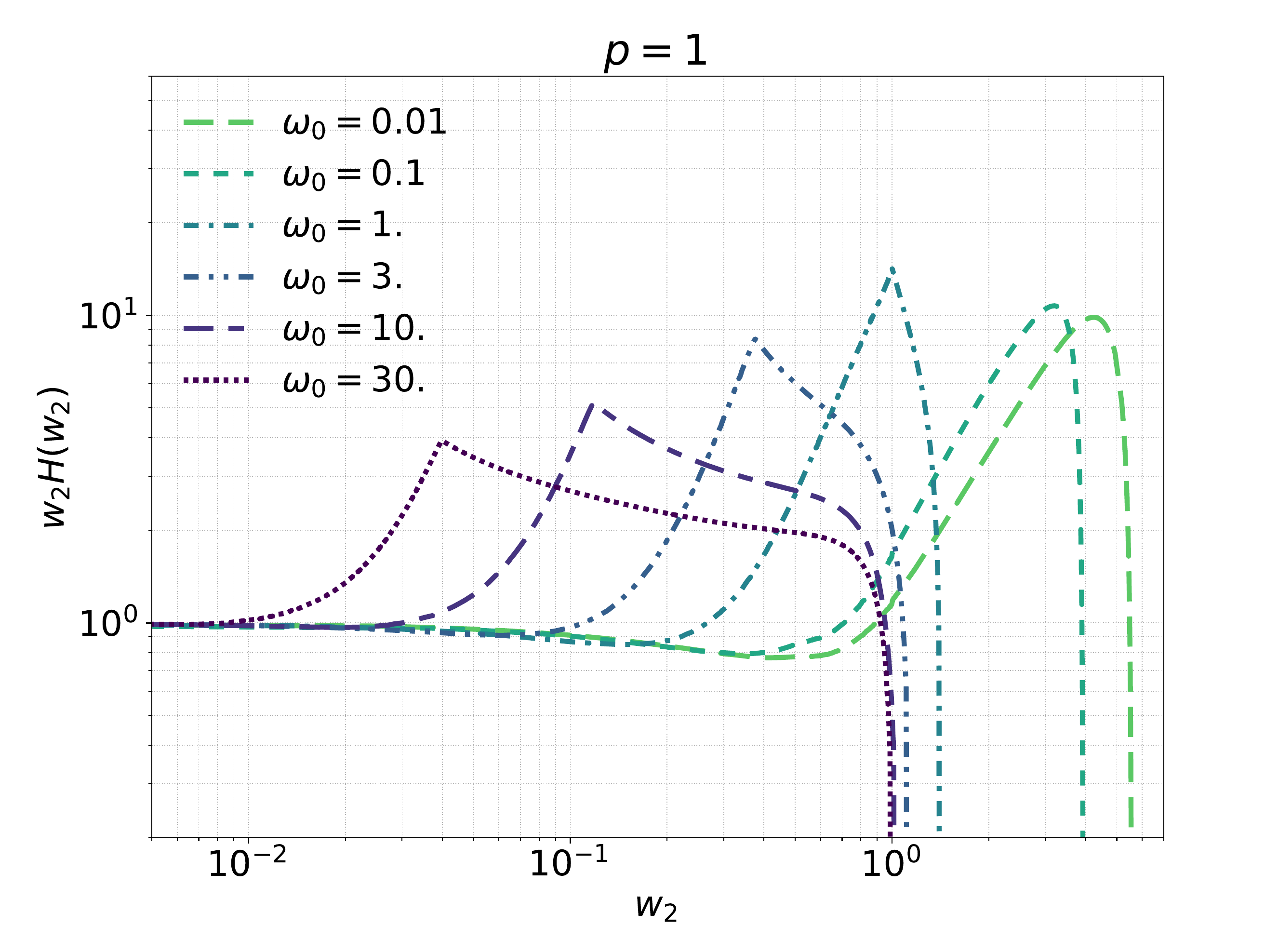}
\begin{center}
\includegraphics[width=0.497\columnwidth]{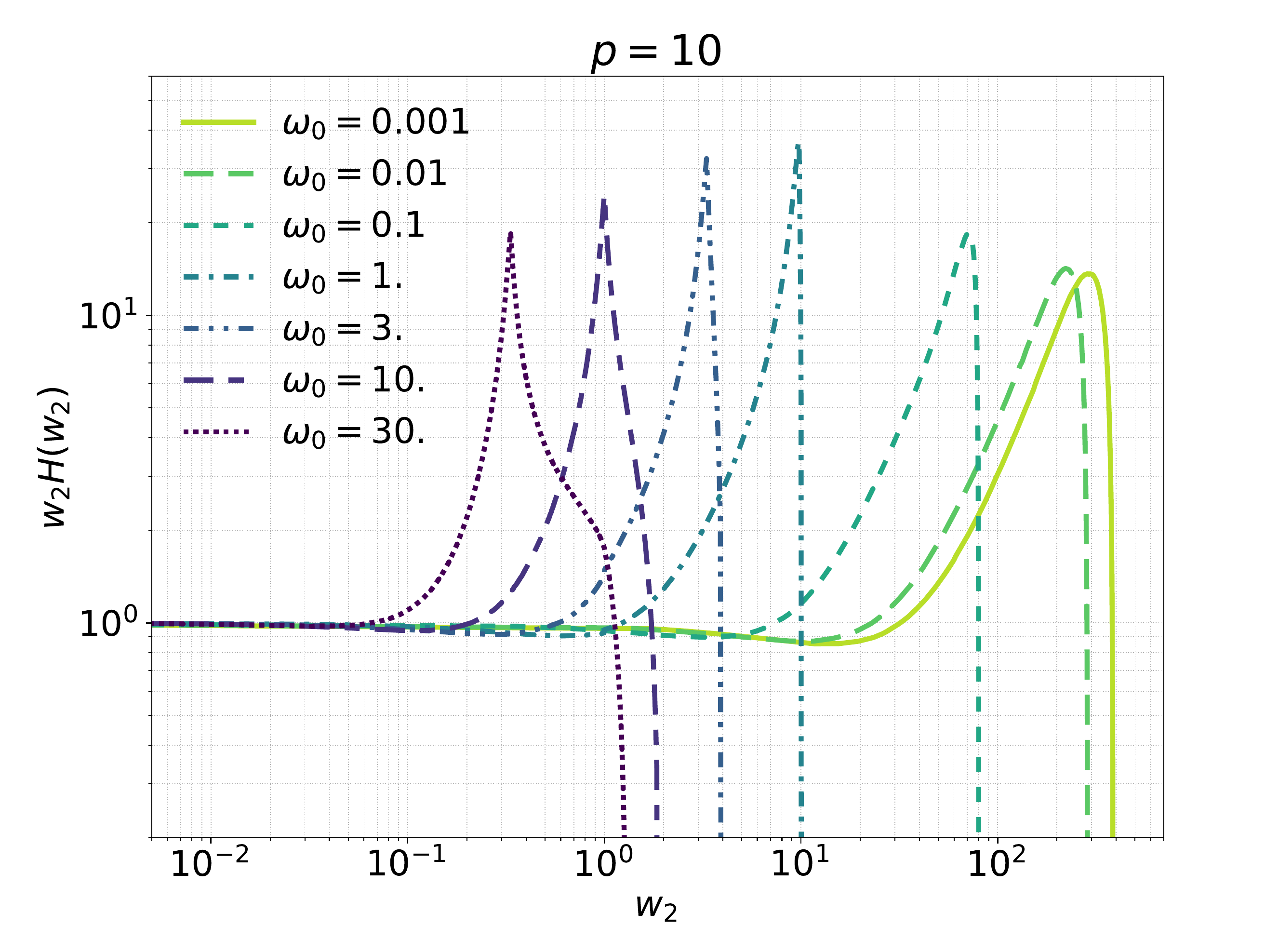}
\end{center}
\caption{Same as figure \ref{fig:HG_varying_p}, but each panel shows the impact of photons with different energies on the same distribution of electrons.}
\label{fig:HG_varying_Om0}
\end{figure}

An illuminating selection of cases is displayed in figure \ref{fig:HG_varying_p}, where in each panel we fix a value of $\omega_0$ and show the effect of increasing the electron momentum $p$,
and in figure \ref{fig:HG_varying_Om0}, where we do the opposite.
The kernel retains the main features it had in the resting electrons limit,
namely, it is correctly normalized to 1 for small $w_2$, it does not display the $w \leftrightarrow 1-w$ symmetry of the \Gould limit, and in the configurations for which $\wc < 1$ a cusp in the back-scattering peak.
Notice that in the right column of figure \ref{fig:HG_varying_p} the peak appears as a cusp for $\wc > 1$ because of the logarithmic scale, but it is actually continuous with continuous derivative.
The effect of higher electron energies is a Doppler broadening of the kernel.
We highlight how we limit ourselves to a few orders in magnitude of energy because of the high computational demand of this calculation.
In particular, for high electron energies the integrand becomes sharply peaked and very difficult to evaluate even with modern computers and algorithms.

\subsection{Compton scattering radiative corrections and IR divergences}
\label{sec:RadiativeCorrections}
The problem of divergent cross sections arises when one consider a double Compton event with a soft emitted photon or next to leading order corrections to a Compton event, as it is well known in the literature \cite{Jauch1976}.
Already in the first half of the last century, even before modern quantum field theory was properly formalized, tentative solutions where discussed.
It was later understood that the solution to these problems lies in the relation between the two processes
as they are indistinguishable at energies lower than the system resolution.
Finally its solution has been formulated at all orders in a theoretically sound framework in the quantum field theory formulation \cite{Bloch:1937pw, Brown:1952eu, Jauch:1954, Yennie:1961ad, Lee:1964is, Chung:1965zza, Weinberg:1965nx}.
As we are only interested in the order $\alpha^3$ --- we only want to normalize the scattering involving three photons in initial and final states --- we can apply the explicit results for the Compton scattering radiative corrections from \cite{Brown:1952eu, Tsai:1972sg, Dittmaier:1993bj} that we report in appendix \ref{appendix:M2_rad}.
In practice one can calculate the next to leading order correction to the tree level differential Compton cross section $\diff \sigma_C^{LO}$ introducing a regularization parameter in the form of a photon mass $\lambda$.
The Compton cross section up to order $\mathcal{O}(\alpha^3)$ is then \cite{Brown:1952eu}
$
\diff \sigma_C =
\diff \sigma_C^{LO}
\, (1-U_\text{fin}^{NLO}-U_\text{div}^{NLO} \ln \lambda)
$
where $U_\text{fin,div}^{NLO}$ are finite functions of the initial particle momenta, of which we give an explicit expression in eq. \eqref{eq:RadDiv_SigmaCompton}.
The cross section is therefore divergent for massless photon, but the divergent part can be encapsulated in the term $U_\text{div}^{NLO} \ln \lambda$.
The double Compton scattering can be similarly expressed as \cite{Brown:1952eu}
$
\diff \sigma_{DC} =
\diff \sigma_C^{LO}
\, [V_\text{fin}^{NLO}+U_\text{div}^{NLO} (\ln \lambda - \ln 2\omega_\text{min})]
$.
If our detector were to have a energy resolution $\omega_\text{min} = \omega_\text{res}$, we could sum the cross sections for the two undistinguishable process (Compton scattering and double Compton with a photon below the detector resolution)
to obtain a total cross section which is finite in the limit $\lambda\rightarrow0$:
$
\diff \sigma_C +
\diff \sigma_{DC} (\omega_2 < \omega_\text{res}) =
\diff \sigma_C^{LO}
\, [1-U_\text{fin}^{NLO} + V_\text{fin}^{NLO}+U_\text{div}^{NLO} ( - \ln 2\omega_\text{min})]
$.
This argument, valid for actual detectors can be extended to other physical systems, such as the electron-photon plasma we are interested in, as mentioned above.

\begin{figure}
\includegraphics[width=0.497\columnwidth]{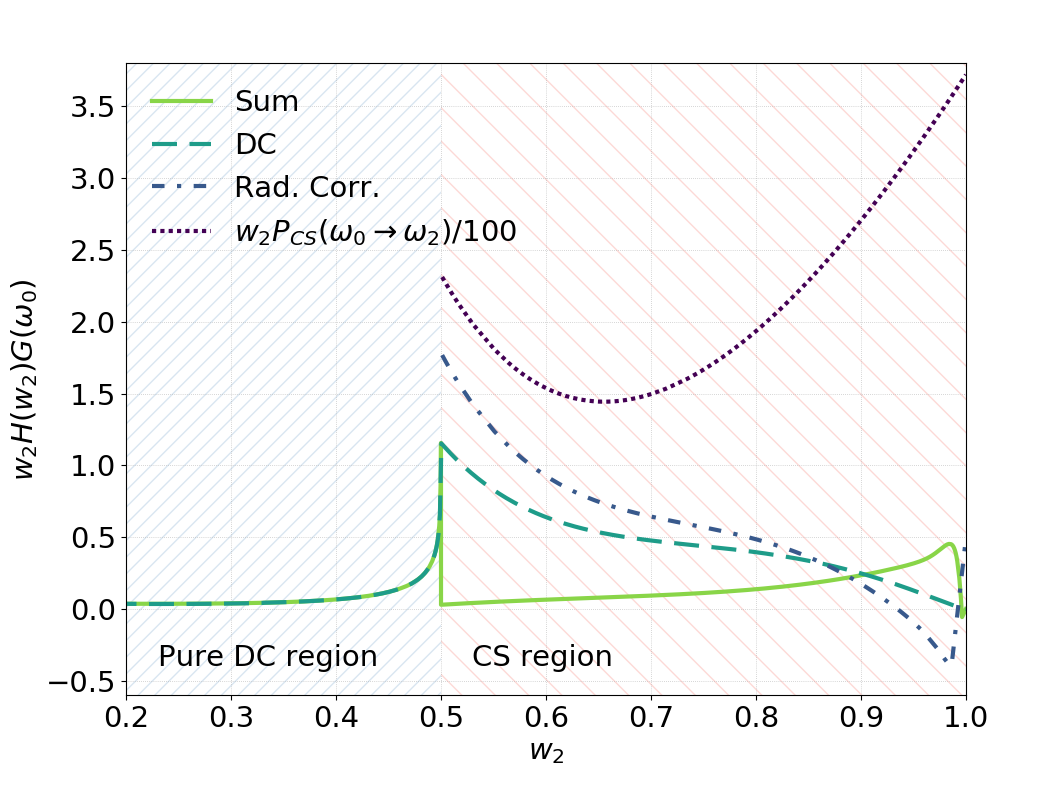}
\includegraphics[width=0.497\columnwidth]{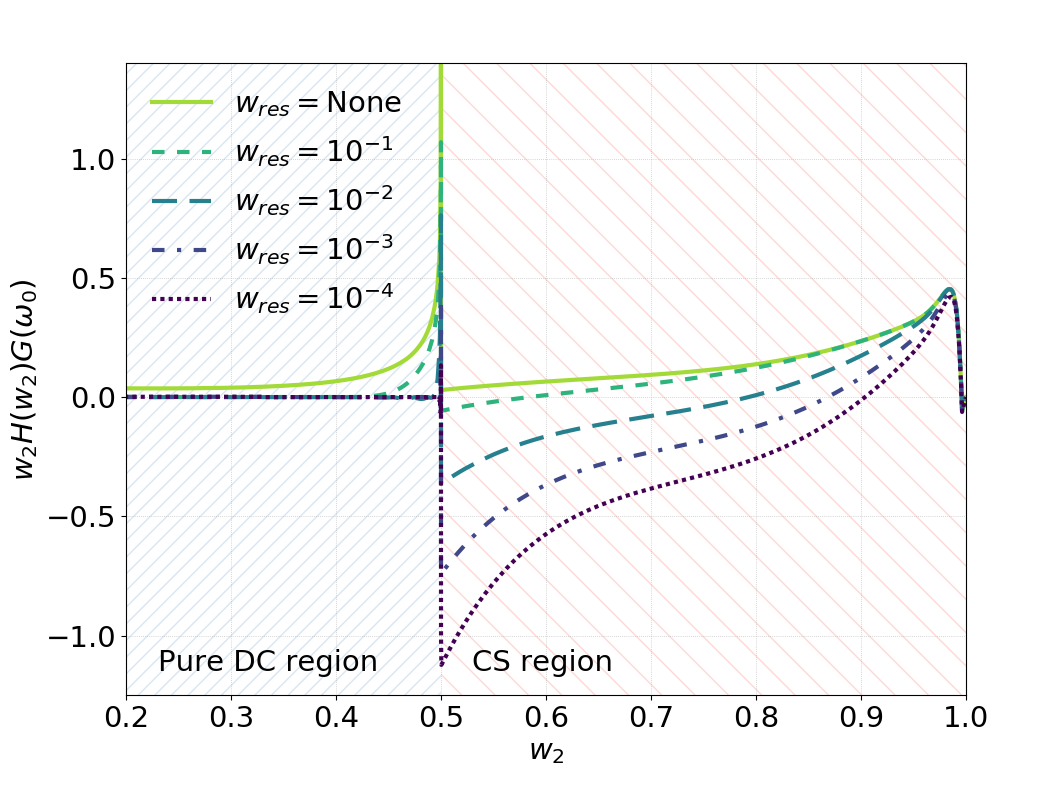}
\caption{Left --- comparison of the scattering kernels for Compton scattering at tree and 1 loop level (in absolute value), DC scattering, and sum of the latter two contributions.
Right --- dependence of the total order $\alpha^3$ contribution to the Compton scattering as a function of the energy resolution.}
\label{fig:Non-divergent-sum}
\end{figure}

In figure \ref{fig:Non-divergent-sum} we display the leading and next to leading order contributions to the Compton kernel and the double Compton kernel for a specific choice of initial conditions ($p=0$, $\omega_0= 0.5$), but the result generalises to any other.
In the left panel we show the interplay among the various contributions that cancel each other.
The Compton scattering emissivity is shown for reference (dotted line, rescaled by a factor 100).
The dashed line is the DC kernel, which depends on the cut-off (cf. figure \ref{fig:H_DC_rest}, right panel), here displayed for $\wmin=10^{-4}$.
The dash-dotted line is the Compton scattering 1-loop correction, that has exactly the same dependence on the cut-off.
The sum of the two is the solid line and is independent of $\wmin=10^{-4}$ in the Compton scattering region, as expected and discussed below.
Since we are integrating both order $\alpha^3$ corrections over all energies of the other photon,  there is no dependence on any energy resolution of the system, a fact that will play a role in a future work where we will discuss the contribution of DC scattering to the net energy transfer.

In the right panel we show instead the dependence of the order $\alpha^3$ correction on the system energy resolution $w_\text{res}$, therefore assuming that resolved DC events do not contribute to the Compton scattering probability.
Considering resolved and unresolved processes separately forces us to define the system energy resolution on a case by case basis.

\begin{figure}
\centering
\includegraphics[width=0.55\columnwidth]{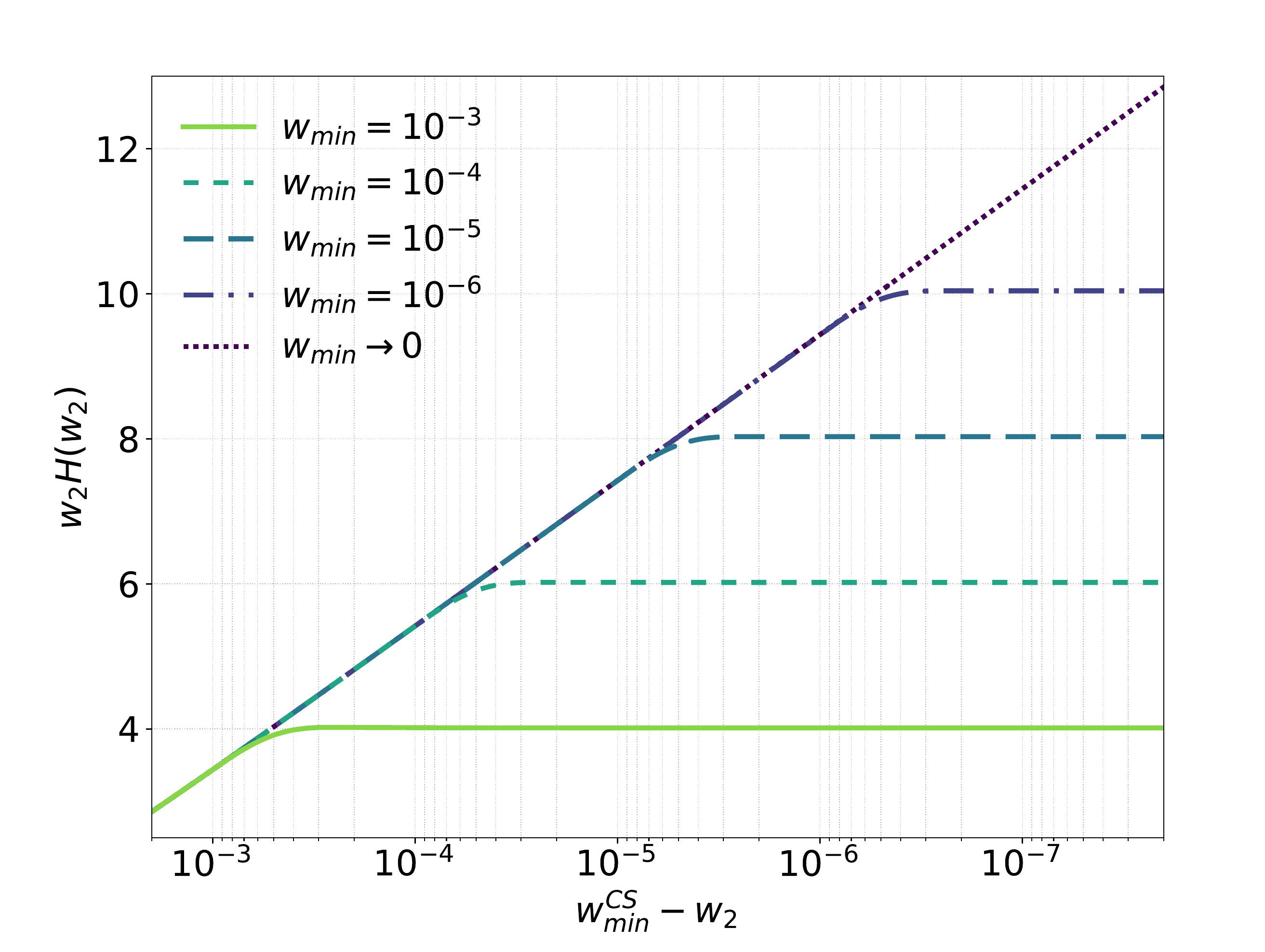}
\caption{Focused view of the DC kernel, \emph{considering the CS radiative corrections}, just below the CS minimal energy $\omegamin^\text{CS}$.
Since the CS kernel is zero on this range of emitted particle energy, the DC scattering probability is not compensated, and depends on $\wmin$, eventually diverging when $\wmin \rightarrow 0$.}
\label{fig:LeftSideDivergence}
\end{figure}

We notice that the normalization approach relies on the fact that all divergent contributions at a given order are proportional to the tree level cross section, and as such are physical on the same initial and final real particle configurations.
However, as we discussed, the double Compton displays a low energy tail \emph{below} $\omegamin^\text{C}$, the minimal energy allowed in Compton scattering.
Figure \ref{fig:LeftSideDivergence} shows how, since below this energy the probability is not compensated by the associated 1-loop Compton scattering counterpart, it still depends on the cut-off, and furthermore logarithmically diverges in  $\omega \rightarrow \omega_\text{min}^\text{CS}$ when $\wmin \rightarrow 0$, unless a finite system energy resolution $w_\text{res}$ is taken into account.
Considering a finite value of $w_\text{res}$, is in practice equivalent to just taking $\wmin=w_\text{res}$.
Beside highlighting the role of the system energy resolution, we do not propose a full and consistent solution to this residual divergence,
but we point out that a more sophisticated treatment that fully accounts for plasma effects, and does not relies on asymptotic free particle initial and final states is probably needed to fully understand the physics.
The residual divergence does not constitute a problem as any physical quantity (e.g., the transferred energy, or the number of emitted photon in a finite energy interval) is finite as they involve the frequency integral of the logarithmic divergence, which converges.
On top of this we have to consider that in any physical system a lower cutoff is always present \cite{Heitler:1936jqw} --- the finite size of the laboratory, the finite size of the horizon in an empty universe, and the aforementioned plasma effects in the case that is relevant to us.

\section{Isotropic DC Emissivity}
\label{sec:emissivity}
In cosmology, the main interest in double Compton scattering comes from the fact that it is not a photon number conserving process.
In the early Universe --- as already explained this applies to other astrophysical environments too --- the photon bath thermalizes through the interaction with the free electrons
\cite{Sunyaev:1970er, 1970CoASP...2...66S, Hu:1992dc, Chluba:2011hw}.
At the energies we are interested in (below the pair production threshold)
any energy injected in the photon bath is redistributed among the photons with different frequency by Compton scattering.
Since it is a photon-number conserving scattering, through this process the photon bath can only relax to a Bose-Einstein distribution \citep{Sunyaev:1970er}.
However, double Compton and Bremsstrahlung cooperate by adjusting the photon number density ---they do so mainly emitting or absorbing low frequency photons--- so that, given enough time, the photons can return to Planckian.
In this section we are therefore interested in quantifying the amount and distribution of the photons newly emitted or absorbed by double Compton scattering.

In order to simplify the problem both from a conceptual and analytical point of view, and to make the numerical calculation more manageable, it is helpful to separate its two aspects: The double Compton process is intrinsically a correction to a Compton event, and as such it participates to the energy redistribution, but differs from it in that it emits extra photons, thereby violating number conservation.

We thus divide the scattering in two parts: a pure emission event and a Compton-like scattering (CLS) event.
The two indistinguishable emitted photons can always be labelled by their final energy.
By tracking separately the evolution of the softer and of the harder photon, the two parts are automatically distinguished.
In practice
\begin{equation}
	\left.
		\pd{n(\vec k_2)}{t}
	\right|_\text{DC}
	=
	\left.
		\pd{n(\vec k_2)}{t}
	\right|_\text{DC} ^\text{em/abs}
	+
	\left.
		\pd{n(\vec k_2)}{t}
	\right|_\text{DC} ^\text{CLS} , 
\label{eq:dn2dt_split_Em_CLS}
\end{equation}
where we defined
\begin{gather}
	\left.
		\pd{n(\vec k_2)}{t}
	\right|_\text{DC} ^\text{em/abs}
	\equiv
	\left.
		\pd{n(\vec k_2 \, | \, \omega_2 < \omega_1)}{t}
	\right|_\text{DC} ^\text{EP $\rightarrow$}
	-
	\left.
		\pd{n(\vec k_2 \, | \, \omega_2 < \omega_1)}{t}
	\right|_\text{DC} ^\text{EP $\leftarrow$}
	\label{eq:dn2dt_split_Em}
\\
	\begin{split}
		\left.
			\pd{n(\vec k_2)}{t}
		\right|_\text{DC} ^\text{CLS}
		\equiv
		&
		\left.
			\pd{n(\vec k_2 \, | \, \omega_2 > \omega_1)}{t}
		\right|_\text{DC} ^\text{EP $\rightarrow$}
		-
		\left.
			\pd{n(\vec k_2 \, | \, \omega_2 > \omega_1)}{t}
		\right|_\text{DC} ^\text{EP $\leftarrow$}
	\\
		&
		-
		\left.
			\pd{n(\vec k_2)}{t}
		\right|_\text{DC} ^\text{PP $\rightarrow$}
		+
		\left.
			\pd{n(\vec k_2)}{t}
		\right|_\text{DC} ^\text{PP $\leftarrow$}\,.
	\end{split}
	\label{eq:dn2dt_split_CLS}
\end{gather}
This is a trivial regrouping of terms, but allows us to take a big step forward in the derivation of kinetic equations that account for the exact double Compton scattering probability. It also naturally appears in the Gould limit by integrating only half of the emission profile.

\begin{figure}
\includegraphics[width=\columnwidth]{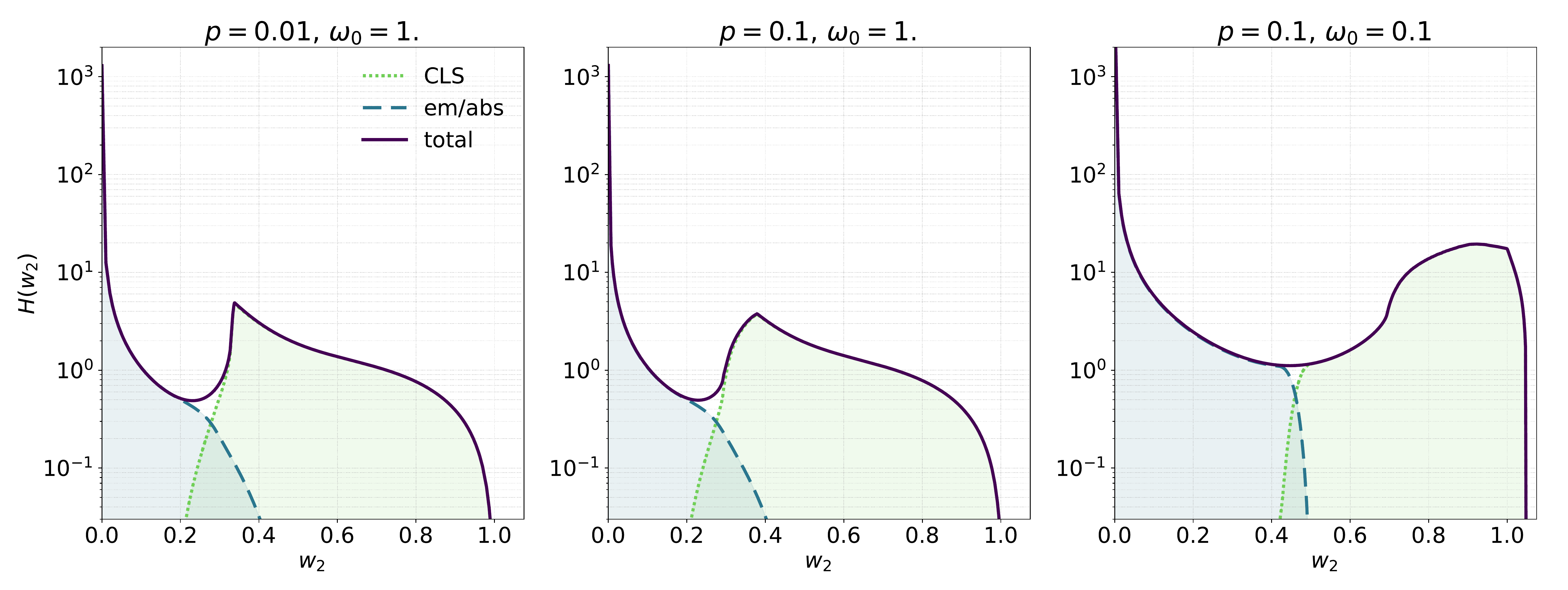}
\caption{The em/abs and CLS contributions to the total EP forward emissivity profile, according to eq. \eqref{eq:dn2dt_split_Em} and eq. \eqref{eq:dn2dt_split_Em_CLS}.
The em/abs term captures the emission of an extra photon at low energies (dark green region), while the CLS term the energy redistribution of the photon in the initial state (light green region).}
\label{fig:Label_hi_low_split}
\end{figure}

In figure \ref{fig:Label_hi_low_split} we show how the total EP$\,\rightarrow$ emissivity is divided in the em/abs and CLS terms.
We checked that the number density of emitted photon, i.e. the integral over $\vec{k_2}$, for both processes amounts to half of the total (within numerical accuracy) as it should be since the two photons are indistinguishable and the system is symmetric when the two emitted photons energies are swapped.
This implies that either part of the EP kernel account for the emission of {\it one single photon}.
Combined with the fact that the PP terms describe the disappearance/reappearance of one of the photons present in the original distribution, we can infer that, as wanted, the CLS term describes the scattering of one photon from one portion to the phase space to another one, without adding or subtracting additional ones, whereas the em/abs term is responsible for the latter effect.
The em/abs term can be easily added to thermalization/Boltzmann solvers such as \texttt{CosmoTherm} \cite{Chluba:2011hw}, as a pure source or sink term that creates or destroys photons without energetically redistributing the existing ones.
The CLS term instead is number conserving and its kernel is (often narrowly) distributed around the energy of the initial particle, in a similar fashion to the Compton scattering.
In that sense, it can be described by a Kompaneets-like evolution equation, however, we leave the development of such a procedure to a future work.
As pointed out also in \cite{Dittmaier:1993bj}, considering the CLS term together with the Compton scattering, by assuming no resolution of the second photon, will automatically prevent any divergence.
We further note that the em/abs and CLS terms do not conserve energy separately, but do when considered together.

In order to talk about a photon belonging to the em/abs term, we have to implicitly assume that it has a high enough energy to be resolved by the system ($\omega_2 > \omega_\text{res}$, for some system resolution if we wish).
Given that $\omega_2 < \omega_1$, the other emitted photon must be likewise resolved.
As such, the process is always distinguished from a (1-loop) Compton scattering, and therefore their probabilities do not interfere.
In fact, the em/abs spectrum is now independent of the cut-off $\wmin$, a fact that we also verified numerically.

On the other hand, considering the CLS term, we have the opposite. The emission of a photon with energy $\omega_2$ is paired with the emission of a photon whose energy can be either above or below the system energy resolution.
In the latter case, we have to assume an agnostic approach and admit the possibility that the observed photon participated to a Compton scattering rather than in a CLS \cite{Brown:1952eu}.
As discussed in section \ref{sec:IR_div}, in this case we can add the 1-loop Compton scattering probability and safely take the limit $\wmin \rightarrow 0$, leaving the probability as a sole function of the system energy resolution (plasma frequency in the astrophysical context).
To circumvent the unsatisfactory problem of having to define the plasma frequency, when we will discuss the CLS contribution to the energy redistribution in a future work, we can consider the total correction to the Compton scattering at $\mathcal{O}(\alpha^3)$, which is automatically integrated from $0$ to $\omega_\text{max}$.

Our findings and approximations can be applied to any distribution of photons and electrons.
However we argue that three limits cover the vast majority of cases needed in Cosmology:
i) thermal electrons interacting with thermal photons (not necessarily with the same temperature and with non-zero photon chemical potential), which is the case one encounter in a vanilla primordial Universe;
ii) highly energetic photons scattering on (low-temperature) thermal electrons; and iii) highly energetic electrons scattering on (low-temperature) thermal photons. The latter two can be relevant in non-standard energy release scenarios caused by particle decay or primordial black hole evaporation \cite{Chluba:2015hma, SandeepEtAl}. We now consider each in turn.

\subsection{Thermal electrons and photons}
To compare our results with the literature \cite{ChlubaThesis, Chluba:2011hw}, and assess what the impact of the correction on the thermal history could be, we use the improved numerical result to calculate the emissivity in the case of thermal electrons, and blackbody photons. While discussing about thermal particles, it is often useful to express frequencies in units of dimensionless electron temperature $\theta_\text{e} \equiv k_B \Te/\me c^{2}$. For this purpose we also introduce $x_i=\omega_i/\theta_\text{e}$. Assuming that $x_0\approx x_1+x_2$ holds, one only needs to consider the forward emission process, while the DC absorption process can be obtained using detailed balance. As stated above, we will also only focus on the emission process, neglecting the CLS.

In \cite{Chluba:2011hw}, it was argued that the emissivity due to DC scattering, at low frequencies and close to thermodynamic equilibrium [$\theta_{\rm e}\approx\theta_\gamma$ and $n\approx n_{\rm Pl} \equiv 1/({\rm e}^x-1)$] is given by
\begin{gather}
	\left. \pd{n(\vec{k_2})}{t} \right|_\text{DC}^\text{em/abs}
	\approx
	\frac{4\alpha}{3\pi} \sigmaT
	\nume
	\frac{\theta_\text{e}^2\,\mathcal{I}^\text{pl}_4}{x_2^3}
	\left[
		1-n(\omega_2)\left(e^{x_{2}}-1\right)
	\right]
	g_\text{th} (x_2, \theta_\text{e}) \, ,
\label{eq:dn2dt_emabs_CosmoTherm}
\\[2mm]
\begin{split}
	g_\text{th} (x_2, \theta_\text{e})
	\approx	
	&\,
	\tilde{G}_\text{soft,th} (\theta_\text{e}) \,
	\tilde{H}_\text{DC} (x_2)\, ,
\\	
	\tilde{G}_\text{soft,th} (\theta_{\rm e})
	\approx
	&\,
	\frac{1}{1+14.16\,\theta_{\rm e}} \, ,
\\[1mm]
	\tilde{H}_\text{DC}(x_2)
	\approx	
	&\,
	\frac{1}{\mathcal{I}^\text{pl}_4}
	\int_{2x_2}^\infty
	\diff x_0 \, x_0^4 \, n_\text{Pl}(x_0)[1+n_\text{Pl}(x_0-x_2)] 
	\left[
		\frac{x_0}{x_2} H_\text{G}\left(\frac{x_0}{x_2}\right)
	\right]
	\, ,
\\[1mm]
    \mathcal{I}^\text{pl}_4 
    \equiv	
	&\,
	\int \diff x \, x^4 \,n_\text{Pl} (x) [1+n_\text{Pl} (x)] =\frac{4\pi^4}{15}\approx 25.98\, .
\end{split}
\end{gather}
Setting $g_\text{th} (x_2, \theta_\text{e})=1$, we obtain the \Lightman-\Thorne approximation. The integral in $\tilde{H}_\text{DC}(x_2)$ can furthermore be approximated as \citep{Chluba:2011hw}
\bsub
\label{eq:Lam_CS}
\begin{align}
\tilde{H}_\text{DC}(x_2)&\approx \expf{-2x}\left[1+\frac{3}{2}x+\frac{29}{24}x^2+\frac{11}{16}x^3+\frac{5}{12}x^4\right],
\end{align}
\esub
which highlights that at high frequencies, the DC emissivity is exponentially suppressed by $\expf{-2x}$, thus dropping faster than for BR.
These expressions provide the main description for the DC process in the computations of {\tt CosmoTherm}.
Here, we expand this result by adding the improved scaling at low frequency, captured by the DC Gaunt factor, $G_\text{soft}(\omega_0, p)$, and the improved Gould function, $H(w,\omega_0, p)$, introduced here, overall yielding
\begin{equation}
	g_{\rm th}(x_2, \theta_\text{e})
	\approx
	\! \int \!
	\frac{\diff p \, p^2 \fe(p, \theta_{\rm e})}{2\pi^2 \nume}
	\! \int \!
	\diff x_0 \, x_0^4 \, n_\text{Pl}(x_0)[1+n_\text{Pl}(x_0-x_2)] 
	\frac{G_\text{soft}(\omega_0, p)	
	w_2 H(w_2|\omega_2 \! < \! \omega_1, p)}{\mathcal{I}^\text{pl}_4} \, .
\label{eq:improved_Gthermal}
\end{equation}
Only one approximation enters here through the presence of stimulated DC scattering terms $\propto (1+n_1)$. In general, the angle integrals cannot be carry out independently, as the electron carries away some of the energy, violating $\omega_1=\omega_0-\omega_2$. However, stimulated terms usually only contribute at the few percent-level and for moderate temperatures the energy of the ``scattered'' photon $\omega_1$ furthermore remains close to $\omega_0$. The approximation should thus suffice for a broad range of temperature and extensions beyond this limit will be left to future work.

\begin{figure}
\includegraphics[width=0.497\linewidth]{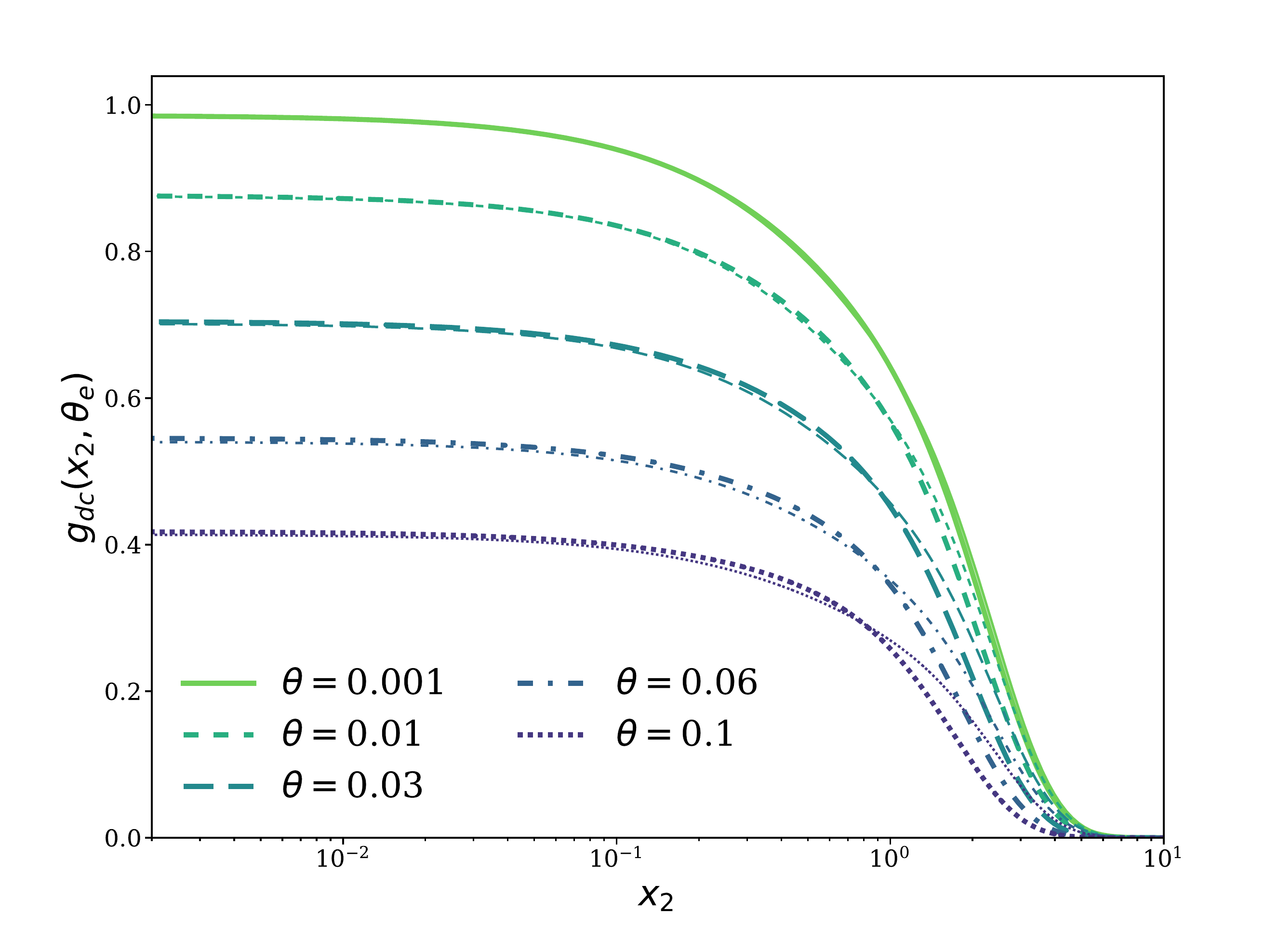}
\includegraphics[width=0.497\linewidth]{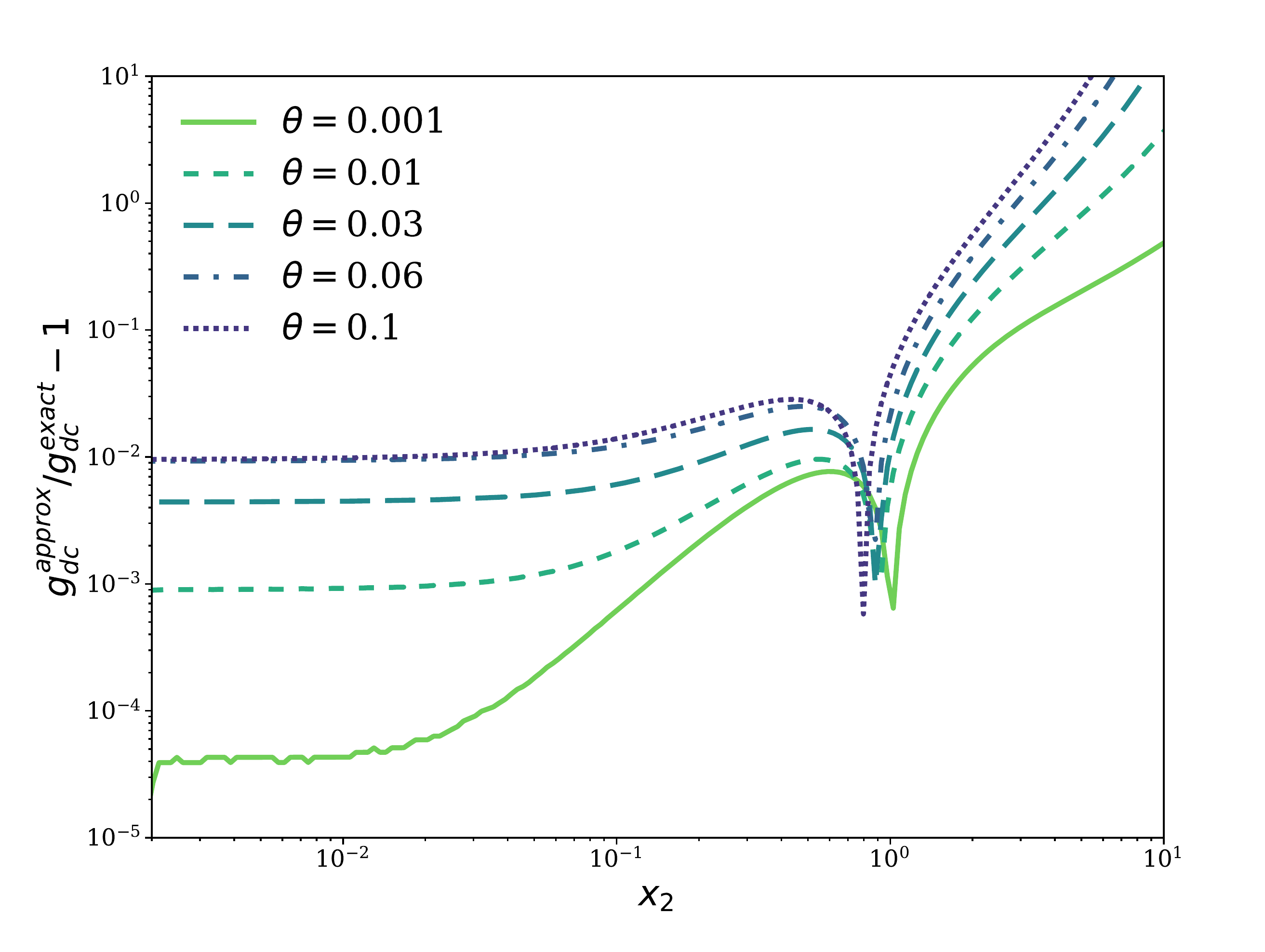}
\caption{Comparison of the exact (this work) and approximated DC emissivity. The electrons are assumed to be thermal with temperature, $\The$, and the photons follow a Planck distribution at the same temperature.
Left -- absolute value of $G_{\rm th}(x_2, \The)=\mathcal{I}^\text{pl}_4\,g_{\rm dc}(x_2, \The)$.
Thin lines represent the analytic approximation, while thick lines are the exact numerical result. Right -- fractional difference.
} 
\label{fig:Thermally_averaged_emissivity}
\end{figure}

In figure \ref{fig:Thermally_averaged_emissivity} we show the results for different electron and photons temperatures, comparing the exact expressions obtained numerically with the approximation.
To accelerate the computations, we tabulated the required functions before carrying out the thermal averages.
We confirm that in the regime relevant in the early universe, assuming a vanilla thermal history, the approximation is good to better than 1\% at low frequency, where most of the photon are indeed created.
On the other hand, the approximation deviates from the exact result at high frequency as the temperature increases. This reflect the more complex structure of $H_G$ with respect to the Gould result, one obtains when increasing $\omega_0$ and $p$ (see figure \ref{fig:HG_varying_p} and \ref{fig:HG_varying_Om0} with respect to figure \ref{fig:H_DC_rest}, left panel). However, the number of photons created at these energies is small compared to the soft photon part, such that the effect on the computations of {\tt CosmoTherm} should remain limited. Nevertheless, {\tt DCpack} now allow us to accurately include all these corrections for precise thermalization computations.

\subsection{Thermal electrons and energetic photons}
The results of the previous paragraph further simplify if we consider the scattering of non-thermal mono-energetic photons on thermal electrons. In this regime, we can generally assume that stimulated effect do not play a role: the energetic photons necessarily have to be few in number to evade current constraints on spectral distortion amplitude. In addition, energy conservation prevents the reverse process from being important. Equation~ \eqref{eq:dn2dt_emabs_CosmoTherm} then becomes
\begin{equation}
\label{eq:DC_nt_gamma}
	\left. \pd{n(\vec{k_2})}{t} \right|_\text{DC}^\text{em}
	\approx 
	\frac{4\alpha}{3\pi} \sigmaT
	\nume
	(2\pi^2)
	N^{\rm nt}_\gamma
	\frac{\omega_0^2}{\omega_2^3}
	\left[ 1+n_2\right]
    \int \frac{\diff p \, p^2 \fe(p)}{2\pi^2 \nume}
	G_\text{soft}(\omega_0, p)\,	
	w_2 H_G(w_2|\omega_2<\omega_1) \, ,
\end{equation}
where $N^{\rm nt}_\gamma$ defines the number density of the non-thermal photons around $\omega_0$. 
To compare with the DC emission of the thermal plasma, we observe that the non-thermal DC emissivity increases as
\begin{align}
\left. \pd{n(\vec{k_2})}{t} \right|_\text{DC}^\text{em}
&\propto
\frac{\omega^2_0}{\omega_2^3}\,(2\pi^2)N^{\rm nt}_\gamma
= \frac{x_0^2}{\The x_2^3}\,(2\pi^2)N^{\rm nt}_\gamma
=\frac{\mathcal{I}^\text{pl}_4\The^2}{x_2^3}\,\left[\frac{x_0^2}{x_{\rm th}^2}\,\frac{N^{\rm nt}_\gamma}{N^{\rm Pl}_\gamma}\right],
\end{align}
with the photon energy. Here, we used $G_2^{\rm Pl}=\int x^2 n_{\rm Pl}(x)\id x\approx 2.404$, $N^{\rm Pl}_\gamma=\The^3G_2^{\rm Pl}/(2\pi^2)$ and $x_{\rm th}=(\mathcal{I}^\text{pl}_4/G_2^{\rm Pl})^{1/2}\approx 3.287$. The ratio $x_0^2/x_{\rm th}^2$ can boost the non-thermal DC emissivity by a large factor, however, in the remaining integral of eq.~\eqref{eq:DC_nt_gamma}, $G_\text{soft}(\omega_0, p)$ causes a net suppression at large $\omega_0$. For energy release scenarios in the early Universe and due to the small number of thermal target electrons, this process is thus not expected to be able to compete with the thermal DC production rate, given that $N^{\rm nt}_\gamma\ll N^{\rm Pl}_\gamma$ \citep[see discussion in][]{Chluba:2015hma}.

\subsection{Thermal photons and energetic electrons}
The same procedure can be employed in the opposite case of thermal photons and mono-energetic non-thermal electrons. From this we obtain
\bsub
\begin{gather}
	\left. \pd{n(\vec{k_2})}{t} \right|_\text{DC}^\text{em}
	\approx
	\frac{4\alpha}{3\pi} \sigmaT
	\nument
	\frac{\mathcal{I}^\text{pl}_4\,\The^2}{x_2^3}
	\left[
		1+n(\omega_2)
	\right]\,g_{\rm dc}^\text{nt-e}(x_2, p)
\\[1mm]
g_{\rm dc}^\text{nt-e}(x_2, p)=
\,\int \diff x_0 \, x_0^4 \, n_0 (1+n_1)
	\frac{G_\text{soft}(\omega_0, p)	
	\,w_2 H_G(w_2|\omega_2<\omega_1)}{\mathcal{I}^\text{pl}_4} \, ,
\end{gather}
\esub
where $\nument$ is the number density of the non-thermal electrons and we assumed Planckian ambient photons at a temperature $\Thg=k\Tg/\me c^2=\The$ of the thermal electrons. Setting $g_{\rm dc}^\text{nt-e}(x_2, p)=1$, we essentially recover the thermal DC emissivity but suppressed by the ratio $\nument/\nume\ll 1$. No non-thermal absorption term is present, but DC emission is stimulated by the presence of thermal photons around $\omega_2$. 

In contrast to non-thermal photons considered above, the DC Gaunt-factor $g_{\rm dc}^\text{nt-e}(x_2, p)$ rises strongly with the momentum of the non-thermal electron until $4\left<\omega_0\right>_{\rm th} p \approx 11 \The \,p\gtrsim 1$.
Here, $\left<\omega_0\right>_{\rm th}\approx 2.7\,\The$ is the average energy of the blackbody photon field. Before the critical momentum $p_{\rm cr}\simeq 0.09/\The$ is reached, we have $g_{\rm dc}^\text{nt-e}(x_2, p)\approx (24.89/\mathcal{I}^\text{pl}_4)(2p^2+1)$ and hence an enhancement of the DC emissivity by $\simeq 2p^2$ for large $p$.
Here we set the integral $\int \diff x_0 \, x_0^4 \, n_0 (1+n_1)\approx \int \diff x_0 \, x_0^4 \, n_0\approx 24.89$, since the up-scattered photon ($\omega_1$) will be moved far into the Wien-tail of the ambient blackbody, such that stimulated effects become tiny.
For typical temperatures relevant to the cosmological thermalization problem, even momenta up to $p_{\rm cr}\simeq 100$, can be expected. While this is unlikely to overcome the direct thermal rate, because no absorption term is present this can still create a noticeable net contribution.
In addition, the inverse Compton up-scattering by the non-thermal electrons removes photons from the ambient blackbody field \citep{Ensslin2000, Slatyer2015, Acharya2020JCAP} and the non-thermal DC photons may at least partially replenish them. We leave a more detailed investigation to the future.

\section{Conclusions}
\label{sec:conclusions}

In this paper, we carried out a thorough analysis of the double Compton scattering process covering a wide range of energies that are of interest to applications in cosmology and astrophysical plasmas, developing the code {\tt DCpack}. This brings us one step closer to an exact description of the DC mechanism using modern numerical methods, as already envisioned by Mandl and Skyrme in their original work on this process \cite{Mandl52}.

We derived general expressions for the soft photon emissivity, that allows us to quickly and reliably calculate the double Compton Gaunt factor for any initial particle state. This expression can readily replace the approximations commonly used in any application that requires the modelling of the DC scattering.

We accurately described the properties of the DC kernel beyond the soft photons approximation running a suite of numerical calculation. In this way, we assessed the deviations from the standard approximation, that are sizeable at mildly relativistic energies. We also review the role of Compton scattering divergences and radiative corrections, from the unusual perspective of being interested in the emission of extra photons inside astrophysical plasmas. In this context, we pointed out some problems related to residual divergences and the impact of stimulated terms in 1-loop QFT calculations, that are generally overlooked in the astrophysics literature. We offer possible solutions related to the presence of plasma effect, that have also been discussed by the QFT and plasma physics communities.

We introduced a procedure that allows us to separately describe the two aspects of the DC scattering: the emission of extra photons and the energy redistribution of the existing ones.
With this, we find improved description for the DC emissivity, that we compare with the standard approximations used for the cosmological thermalization problem \citep{Chluba:2011hw}, finding good agreement for the conditions encountered in vanilla thermal histories, with small energy release.
On the other hand, we point to sizeable deviations that are found when considering non-standard high energy injection scenarios, as for example discussed in \cite{Chluba:2015hma}, \cite{Acharya2020JCAP} and \cite{SandeepEtAl}.
With {\tt DCpack}, we can hope to explore these cases in the future.

\acknowledgments
This work was supported by the ERC Consolidator Grant {\it CMBSPEC} (No.~725456) as part of the European Union's Horizon 2020 research and innovation program.
JC was supported by the Royal Society as a Royal Society URF at the University of Manchester, UK.

{\small
\appendix
\section{Squared matrix element for double Compton}
\label{appendix:M2}
To obtain the differential DC cross section we need the squared matrix
element, $|\mathcal{M}|^2=e^6\,X$, of the process. It was first obtained by \cite{Mandl52} and is also given in \cite[][pp. 235]{Jauch1976}:
\begin{subequations}
\begin{equation}
\begin{split}
X
&=2\left( a\,b - c\right)\big[ \left( a + b \right)
\left( 2+ x \right) -(a\,b - c) -8\big] - 2\,x\left[ a^2 + b^2 \right] 
-2\,\left[a\,b + c\left( 1 - x \right) \right]\rho
\\ 
& -8\,c +\frac{4\,x}{A\,B}\,\left[ \left( A + B \right) \,\left( 1 + x
\right) + x^2\,\left( 1 - z \right) + 2\,z - \left( a\,A + b\,B \right)
\,\left( 2 + \frac{\left( 1 - x \right) \,z}{x} \right) \right]
,
\end{split}
\end{equation}
where the following abbreviations were used
\begin{gather}
		a
		=
		\frac{1}{\kappa_0}+\frac{1}{\kappa_1} + \frac{1}{\kappa_2} \, ,
	\quad
		b
		=
		\frac{1}{\kappa'_0} + \frac{1}{\kappa'_1} + \frac{1}{\kappa'_2} \, ,
	\quad
		c
		=
		\frac{1}{\kappa_0\,\kappa'_0}+ \frac{1}{\kappa_1\,\kappa'_1} + \frac{1}{\kappa_2\,\kappa'_2} \, ,
\\[1mm]
		x
		=
		{\kappa_0} + {\kappa_1} + {\kappa_2} \, 
		\equiv
		{\kappa'_0} + {\kappa'_1} + {\kappa'_2} 
		= y\, ,
	\quad
		z
		=
		\kappa_0\,\kappa'_0+ \kappa_1\,\kappa'_1 + \kappa_2\,\kappa'_2  \, ,
\\[1mm]
		A
		=
		{{\kappa}_0}\,{{\kappa}_1}\,{{\kappa}_2} \, ,
	\quad
		B
		=
		\kappa'_0\,\kappa'_1\,\kappa'_2 \, ,
	\quad
	\rho
	=
	\frac{\kappa_0}{\kappa'_0}+\frac{\kappa'_0}{\kappa_0}
	+\frac{\kappa_1}{\kappa'_1}+\frac{\kappa'_1}{\kappa_1} 
	+\frac{\kappa_2}{\kappa'_2} + \frac{\kappa'_2}{\kappa_2} \, .
\label{eq:MatAbr}
\end{gather}
For the definitions of $\kappa_i,\,\kappa'_i$ we adopt those of the original paper
from \cite{Mandl52}:
\begin{align}
\label{eq:KAPPAs}
		\me^2\,\kappa_0&= -P\cdot K_0 \, , &
	\quad
		\me^2\,\kappa_1&= +P\cdot K_1 \, , &
	\quad
		\me^2\,\kappa_2&= +P\cdot K_2 \, ,
\\
		\me^2\,\kappa'_0&= +P'\cdot K_0 \, , &
	\quad
		\me^2\,\kappa'_1&= -P'\cdot K_1 \, , &
	\quad
		\me^2\,\kappa'_2&= -P'\cdot K_2 \, ,
\end{align}
\end{subequations}
with the standard signature of the Minkowski-metric $(+\,-\,-\,-)$. Here $P=(E,\vec{p}),\,P'=(E',\vec{p}')$ and $K_i=(\nu_i,\vec{k}_i)$ denote the relevant electron ($P$) and photon ($K$) four-momenta.%
\footnote{Henceforth bold font denotes 3-vectors.}
Introducing the products $\alpha_{ij}= \hat{K}_i \cdot \hat{K}_j=1-\vech{k}_i\cdot \vech{k}_j=1-\mu_{ij}$ and $\lambda_i=P \cdot \hat{K}_i=\gamma-p\mu_{{\rm e}i}$ with $\mu_{\text e i} = \vers p \cdot \vers k_i$, for $P'=P+K_0-K_1-K_2$ we then have%
\footnote{Note that in the following an additional hat above 3- and 4- vectors indicates that they are normalized to the time-like coordinate of the corresponding 4-vector.}
\begin{align}
\label{eq:KAPPAs_rewrite}
		\kappa_0&= -\omega_0 \lambda_0\, , &
	\,
		\kappa_1&= +\omega_1 \lambda_1 \, , &
	\,
		\kappa_2&= +\omega_2 \lambda_2 \, ,
\\\nonumber
		\kappa'_0&= \omega_0 (\lambda_0 - \omega_1 \alpha_{01} - \omega_2 \alpha_{02}) \, , &
	\,
		\kappa'_1&= -\omega_1 (\lambda_1 + \omega_0 \alpha_{01} - \omega_2\alpha_{12}) \, , &
	\,
		\kappa'_2&= -\omega_2 (\lambda_2 + \omega_0 \alpha_{02} - \omega_1\alpha_{12}) \,.
\end{align}
These relations also immediately prove $x+y\equiv 2x\equiv 2y \equiv 2(\omega_0\omega_1\alpha_{01}+\omega_0\omega_2\alpha_{02}-\omega_1\omega_2\alpha_{12})$, which can be helpful in analytic derivations.


\section{Squared matrix element for Compton scattering radiative corrections}
\label{appendix:M2_rad}
The matrix element for Compton scattering radiative corrections was first evaluated by Brown and Feynman \cite{Brown:1952eu} and 20 years later confirmed by Tsai and coworkers \cite{Tsai:1972sg}. The process was also studied in detail by \cite{Mork1965, Mork1971}, using the results for the matrix element from \citep{Brown:1952eu}.%
\footnote{Note a typo in eq.~(III.5) of \citep{Mork1965}, where $4y\sinh(y) (\kappa\tau)^{-1} \rightarrow 4y\sinh(2y) (\kappa\tau)^{-1}$. Also, in the forth line in eq.~(30) of \cite{Brown:1952eu} a spurious parenthesis is present.}
The expressions can be simplified by separating contributions according to their properties. We shall use the same notation as in Appendix.~\ref{appendix:M2}.
First we introduce a few auxiliary functions following Brown and Feynman:
\begin{align}
\label{eq:M2_rad_def_aux_funcs}
y(\kappa_0,\kappa_1)&=y(\kappa_1,\kappa_0)={\rm arcsinh}\left(\sqrt{-\frac{(\kappa_0 + \kappa_1)}{2}}\right)\, ,
\\
h(y)&=\frac{1}{y}\int_0^y u \coth u  \id u =\frac{\pi^2+6y^2+12 y \ln(1-\expf{-2y})-6{\rm Li}_2(\expf{-2y})}{y}\, ,
\\
G_0(x)&=-\frac{2}{x}\int_{1-x}^1 \frac{\ln(1-u)}{u} \id u =\frac{\pi^2-6{\rm Li}_2(1-x)}{3 x}\, ,
\end{align}
where $-(\kappa_0 + \kappa_1)=\omega_0\lambda_0-\omega_1\lambda_1>0$ and ${\rm Li}_2(z)$ is the Dilogarithm.
The matrix element for the process then takes the form $|\mathcal{M}|^2_{\rm rad}=e^6\,X_{\rm rad}$, where
\begin{align}
\label{eq:M2_rad_def}
X_{\rm rad}(\kappa_0,\kappa_1,\lambda)&=-\left[2X^{\rm div}_{\rm rad}(\kappa_0,\kappa_1,\lambda)+2X^{\rm sym}_{\rm rad}(\kappa_0,\kappa_1)+X^{\rm asym}_{\rm rad}(\kappa_0,\kappa_1)+X^{\rm asym}_{\rm rad}(\kappa_1,\kappa_0)\right].
\end{align}
with three main pieces, a (symmetric) divergent part that depends on the 'photon mass', $\lambda$
\begin{align}
\label{eq:M2_rad_def_div}
X^{\rm div}_{\rm rad}(\kappa_0,\kappa_1,\lambda)
&=[1-2y \coth (2y)] X_{\rm CS}(\kappa_0,\kappa_1)\,\ln \lambda,
\\
X_{\rm CS}(\kappa_0,\kappa_1)
\label{eq:RadDiv_SigmaCompton}
&=
\left(\frac{1}{\kappa_0}+\frac{1}{\kappa_1}\right)^2
-2\left(\frac{1}{\kappa_0}+\frac{1}{\kappa_1}\right)
-\left(\frac{\kappa_0}{\kappa_1}+\frac{\kappa_1}{\kappa_0}\right)
\end{align}
where $X_{\rm CS}$ is related to the matrix element for the usual Compton process (cf. eq. \eqref{eq:sigma_Compton}), which is symmetric when switching variables, $\kappa_0\leftrightarrow\kappa_1$. Defining $a=\frac{1}{\kappa_0}+\frac{1}{\kappa_1}$, $\rho=\frac{\kappa_0}{\kappa_1}+\frac{\kappa_1}{\kappa_0}$ and $A=\kappa_0\kappa_1$we then may express the symmetric and asymmetric parts of the matrix element as
\begin{gather}
\begin{split}
\label{eq:M2_rad_def_sym_asym}
	X^{\rm sym}_{\rm rad}(\kappa_0,\kappa_1)
	&
	=\left(
		\frac{3}{4}-2y\coth(2y)[2h(y)-h(2y)]
	\right)
	X_{\rm CS}+y\coth(y)\left(4+a +\rho\right) h(y)
\\
	&
	-2y\tanh(y)\left(1-\frac{a}{2}\right)
	+y^2
	\left(
		2-\frac{1}{A}+\frac{3}{2}\rho
	\right)
	+\frac{a^2}{4}(1-2Aa)
\\
	&
	+ 
	\ln(4|A|)
	\left[
		y\, {\rm csch}(2y)\left(1-3a +\rho\right)
		+\frac{1}{2}\left(1+a-\frac{2}{A}\right)
	\right.
\\
	&
	\qquad\qquad\quad
	\left.
		-2y\coth(2y)\left(1+\frac{a}{2}+\rho -\frac{1}{A}\right)
	\right] ,
\end{split}
\\[1mm]
\begin{split}
	X^{\rm asym}_{\rm rad}(\kappa_0,\kappa_1)
	&
	=
	-\frac{2\kappa_0+\kappa_1}{2\kappa_1(1-2\kappa_0)}
	+G_0(\kappa_0)
	\left(
		\frac{2\kappa_0^2}{\kappa_1}+\frac{\kappa_1}{2\kappa_0^2}
		+2\kappa_0+\kappa_1+\frac{1}{\kappa_0}-\frac{3}{2\kappa_1}
		+\frac{\kappa_0}{\kappa_1}-1
	\right)
\\
	&
	+
	\ln(2|\kappa_0|)
	\left\{
		2y \,{\rm csch}(2y)\left[\left(\frac{2}{\kappa_0}
		+\kappa_1\right)\frac{1-\kappa_1}{\kappa_0}-\kappa_1\right]
		+\frac{3+2\kappa_1}{\kappa_0}-\frac{2-\kappa_1}{\kappa_0^2}
	\right.
\\
	&
	\qquad\qquad\qquad
	\left.
		+\frac{\kappa_1^2-\kappa_0(1+\kappa_1+2\kappa_1^2)}{\kappa_1(1-2\kappa_0)^2}
	\right\}
	.
\end{split}
\end{gather}
We note the identity $2y\coth(2y)=\frac{\Delta_0}{p_0}\ln\left(\Delta_0 +p_0\right)$ with $\Delta_0=1-\kappa_0-\kappa_1=1+\lambda_0\omega_0-\lambda_1\omega_1$, which is important when discussing the cancellation of infrared divergences (section~\ref{sec:IR_div}).


\section{Integration boundaries}
\label{app:IntDom}
%
In order to perform an efficient numerical integration, it is useful to use properly defined integration boundaries, instead of integrating a function which is equal to zero in a vast portion of the domain.
Naively assuming that in a double Compton scattering all particles can scatter in any direction, one would fall exactly in this unfortunate case.
Let us recall that to obtain the DC collision term we have to integrate over 6 angular variables $\mue, \muuno, \mutwo, \phie, \phione, \phitwo$.
In principle the polar angles vary in $[-1,1]$ and the azimuthal angles in $[0, 2\pi)$.
However, the considerations about energy made in section~\ref{sec:beyond_soft} imply that the integration domain is effectively reduced.
We start by noticing that the system has to be invariant under rotations around $\vers{z}$.
This is corroborated by the fact that the integrand do not have an absolute dependence on the azimuthal angles, but rather to their difference.
Hence, we can assume to align the coordinate system is such a way that $\phitwo = 0$.
The integration over this variable is therefore trivial, and can be performed right away.
We choose this specific frame rather then the one with $\phie=0$ because in the electron rest frame $\mue$ and $\phie$ are indeterminate.
Since our rest frame is independent of the electron speed, all the equations we present are valid for any combination of initial energies.
Since the electromagnetic interaction is parity invariant, we can further restrict $\phie$ in $[0, \pi]$, arguing that the other half of the circle can be obtained with a parity transformation.
The result has then to be multiplied by $2 \times 2\pi$ to account for the trivial integration just discussed.
In summary, exploiting its symmetries we reduced the problem to a 5 dimensional integral over $\mue, \muuno, \mutwo, \phie, \phione$, with the caveat that $\phie \in [0, \pi]$.

The boundaries on the other variables can be obtained imposing eq.\eqref{eq:w1condition}, and the energy conservation constraint.
By an explicit algebraic calculation, it can be shown that these two conditions are equivalent to 
\begin{equation}
\begin{split}
&
 	  \gamma \, (1 - \beta \mue)
	- \gamma \, w_2 \, (1 - \beta \mu_{\text e2})
	- \omega_0 \, w_2 \, (1 - \mutwo) 
\\
	-
&
	\wmin
	\left[
		  \gamma \, (1 - \beta \mu_{\text e 1})
		+ \omega_0 \, (1 - \beta \mu_{01})
		- w_2 \, \omega_0 \, (1 - \mu_{12})
	\right]
	\geq 0 \, ,
\end{split}
\label{eq:angle_condition}
\end{equation}
where, for $i,j = \text e, 1, 2$,
\begin{equation}
	\mu_{ij} = \mu_{0i} \, \mu_{0j} + \cos(\phi_i - \phi_j) \sqrt{1-\mu_i^2}\sqrt{1-\mu_j^2} \, .
\end{equation}
In principle we can find all the extreme points numerically, or even with an exact analytical analysis, but while the first is computationally expensive, the second is rather unstable when implemented in the numerical code.
We find more cost-effective to use an hybrid approach that relies on some approximations.
Eq. \eqref{eq:angle_condition} is solved analytically for $\mutwo$, assuming fixed all the other parameters, in order to find the exact $\mutwo$ boundaries.
With this approach we will end up with a handful of combination $\mue, \muuno, \phie, \phione$ that do not allow any valid $\mutwo$ value.
Those unphysical configurations are therefore discarded later on.
Since all terms where $\muuno$ and $\phione$ appear are suppressed by a factor $\wmin$, both parameters have little influence on the ``bulk'' of the integration domain. For this reason we do not restrict their domain.
To restrict the domains of $\mue$ and $\phie$ instead of solving eq. \eqref{eq:angle_condition}, we study 
\begin{equation}
	\text{Num} 
	\equiv
 	  \gamma \, (1 - \beta \mue)
	- \gamma \, w_2 \, (1 - \beta \mu_{\text e2})
	- \omega_0 \, w_2 \, (1 - \mutwo) 
	\geq 0 \, .
\label{eq:Numerator_condition}
\end{equation}
Notice how this is equivalent to taking $\wmin = 0$.
The advantage is that eq. \eqref{eq:Numerator_condition} only depends on $\mue$, $\mutwo$, and $\phie$, making the problem more manageable both from the analytical and numerical point of view.
We start from $\mue$. After finding the extreme $\mutwo$ and $\phie$ analytically solving $\partial_{\mutwo} \text{Num} = \partial_{\phie} \text{Num} = 0$, since Num is a continuous function, we can find the $\mue$ boundaries solving numerically $\text{Num} = 0$.
For fixed $\mue$, the same can be done for $\phie$:
we find analytically the extreme $\mutwo$, and then find numerically $\phie$.
Since the domain of all the other parameters is determined independently from $\mutwo$, we can treat the latter as the innermost integration variable, and assume all the other parameters are fixed.
Hence we can recast eq. \eqref{eq:angle_condition} in the form
\begin{equation}
	a + b \, \mutwo
	\geq c \, \sqrt{1- \mutwo^2} \, ,
\label{eq:mutwo_condition}
\end{equation}
where
\begin{equation}
\nonumber
\begin{split}	
		a \equiv 
		&
		\ \gamma
		- \beta \, \mue \, \gamma
		- w_2 \, \omega_0 - w_2 \, \gamma
		+ \muuno \, \wmin \, \omega_0 
		+ \beta \, \muuno \, \mue \, \wmin \, \gamma \, ,
\\
		&
		+ w_2 \, \wmin \, \omega_0 
		- \wmin \, \omega_0 
		- \wmin \, \gamma
		+ \beta  \sqrt{1-\muuno^2} \sqrt{1-\mue^2} \, \wmin \, \gamma  \cos (\phi_1-\phie)
	\, ,
\\
	b \equiv
	&\ 
	\beta \, \mue \, w_2 \, \gamma -\muuno \, w_2 \, \wmin \, \omega_0 +w_2 \, \omega_0
	\, ,
\\
	c \equiv
	&\ 
	  \beta \sqrt{1-\mue^2} \, w_2 \, \gamma  \cos (\phie)
	- \sqrt{1-\muuno^2} \, w_2 \, \wmin \, \omega_0  \cos \phi_1
	\, .
\end{split}
\end{equation}
The solutions therefore are
\begin{equation}
\begin{split}
	&
	\begin{cases}
    	c = 0 \ \land \ b\neq 0
    	\\
    	\mutwo \geq -\frac{a}{b}
	\end{cases}
\\
	\vee
	&
	\begin{cases}
    	c > 0 \\
    	(b < 0 \ \land \ \mutwo \leq -\frac{a}{b})
    	\ \vee \ 
    	(b > 0 \ \land \ \mutwo \geq -\frac{a}{b})
    	\ \vee \ 
    	b = 0
    	\\
    	\mutwo \leq \mutwo^- \ \vee \ \mutwo \geq \mutwo^+ \ \vee \ \Delta < 0
	\end{cases}
\\
	\vee
	&
	\begin{cases}
    	c < 0 \\
    	[(b < 0 \ \land \ \mutwo \leq -\frac{a}{b})
    	\ \vee \ 
    	(b > 0 \ \land \ \mutwo \geq -\frac{a}{b})
    	\\
    	\ \ \vee \ 
    	(b = 0 \ \land \ \Delta \geq 0)
    	\ \vee \ 
    	(\mutwo^- \leq \mutwo \leq \mutwo^+ \ \land \ \Delta \geq 0)]
	\end{cases}
\end{split}
\end{equation}
where
\[
	\mutwo^{\pm} \equiv 
	\frac{-a \, b \pm \sqrt{\Delta/4}}{b^2 + c^2}
	\, ,
\quad
	\frac{\Delta}{4} \equiv c^2\left(b^2 + c^2 - a^2 \right)
	\, .
\]
In figure \ref{fig:IntegDom_OmegaLess} and \ref{fig:IntegDom_OmegaMore} we show how the exact integration domain changes for various $w_2$ values, respectively in the case $\omega_0 < \omegazerocrit$ and $\omega_0 > \omegazerocrit$.
It can be appreciated how the critical points coincide with changes in the integration domain.

\begin{figure}
\centering
\begin{subfigure}{.25\textwidth}
  \centering
  \includegraphics[width=\linewidth]{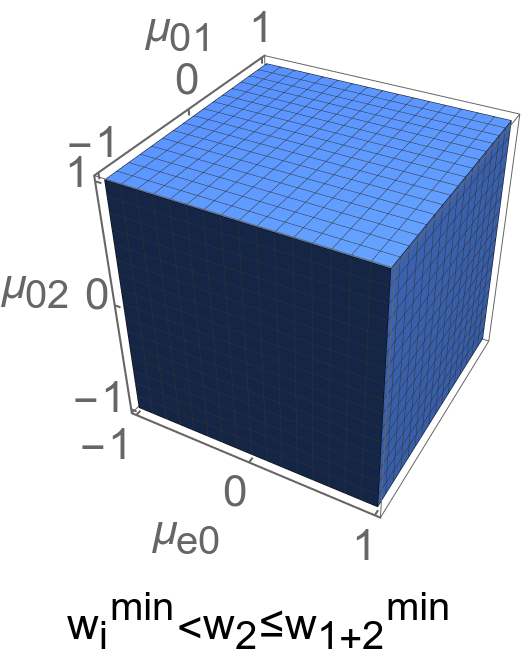}
\end{subfigure}%
\begin{subfigure}{.25\textwidth}
  \centering
  \includegraphics[width=\linewidth]{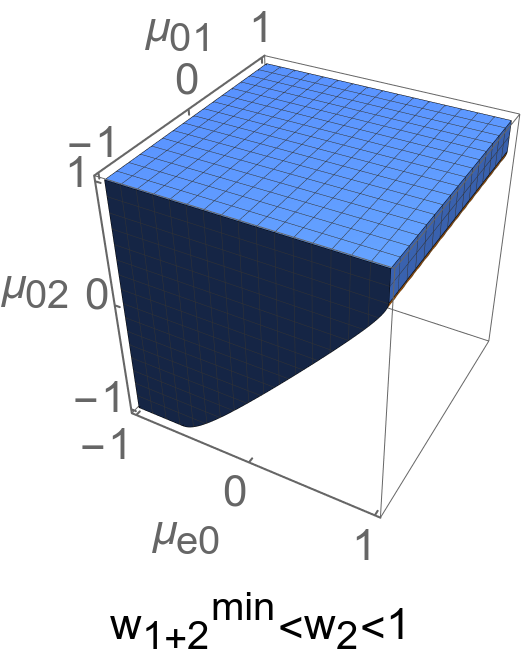}
\end{subfigure}%
\begin{subfigure}{.25\textwidth}
  \centering
  \includegraphics[width=\linewidth]{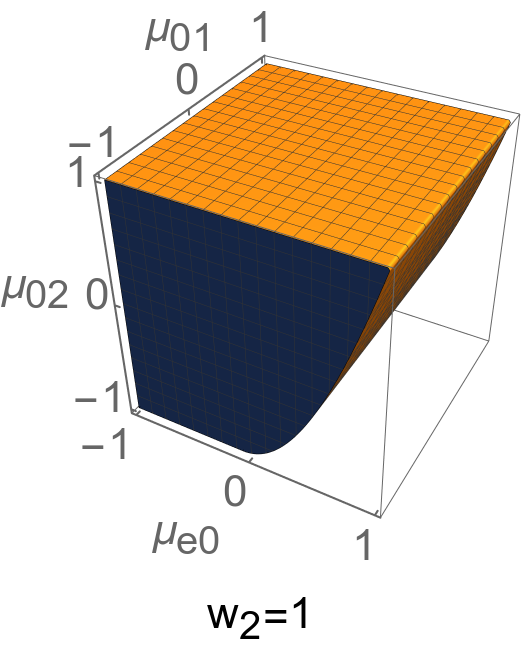}
\end{subfigure}%
\begin{subfigure}{.25\textwidth}
  \centering
  \includegraphics[width=\linewidth]{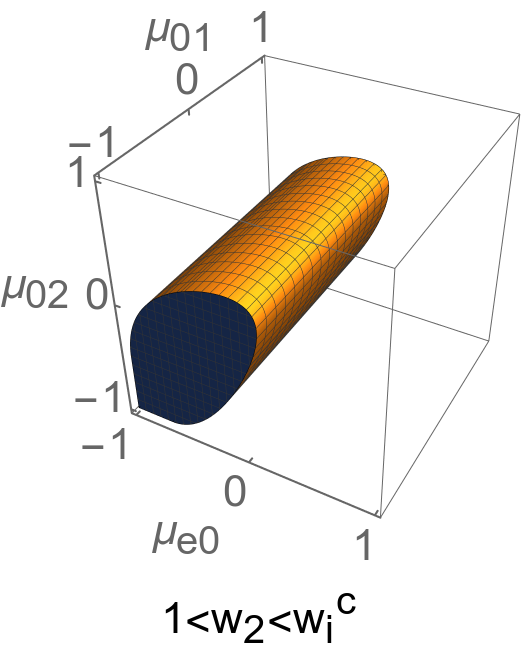}
\end{subfigure}
\caption{Polar angles integration domain for different $w_2$. $\omega_0 < \omega_0^\text{crit}$.}
\label{fig:IntegDom_OmegaLess}
\end{figure}

\begin{figure}
\centering
\begin{subfigure}{.25\textwidth}
  \centering
  \includegraphics[width=\linewidth]{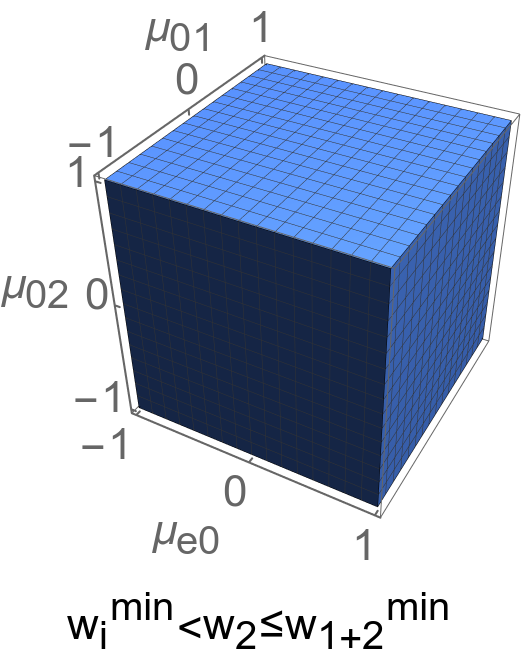}
\end{subfigure}%
\begin{subfigure}{.25\textwidth}
  \centering
  \includegraphics[width=\linewidth]{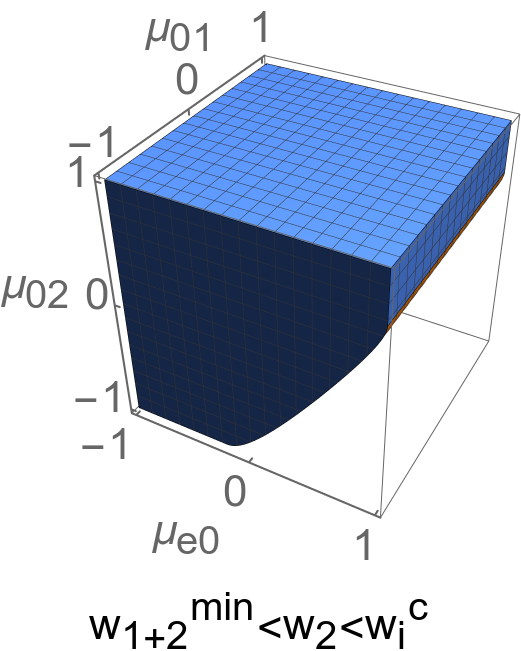}
\end{subfigure}%
\begin{subfigure}{.25\textwidth}
  \centering
  \includegraphics[width=\linewidth]{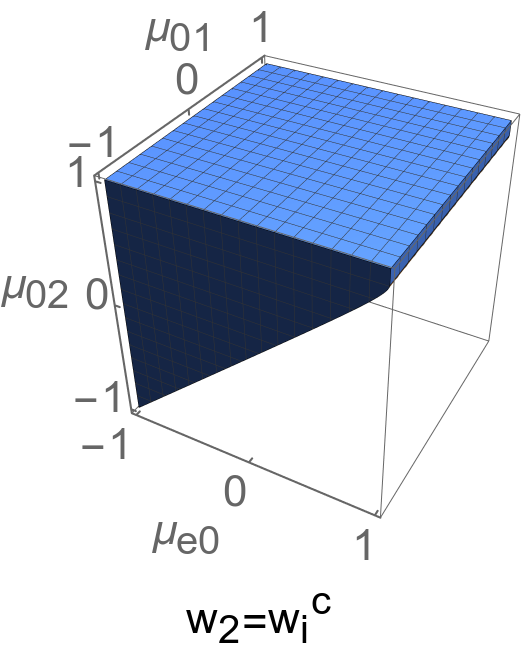}
\end{subfigure}\\
\begin{subfigure}{.25\textwidth}
  \centering
  \includegraphics[width=\linewidth]{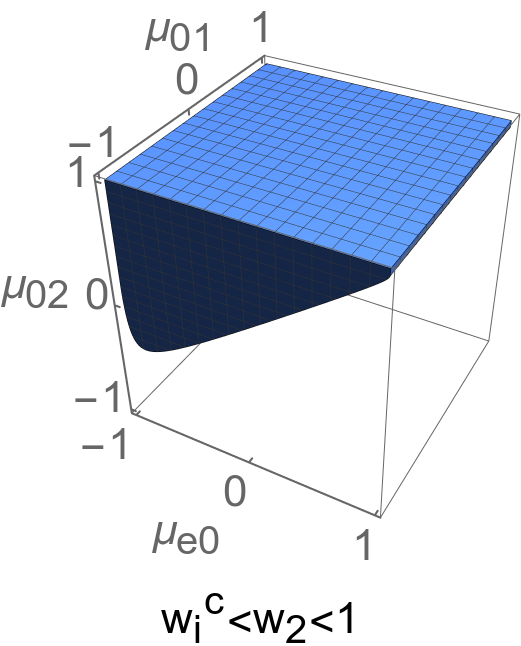}
\end{subfigure}%
\begin{subfigure}{.25\textwidth}
  \centering
  \includegraphics[width=\linewidth]{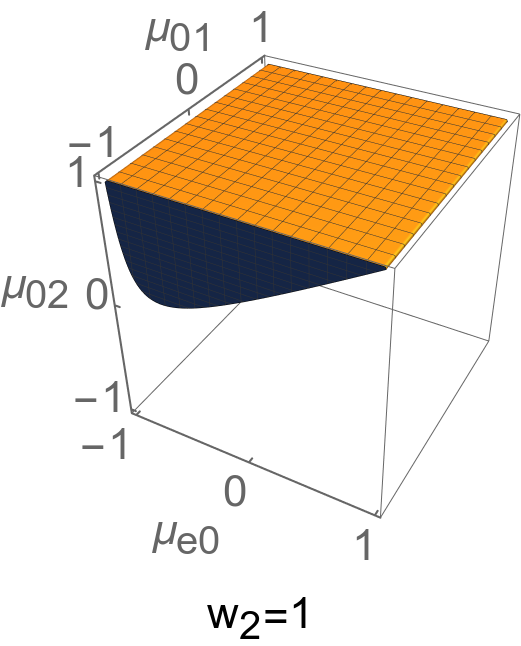}
\end{subfigure}%
\begin{subfigure}{.25\textwidth}
  \centering
  \includegraphics[width=\linewidth]{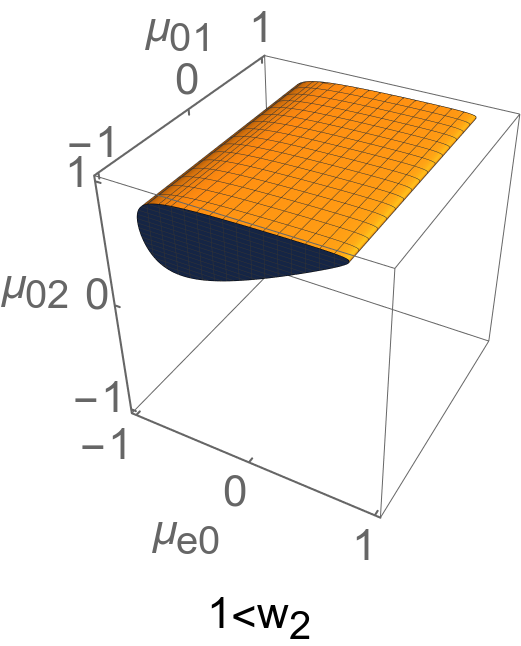}
\end{subfigure}
\caption{Same as figure \ref{fig:IntegDom_OmegaLess} but with $\omega_0 > \omega_0^\text{crit}$.}
\label{fig:IntegDom_OmegaMore}
\end{figure}




\bibliographystyle{JHEP}
\bibliography{bibliografia}

\providecommand{\href}[2]{#2}\begingroup\raggedright\begin{thebibliography}{10}

\bibitem{Mandl52}
F.~Mandl and T.~H.~R. Skyrme, {\it {The Theory of the Double Compton Effect}},
  {\em Proceedings of the Royal Society of London Series A} {\bf 215} (Dec.,
  1952) 497--507.

\bibitem{Lightman1981}
A.~P. {Lightman}, {\it {Double Compton emission in radiation dominated thermal
  plasmas}},  {\em \apj} {\bf 244} (Mar., 1981) 392--405.

\bibitem{Thorne1981}
K.~S. {Thorne}, {\it {Relativistic radiative transfer - Moment formalisms}},
  {\em \mnras} {\bf 194} (Feb., 1981) 439--473.

\bibitem{Svensson:1984MNRAS.209..175S}
R.~{Svensson}, {\it {Steady mildly relativistic thermal plasmas - Processes and
  properties}},  {\em \mnras} {\bf 209} (July, 1984) 175--208.

\bibitem{Danese:1982A&A...107...39D}
L.~{Danese} and G.~{de Zotti}, {\it {Double Compton process and the spectrum of
  the microwave background}},  {\em \aap} {\bf 107} (Mar., 1982) 39--42.

\bibitem{Burigana:1991A&A...246...49B}
C.~{Burigana}, L.~{Danese}, and G.~{de Zotti}, {\it {Formation and evolution of
  early distortions of the microwave background spectrum - A numerical study}},
   {\em \aap} {\bf 246} (June, 1991) 49--58.

\bibitem{Hu:1992dc}
W.~Hu and J.~Silk, {\it {Thermalization and spectral distortions of the cosmic
  background radiation}},  {\em Phys. Rev. D} {\bf 48} (1993) 485--502.

\bibitem{Chluba:2011hw}
J.~Chluba and R.~Sunyaev, {\it {The evolution of CMB spectral distortions in
  the early Universe}},  {\em Mon.Not.Roy.Astron.Soc.} {\bf 419} (2012)
  1294--1314, [\href{http://arxiv.org/abs/1109.6552}{{\tt arXiv:1109.6552}}].

\bibitem{McKinney:2016voq}
J.~C. McKinney, J.~Chluba, M.~Wielgus, R.~Narayan, and A.~Sadowski, {\it
  {Double Compton and cyclo-synchrotron in super-Eddington discs, magnetized
  coronae and jets}},  {\em Mon. Not. Roy. Astron. Soc.} {\bf 467} (2017),
  no.~2 2241--2265, [\href{http://arxiv.org/abs/1608.08627}{{\tt
  arXiv:1608.08627}}].

\bibitem{Sarkar:2019har}
A.~Sarkar, J.~Chluba, and E.~Lee, {\it {Dissecting the Compton scattering
  kernel I: Isotropic media}},  {\em Mon. Not. Roy. Astron. Soc.} {\bf 490}
  (2019), no.~3 3705--3726, [\href{http://arxiv.org/abs/1905.00868}{{\tt
  arXiv:1905.00868}}].

\bibitem{Chluba:2019ser}
J.~Chluba, A.~Ravenni, and B.~Bolliet, {\it {Improved calculations of
  electron-ion Bremsstrahlung Gaunt factors for astrophysical applications}},
  {\em Mon. Not. Roy. Astron. Soc.} {\bf 492} (2020), no.~1 177--194,
  [\href{http://arxiv.org/abs/1911.08861}{{\tt arXiv:1911.08861}}].

\bibitem{Sunyaev:1970er}
R.~Sunyaev and Y.~Zeldovich, {\it {The Interaction of matter and radiation in
  the hot model of the universe}},  {\em Astrophys. Space Sci.} {\bf 7} (1970)
  20--30.

\bibitem{1970CoASP...2...66S}
R.~A. {Sunyaev} and Y.~B. {Zeldovich}, {\it {The Spectrum of Primordial
  Radiation, its Distortions and their Significance}},  {\em Comments on
  Astrophysics and Space Physics} {\bf 2} (Mar., 1970) 66.

\bibitem{ChlubaThesis}
J.~{Chluba}, {\em {Spectral Distortions of the Cosmic Microwave Background}}.
\newblock PhD thesis, LMU M{\"u}nchen, Mar., 2005.

\bibitem{Chluba:2006kg}
J.~Chluba, S.~{\relax Yu}. Sazonov, and R.~A. Sunyaev, {\it {The double Compton
  emissivity in a mildly relativistic thermal plasma within the soft photon
  limit}},  {\em Astron. Astrophys.} (2006)
  [\href{http://arxiv.org/abs/astro-ph/0611172}{{\tt astro-ph/0611172}}].

\bibitem{Gould1984}
R.~J. {Gould}, {\it {The cross section for double Compton scattering}},  {\em
  \apj} {\bf 285} (Oct., 1984) 275--278.

\bibitem{Hahn:2004fe}
T.~Hahn, {\it {CUBA: A Library for multidimensional numerical integration}},
  {\em Comput. Phys. Commun.} {\bf 168} (2005) 78--95,
  [\href{http://arxiv.org/abs/hep-ph/0404043}{{\tt hep-ph/0404043}}].

\bibitem{Brown:1952eu}
L.~M. Brown and R.~P. Feynman, {\it {Radiative corrections to Compton
  scattering}},  {\em Phys. Rev.} {\bf 85} (1952) 231--244.

\bibitem{Jauch1976}
J.~M. {Jauch} and F.~{Rohrlich}, {\em {The theory of photons and electrons. The
  relativistic quantum field theory of charged particles with spin one-half}}.
\newblock Springer Science \& Business Media, 1976.

\bibitem{Jost1947}
R.~{Jost}, {\it {Compton Scattering and the Emission of Low Frequency
  Photons}},  {\em Physical Review} {\bf 72} (Nov, 1947) 815--820.

\bibitem{Jones1968}
F.~C. {Jones}, {\it {Calculated Spectrum of Inverse-Compton-Scattered
  Photons}},  {\em Physical Review} {\bf 167} (Mar., 1968) 1159--1169.

\bibitem{SandeepEtAl}
J.~Chluba, S.~Acharya, and A.~Ravenni, {\it {Thermalization of large single
  energy release in the early Universe}},  {\em {\rm in preparation}}.

\bibitem{RavenniChlubaDCAniso}
A.~Ravenni and J.~Chluba, {\it {The anisotropic double Compton emissivity}},
  {\em {\rm in preparation}}.

\bibitem{Mork1965}
K.~{Mork} and H.~{Olsen}, {\it {Radiative Corrections. I. High-Energy
  Bremsstrahlung and Pair Production}},  {\em Physical Review} {\bf 140} (Dec.,
  1965) 1661--1674.

\bibitem{Mork1971}
K.~J. {Mork}, {\it {Radiative Corrections. II. Compton Effect}},  {\em Physical
  Review A} {\bf 4} (Sept., 1971) 917--927.

\bibitem{Tsai:1972sg}
W.-Y. Tsai, L.~L. Deraad, and K.~A. Milton, {\it {Compton scattering. II.
  differential cross-sections and left-right asymmetry}},  {\em Phys. Rev.}
  {\bf D6} (1972) 1428--1438. [Erratum: Phys. Rev.D11,703(1975)].

\bibitem{Heitler:1936jqw}
W.~Heitler, {\em {The quantum theory of radiation}}, vol.~5 of {\em
  International Series of Monographs on Physics}.
\newblock Oxford University Press, Oxford, 1936.

\bibitem{Rybicki1979}
G.~B. {Rybicki} and A.~P. {Lightman}, {\em {Radiative processes in
  astrophysics}}.
\newblock New York, Wiley-Interscience, 1979.~393 p., 1979.

\bibitem{Chluba:2013kua}
J.~Chluba, {\it {Refined approximations for the distortion visibility function
  and $\mu$-type spectral distortions}},  {\em Mon. Not. Roy. Astron. Soc.}
  {\bf 440} (2014), no.~3 2544--2563,
  [\href{http://arxiv.org/abs/1312.6030}{{\tt arXiv:1312.6030}}].

\bibitem{Weldon:1991eg}
H.~A. Weldon, {\it {Bloch-Nordsieck cancellation of infrared divergences at
  finite temperature}},  {\em Phys. Rev.} {\bf D44} (1991) 3955--3963.

\bibitem{Bloch:1937pw}
F.~Bloch and A.~Nordsieck, {\it {Note on the Radiation Field of the electron}},
   {\em Phys. Rev.} {\bf 52} (1937) 54--59.

\bibitem{Jauch:1954}
J.~M. Jauch and F.~Rohrlich, {\it {The infrared divergence}},  {\em Helvetica
  Physica Acta} {\bf 27} (1954).

\bibitem{Yennie:1961ad}
D.~R. Yennie, S.~C. Frautschi, and H.~Suura, {\it {The infrared divergence
  phenomena and high-energy processes}},  {\em Annals Phys.} {\bf 13} (1961)
  379--452.

\bibitem{Lee:1964is}
T.~D. Lee and M.~Nauenberg, {\it {Degenerate Systems and Mass Singularities}},
  {\em Phys. Rev.} {\bf 133} (1964) B1549--B1562.

\bibitem{Chung:1965zza}
V.~Chung, {\it {Infrared Divergence in Quantum Electrodynamics}},  {\em Phys.
  Rev.} {\bf 140} (1965) B1110--B1122.

\bibitem{Weinberg:1965nx}
S.~Weinberg, {\it {Infrared photons and gravitons}},  {\em Phys. Rev.} {\bf
  140} (1965) B516--B524.

\bibitem{Dittmaier:1993bj}
S.~Dittmaier, {\it {Full O(alpha) radiative corrections to high-energy Compton
  scattering}},  {\em Nucl. Phys.} {\bf B423} (1994) 384--404,
  [\href{http://arxiv.org/abs/hep-ph/9311363}{{\tt hep-ph/9311363}}].

\bibitem{Chluba:2015hma}
J.~Chluba, {\it {Green's function of the cosmological thermalization problem --
  II. Effect of photon injection and constraints}},  {\em Mon. Not. Roy.
  Astron. Soc.} {\bf 454} (2015), no.~4 4182--4196,
  [\href{http://arxiv.org/abs/1506.06582}{{\tt arXiv:1506.06582}}].

\bibitem{Ensslin2000}
T.~A. {En{\ss}lin} and C.~R. {Kaiser}, {\it {Comptonization of the cosmic
  microwave background by relativistic plasma}},  {\em \aap} {\bf 360} (Aug.,
  2000) 417--430, [\href{http://arxiv.org/abs/astro-ph/0}{{\tt astro-ph/0}}].

\bibitem{Slatyer2015}
T.~R. {Slatyer}, {\it {Indirect Dark Matter Signatures in the Cosmic Dark Ages
  II. Ionization, Heating and Photon Production from Arbitrary Energy
  Injections}},  {\em ArXiv e-prints} (June, 2015)
  [\href{http://arxiv.org/abs/1506.03812}{{\tt arXiv:1506.03812}}].

\bibitem{Acharya2020JCAP}
S.~K. {Acharya} and R.~{Khatri}, {\it {CMB spectral distortions constraints on
  primordial black holes, cosmic strings and long lived unstable particles
  revisited}},  {\em \jcap} {\bf 2020} (Feb., 2020) 010,
  [\href{http://arxiv.org/abs/1912.10995}{{\tt arXiv:1912.10995}}].

\end{thebibliography}\endgroup

\end{document}